\documentclass[default,iicol]{sn-jnl}

\usepackage{natbib}

\usepackage{graphicx}%
\usepackage{multirow}%
\usepackage{amsmath,amssymb,amsfonts}%
\usepackage{amsthm}%
\usepackage{mathrsfs}%
\usepackage[title]{appendix}%
\usepackage{xcolor}%
\usepackage{textcomp}%
\usepackage{manyfoot}%
\usepackage{booktabs}%
\usepackage{algorithm}%
\usepackage{algorithmicx}%
\usepackage{algpseudocode}%
\usepackage{listings}%
\usepackage{mathtools, cuted}

\begin{document}
\title{Lift at low Reynolds number}

\author*[1]{\fnm{Lionel} \sur{Bureau}}\email{lionel.bureau@univ-grenoble-alpes.fr}
\equalcont{These authors contributed equally to this work.}

\author*[1]{\fnm{Gwennou} \sur{Coupier}}\email{gwennou.coupier@univ-grenoble-alpes.fr}
\equalcont{These authors contributed equally to this work.}

\author*[2]{\fnm{Thomas} \sur{Salez}}\email{thomas.salez@cnrs.fr}
\equalcont{These authors contributed equally to this work.}

\affil[1]{Univ. Grenoble Alpes, CNRS, LIPhy, 38000 Grenoble, France}

\affil[2]{Univ. Bordeaux, CNRS, LOMA, UMR 5798, 33400 Talence, France}

\abstract{
Lift forces are widespread in hydrodynamics. These are typically observed for big and fast objects, and are often associated with a combination of fluid inertia (\textit{i.e.} large Reynolds numbers) and specific symmetry-breaking mechanisms. In contrast, the properties of viscosity-dominated  (\textit{i.e.} low Reynolds numbers) flows make it more difficult for such lift forces to emerge. However, the inclusion of boundary effects qualitatively changes this picture. Indeed, in the context of soft and biological matter, recent studies have revealed 
the emergence of novel lift forces generated by boundary softness, flow gradients and/or surface charges. The aim of the present review is to gather and analyse this corpus of literature, in order to identify and unify the questioning within the associated communities, and pave the way towards future research.
}
\maketitle


\section{Introduction} \label{sec:intro}

  \begin{figure*}
	\centering 		
\includegraphics[width=\textwidth]{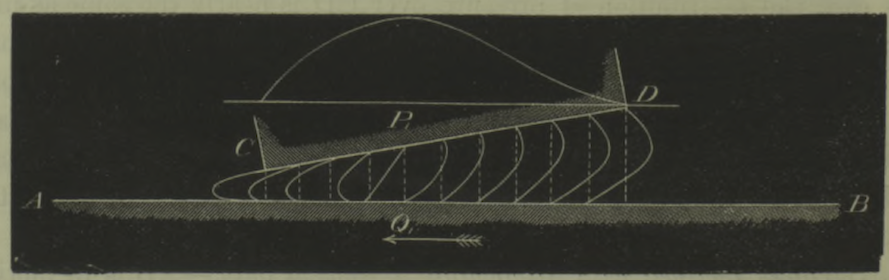}
	\caption{\textit{Reynolds' slider. The fore-aft asymmetry of a
tilted immersed slider that moves tangentially to a neighbouring rigid wall generates a difference in the hydrodynamic pressure magnitudes at the front and at the back of the slider. As a consequence, the slider experiences a net normal force. This seminal system conceived by Reynolds highlights the leitmotiv of this review: at low Reynolds numbers, a symmetry breaking in the transverse direction may generate normal forces. Figure taken from~\cite{Reynolds1886}.}\label{fig:slider}}
\end{figure*}

\subsection{Context}

We are all familiar with dynamically-induced lift forces in hydrodynamics. These are typically observed for big and fast objects, \textit{e.g.} in aeronautics or ball sports, and result from fluid inertia (\textit{i.e.} large Reynolds numbers) and a symmetry-breaking mechanism, such as wing shape or ball rotation. We define here a lift force as a force acting perpendicularly to the initial motion of the object, which is generally due to its initial acceleration or to its advection by external flows.

  While less visible in everyday life, lift effects do also exist in low-Reynolds-number flows, and often result from a key role played by the flow boundaries. Indeed, the confined hydrodynamic interaction between two objects (\textit{e.g.} a particle, a wall, etc.) or the bulk fluid-structure interaction may break the flow symmetry. This was already understood by  \citet{Reynolds1886} through his famous tilted slider. In the latter example, a lift force exists due to a fore-aft geometrical asymmetry between two immersed rigid objects in sliding relative motion (see Fig.~\ref{fig:slider}). However, for a rigid sphere moving along a rigid wall, the time reversal-symmetry of the steady Stokes equations coupled to the fore-aft symmetry of the contact warrants the absence of any emergent normal force in the problem. To overcome this impossibility, in the absence of any inertial effects, other symmetry-breaking mechanisms are thus required.

A prominent example is that of the lift induced by elastic deformations, for which we wish to bridge the gap between two aspects of the
phenomenon that were historically studied in independent contexts. These are: on the one hand, the small-gap limit where the key mechanism is the deformation of elastic surfaces mediated by the fluid separating the objects in relative motion; and, on the other hand, the large-gap limit where the deformation is directly set by an externally applied flow. In the next subsection, we shall give a flavour of these regimes, and of the transition between them, through a qualitative description of some selected works, before entering in the details in sections \ref{sec:softlub} and \ref{sec:extlift}. Interactions between soft particles, that may be brought in close contact by the flow, lead to situations where both short-range and long-range effects come into play. This situation, which is relevant to understand the structure and rheology of suspensions of soft particles, is reviewed in section \ref{sec:particleparticlelift}.

Among other mechanisms giving rise to lift, electrokinetic effects are dominant in the literature and are presented in section \ref{sec:eklift}. They involve ionic currents taking place when fluid-immersed rigid objects carrying surface charges are in relative motion. Such phenomena are typically of interest in microparticle-sorting applications, and embody an illustration of cases where non-inertial lift forces emerge between objects that are not necessarily deformable.

 Those are important mechanisms since they demonstrate that lift effects can be triggered at microscopic and biological scales through a smart role of boundaries. The three mechanisms discussed so far will be the topic of the main sections of the present review. Other mechanisms, unexplored yet or marginally explored, will be also briefly mentioned in the perspectives of the concluding section.  Furthermore, we will only deal with passive mechanisms here, although fluid-structure interactions involving deformations under active forces are of growing interest in the blooming field of microswimming mechanics \cite{Trouilloud2008,Nambiar2022}.

  \begin{figure}
	\centering 		
\includegraphics[width=\columnwidth]{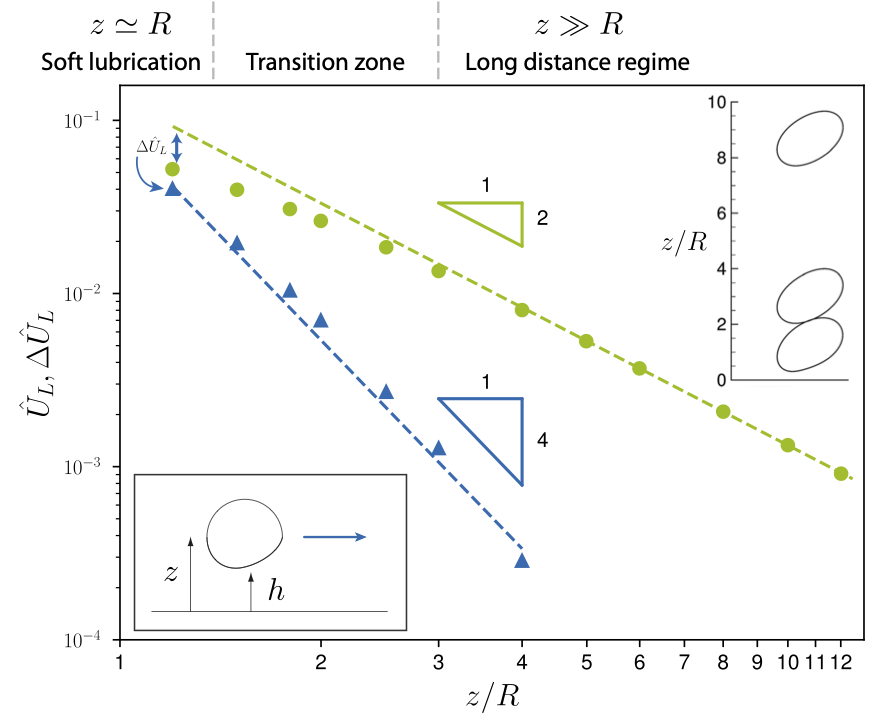}
	\caption{\textit{Transition between the long-distance regime and the near-wall regime. The lift velocity $\hat{U}_L$ of a lipid vesicle sheared above a wall is plotted against its distance $z$ from the wall, located at $z=0$ (green dots). The scale is set by the typical size $R$ of the particle. Blue triangles show the near-wall correction $\Delta \hat{U}_L$ that measures the difference between the  $\sim 1/z^2$  long distance contribution and the velocity $\hat{U}_L$. This contribution scales as $1/z^4$. Top inset shows three shapes at three different positions. Data from \cite{zhao11}, Fig.2, for a vesicle with reduced volume of 0.95 and no viscosity contrast -- see section \ref{sec:extlift} for definitions. Bottom inset shows the notations $z$ and $h$ used along this review. They denote respectively the wall to center of mass distance and the wall to particle enveloppe distance.\label{fig:miseenbouche}}}
\end{figure}

\subsection{Prelude: from soft lubrication to bulk elastohydrodynamics}

Lipid vesicles, which are drops enclosed by a lipidic membrane, have been studied both in the vicinity of a wall and far from it, when deformed by a shear flow.  We highlight here a selection of studies on these objects that explored in particular the transition between the near-wall and the far-wall regimes, that are identified by comparing the particle-to-wall distance with the typical particle size.

The lift force upon detachment acting on a heavy, quasi-spherical, lipid vesicle lying on a substrate was determined experimentally by  \citet{lorz00} using an interferometric technique, that allows one to accurately determine the gap profile between a particle and a substrate, if it is on the order of some hundreds of nanometers. A theoretical interpretation of their results, displaying quantitative agreement with the experimental data, was proposed by  \citet{seifert99}, while in a letter published simultaneously in the same journal \citet{cantat99} also provided an expression for the lift force. Both groups considered a vesicle pinned to a rigid substrate by an adhesive potential, and determined, using similar approaches, the lift force acting on the vesicle as it is pinned to the wall at a given distance $h$ from it (given by the location of the potential energy minimum), while being still able to deform and open an asymmetric gap between its surface and the wall.  Cantat and Misbah considered a 2D vesicle whose asymmetry is essentially described by its front and back curvatures (allowing for feedback between gap shape and flow stress through the curvature energy of the membrane), and found a $h^{-1/2}$ dependence for the lift force. Seifert considered a 3D vesicle but assumed a linearly increasing gap, and used the adhesion energy as a control parameter, rather than the curvature energy. He found a $h^{-1}$ dependence for the lift force. In both models, forces depend quadratically on the shear rate, a signature of soft-lubrication-based mechanisms that will be described in section \ref{sec:softlub}. Interestingly, Cantat and Misbah also ran numerical simulations of the lift force as a function of shear rate and highlighted a transition between this quadratic regime, when the vesicle is pinned, to a regime where the force is linear with the shear rate, when the vesicle is detached. This points to the fact that while the opening of an asymmetric gap is, in the near-wall regime, the result of local balance between flow stress and particle elasticity, the shape of the vesicle that is far from the wall is governed by the sole interaction with the bulk flow. This regime of shear-induced lift will be discussed in section \ref{sec:extlift}. More recently, a similar case of coupling between adhesion forces and shear-induced lift has been studied for drops on a deformable polymer brush, leading to a complex phase diagram \cite{leong21}. We stress here that these shear-induced lift phenomena require the considered object to be deformed, which, in the vanishing-Reynolds-number context that we focus on, implies rather soft objects such as biological materials, drops and artificial capsules. We will exclude here the particular case of filaments, which have generally a very complex shape dynamics even in the absence of walls, thus rendering difficult our quest for universal mechanisms.

 The transition from the near-wall regime to the far-field regime can also be illustrated through the work of Zhao {\it et al.} who simulated a lipid vesicle of typical size $R$ whose center of mass is placed at different positions $z$ above a wall, and sheared by the flow \cite{zhao11}. As shown in Fig. \ref{fig:miseenbouche}, a vesicle sheared far from the wall adopts a tilted shape characterized by a point-wise symmetry, which is reminiscent of the symmetry of the imposed flow, if the presence of the wall is omitted. The vesicle lifts away from the wall with a velocity $\hat{U}_L(z)$ that scales like $1/z^2$, which we will show to be a generic behaviour. Here, the presence of the wall induces a flow that repels the vesicle, whose shape is essentially dictated by the shear flow. If the vesicle is placed closer to the wall, a correction to the far-field lift velocity must be considered, which we show here to scale like $1/z^4$. This higher-order term (with respect to the inverse distance $z^{-1}$) comes together with a change in shape of the vesicle, as a result of the interaction with the wall, mediated by the fluid. Indeed, as seen in the inset of Fig. \ref{fig:miseenbouche}, the bottom surface of the vesicle is deformed. In this regime, the wall not only induces a lift on the vesicle by inducing a mean hydrodynamic stress on it, but it also modifies its shape, which in turn modifies the way the flow pushes away the vesicle. If, far from the wall, a description in terms of center-of-mass-to-wall distance $z$ is relevant, considering the near-wall regime ($z\simeq R$ and below) requires to carefully study the shape of the gap between the particle and the wall. The relevant parameter becomes $h$, the minimal value of the gap thickness. When $h/R \ll1$, the framework to analyze the phenomenology is the soft-lubrication theory, which allows one to solve the elastohydrodynamics (EHD) problem when the hydrodynamic part of it is described in the lubrication approximation.
This soft-lubrication coupling becomes the dominant mechanism giving rise to lift in the absence of imposed flow that deforms the particles.

\section{Soft-lubrication lift}
\label{sec:softlub}
\subsection{Context}

Soft and wet contacts are widespread in nature and technology. Their rich history in science and engineering involves issues and scales as diverse as the lubrication of roller bearings~\cite{Hamrock1977,Greenwood2020} after the industrial revolution, or the catastrophic geological landslides~\cite{Campbell1989}. The properties of these contacts implicate the coupling between the local hydrodynamic pressure induced by the confined fluid flow and the deformation of the confining solids. 

For the high speeds and relatively stiff surfaces associated with industrial devices, non-Newtonian lubricant effects (\textit{e.g.} piezoviscous and thermoviscous behaviours) play an important role on the EHD coupling~\cite{Fillot2011} and require multiscale numerical modeling~\cite{Ewen2021}. In the same spirit, the nonlinear rheology of polymer solutions may itself introduce the necessary symmetry breaking that gives rise to particle lift -- even for rigid boundaries~\cite{Leal1979}. In such a case, particles would \textit{e.g.} migrate to regions of low normal-stress differences (low-shear-rate regions). All these interesting non-Newtonian features may be considered as corrections to the main mechanism at stake here, despite some potential interest for sorting strategies~\cite{Yang12}, and are thus not addressed in details in the following. 
 
Recently, EHD \textit{i.e.} the coupling between flows and elastic boundaries, gained attention in the context of confined, soft and biological matter, where very compliant solids and tiny length scales are common~\cite{Brochard2003}. In fact, this coupling could conceivably play a crucial role in the motion of various physiological and biological entities. Examples include \textit{e.g.} the incredible frictional properties of mammalian joints~\cite{Mow1984} through the fine interplay between soft cartilage and viscous synovial fluid, or the crucial influence of vessel boundaries on the motion of deformable red blood cells
~\cite{goldsmith71}. The normal motion towards a soft wall was investigated in particular~\cite{Balmforth2010,Leroy2011,Wang2017,Karan2018}, with a special attention given to the collision~\cite{Davis1986} and rebound~\cite{Gondret1999,Tan2019} properties.

Furthermore, through surface-forces apparatus (SFA)~\cite{Leroy2011,Leroy2012,Villey2013,Wang2015,Wang2017b,Wang2018b} and atomic-force microscopy (AFM)~\cite{Chan2009,Vakarelski2010,Kaveh2014,Guan2017,Wang2018,Basoli2018}, the near-contact EHD (termed soft-lubrication in the following) coupling offers an alternative strategy for micro and nanorheology of fragile soft materials, with the key advantage of avoiding any solid-solid adhesive contact that could alter their properties. 
\begin{figure*}[t]
\begin{center}
\includegraphics[width=0.9\textwidth]{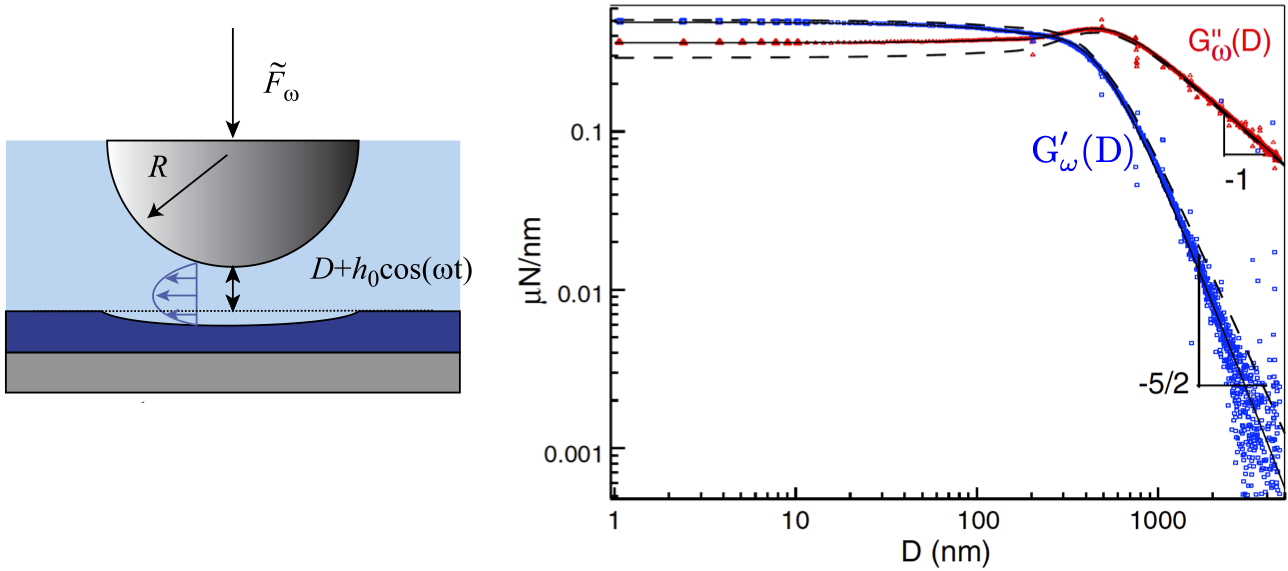}
\caption{\textit{(left) In a surface-forces apparatus (SFA), a flow between an oscillating sphere and an elastic film is created. (right) Real (blue) $G'_{\omega}$ and imaginary (red) $G"_{\omega}$ parts of the force-distance impedance response $G_{\omega}(D)$ obtained for an elastomer (crosslinked PDMS) and compared to soft-lubrication theory (dashed lines). Figure adapted from~\cite{Leroy2012}.}}
\label{charlaix}
\end{center}
\end{figure*}
Indeed, as shown in Fig.~\ref{charlaix}, the motion of a spherical probe in a viscous fluid, and near a soft material, generates a lubrication pressure field that can in turn deform the soft material. As a consequence, the lubrication gap is altered as compared to the rigid case, and the impedance response is directly affected, with the appearance of elastic contributions.  Rallabandi recently reviewed the general case of near-contact fluid elastic interactions \cite{rallabandi2024} while here we focus on the emergence of normal forces.
 
 \subsection{The key mechanism}
Despite the irrelevance of inertia, a soft-lubrication lift force emerges for elastic bodies moving past each other within a viscous fluid. Essentially, any fore-aft-symmetric object moving within such a fluid and along a nearby soft wall is repelled from the latter by a dynamically-generated emergent normal force. 
 \begin{figure}[t!]
	\centering 		
\includegraphics[width=\columnwidth]{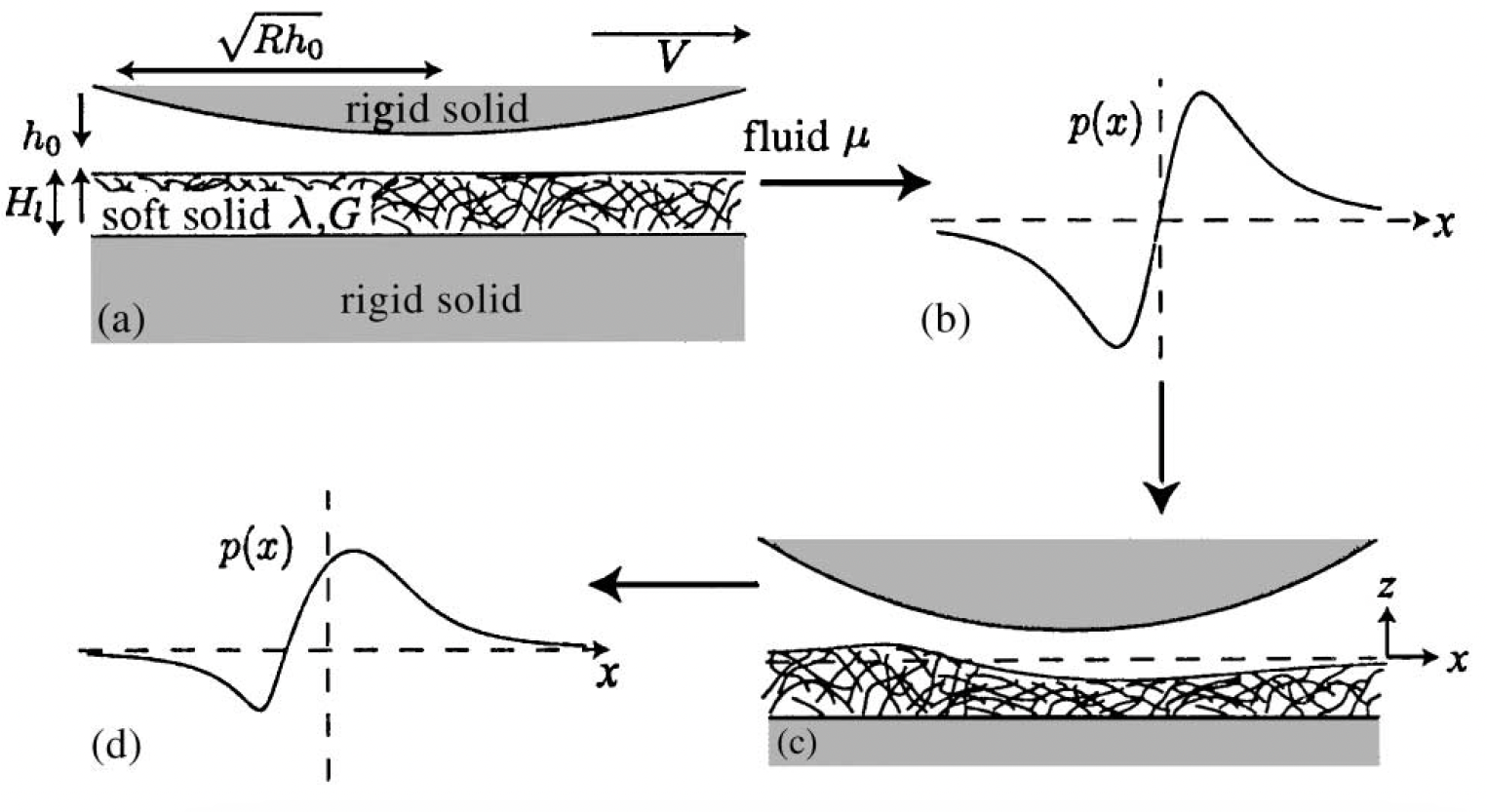}
	\caption{\textit{Principle of the soft-lubrication lift force. For a non-deformed wall (a), the classical lubrication pressure $p(x)$ induced in the viscous fluid by the tangential motion of a sphere at velocity $V$ is antisymmetric in the transverse direction $x$ (b), resulting in a null net force (integral of the pressure along $x$) in the normal direction $z$. In contrast, a soft surface is deformed by the pressure field (c). The latter then loses its symmetry (d), which results in a finite emergent normal force: the soft-lubrication lift force. Figure adapted from~\cite{skotheim2004}.}}.\label{fig:Maha}
\end{figure}
This force intimately arises from a symmetry breaking in the contact shape (see Fig.~\ref{fig:Maha}), and thus the associated flow fields, due to the EHD coupling described above. Qualitatively, the elastic deformation induced by the hydrodynamic pressure generates a self-sustained asymmetric contact similar to the one in Reynolds' rigid slider (see Fig.~\ref{fig:slider}), and thus a normal force. This effect is well known at macroscopic scales, for relatively rigid materials such as car tires undergoing aquaplaning, or industrial roller bearings getting deformed in operating motors and machines. 
 
Moving on to the context of mesoscale physics and soft matter, the earliest theoretical descriptions of such a soft-lubrication lift effect are the ones by~\citet{Coyle1988},
~\citet{Dowson1992}, ~\citet{Lequeux1992}, as well as ~\citet{Sekimoto1993}, to the best of our knowledge. The underlying motivation behind these similar approaches is the calculation of forces between
soft curved surfaces undergoing shear, which are important for the
interpretation of SFA measurements, the physics of cartilage, and the rheology of a variety of complex fluids such as suspensions of colloidal particles protected by grafted or adsorbed
polymer chains, suspensions of gel microparticles, or polymer emulsions and alloys on
certain time scales. 

The general idea can be illustrated from \textit{e.g.}~\cite{Sekimoto1993}, through the calculation of the soft-lubrication interaction between a cylindrical object of radius $R$ moving at transverse velocity $V$ past and nearby a flat wall (as in Fig. \ref{fig:Maha}), within a viscous fluid of dynamic viscosity $\mu$, both solids being covered by polymer brushes. The latter are modeled as identical thin linear-elastic compressible layers. 

\begin{figure}[t!]
	\centering 		
\includegraphics[width=\columnwidth]{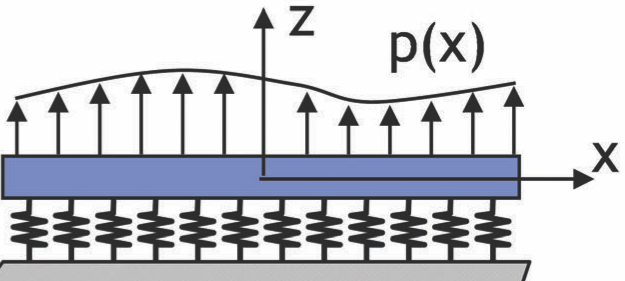}
	\caption{\textit{Schematic involving a Winkler's foundation, \textit{i.e.} a model and simple type of elastic substrate characterized by an assembly of parallel, independent and identical springs, leading to a local and linear response to the external pressure field $p(x)$. For thin enough, compressible elastic layers, such a toy model even provides a quantitative description of their elastic Green's function. Figure adapted from~\cite{Dillard2018}.}}.\label{fig:winkler}
\end{figure}

A thin linear-elastic compressible layer with Lam\'e coefficients of similar magnitudes can be mapped onto a Winkler's foundation, \textit{i.e.} a mattress of independent springs with a local and linear response to the external pressure field $p(x)$ (see Fig.~\ref{fig:winkler}). 
Interestingly, the Winkler's foundation was proven to be of great modelling power~\cite{Dillard2018}. In such a description, the normal deformation field is given by $\delta(x)=-Lp(x)/(2G)$, where we introduced an effective shear modulus $G$ as well as an effective thickness $L$ of the mattress, and where we assumed for simplicity a full compressibility (\textit{i.e.} vanishing Poisson ratio). Near its minimum $h_0$, the steady-state fluid-gap profile $h(x)$, along the transverse direction $x$ of motion, is well approximated by a parabola (\textit{i.e.} second-order development of a spherical contact near the apex) corrected by the elastic deformation of the elastic layer induced by the hydrodynamic pressure. It thus follows that:
\begin{equation}
h(x)\simeq h_0+\frac{x^2}{2R}+\frac{L}{2G} p(x)\ .
\label{gap}
\end{equation}

\paragraph{Scaling analysis}
The main idea is then based on a hierarchical scale separation, by considering that the elastic deformation is small compared to the fluid-gap thickness, which is itself small compared to the cylinder radius. By invoking the steady Stokes equations in the lubrication approximation, it follows that the leading-order pressure magnitude scales as $\sim\mu V\ell/h_0^{\,2}$, where $\ell=\sqrt{2Rh_0}$ is the characteristic horizontal length scale, given by the Hertz-like \textit{hydrodynamic radius} emerging from the parabolic approximation in Eq.~\eqref{gap}.  In such a framework, and in addition to the lubrication parameter $h_0/R\ll1$, one finds $\kappa\sim\mu \sqrt{R}VL/(Gh_0^{5/2})$ -- the dimensionless compliance -- as a second natural small parameter of the problem. Then, an expansion of the soft-lubrication flow problem is performed at order $1$ in $\kappa$. The zeroth-order contribution corresponds to the purely rigid case with gap profile $h^{(0)}(x)\simeq h_0+x^2/(2R)$ and a zeroth-order pressure field $p^{(0)}(x)$ that can be computed analytically, and that is found to be antisymmetric in $x$ (Fig. \ref{fig:Maha}(b)). This is expected in view of the time-reversal symmetry of the steady-Stokes equations and the fore-aft symmetry of the contact shape. As the normal force per unit length $F_z$ exerted on the cylinder is dominated, in the lubrication approximation,  by the pressure contribution (\textit{i.e.} the ratio between the viscous shear stress and the pressure is of order $h_0/\ell\ll1$), the latter antisymmetry of $p^{(0)}(x)$ implies to evaluate the first correction $p^{(1)}(x)$ induced by the elastic deformation. The magnitude of the latter scales as $\sim \kappa (\mu V\ell/h_0^{\,2})$. Therefore, the resulting normal force per unit length reads, at order 1 in $\kappa$:
\begin{equation}
\label{lift}
F_{z}\simeq\int_{-\infty}^{\infty}\textrm{d}x\,p^{(1)}(x)\sim\frac{\mu^2 V^2R^{3/2}L}{Gh_0^{7/2}}\ .
\end{equation}
Interestingly, one sees that this soft-lubrication lift force increases with the viscosity of the fluid, driving velocity, compliance, contact area, and confinement. Then, \citet{Sekimoto1993} confront the theoretical predictions with the result of force
measurements under shear between surfaces covered with grafted polymer chains. While the observed normal forces in these experiments are often attributed to brush swelling due to external flows, the authors quantitatively argue here that the brush-deformation-induced soft-lubrication lift force is instead the dominant mechanism behind the common observations.
 
\paragraph{Soft-lubrication theory}
Here, as an illustration of the typical method for the readers, we aim at retrieving the scaling result above quantitatively. We place ourselves in the rest frame of the cylinder (see Fig.~\ref{fig:Maha}). We introduce the fluid velocity field $u(x,z)$ along $x$, and the dimensionless variables: $z=Z h_0$, $h=H h_0$, $x=X\ell$, $u=UV$, $p=P\mu V\ell/h_0^2$, and $F_z=\mathcal{F}_Z\mu V\ell^2/h_0^2$. Hence, the gap profile given by Eq.~(\ref{gap}) is non-dimensionalized as:
\begin{equation}
\label{NDgap}
H(X)=1+X^2+\kappa P(X)\ , 
\end{equation}
where:
\begin{equation}
\kappa=\frac{L\mu VR^{1/2}}{\sqrt{2}Gh_0^{5/2}}\ .
\end{equation}

In the lubrication approximation where $h_0\ll R$, the incompressible steady Stokes equations reduce to~\cite[]{Reynolds1886,Batchelor1967,Oron1997}:
\begin{equation}
\label{stokes}
\partial_{ZZ}U=\partial_XP,
\end{equation}
with $\partial_ZP=0$.
In addition, we impose no-slip boundary conditions, through $U(X,Z=-\kappa P)=-1$ and $U(X,Z=H-\kappa P)=0$. Solving Eq.~(\ref{stokes}) with these boundary conditions, and invoking the condition of volume conservation yields the Reynolds equation:
\begin{equation} 
\label{eqgen3}
\partial_X\left(H^3\partial_XP+6H\right)=0\ ,
\end{equation}
where we recall that $H$ depends on $\kappa$ (see Eq.~(\ref{NDgap})). Solving Eq.~(\ref{eqgen3}) with vanishing pressure in the far field, one can then calculate the dimensionless normal force (per unit length) exerted on the cylinder, through: 
\begin{equation}
\label{dragp}
\mathcal{F}_Z=\int_{-\infty}^{\infty}\textrm{d}X\ P(X)\ .
\end{equation}
Since $\kappa \ll 1$, perturbation theory~\cite{skotheim2004} using $P\simeq P^{(0)}+\kappa P^{(1)}$, allows one to integrate Eq.~(\ref{eqgen3}) at first order in $\kappa$, eventually leading to the dimensionless lift force:
\begin{equation}
\label{3drag}
\mathcal{F}_Z\simeq\frac{3\pi\kappa}{8}\ ,
\end{equation}
and thus providing Eq.~(\ref{lift}) as well as the missing prefactor therein.

\subsection{Theoretical developments}
Let us make a few comments about Eq.~\eqref{lift}. This typical asymptotic expression of the soft-lubrication lift force per unit length relies on several assumptions: a 2D problem, a pure linear and local compressible elastic rheology, a vanishing compliance, a near-contact/confinement situation, a parabolic contact shape, etc. It is thus expected to find important modifications of the lift force in more complex or realistic situations.  First of all, while the perturbative/asymptotic nature of the approach is expected to hold at small elastic deformations, through the explicit factor $\sim(\mu V)^2/G$ in the force expression, dimensionality and geometry are expected to modify the dependencies on the various length scales of the problem. Similarly, the exact elastic rheology (compressible vs incompressible, thin vs thick) will modify the constitutive response between the pressure $p(x)$ and the elastic deformation $\delta(x)$. Indeed, while a linear response is expected to hold at small deformations, the simple Hookean proportionality relation $\delta(x)/L \sim p(x)/G$ may be replaced by a nonlocal relation of the type: 

 \begin{figure}[t!]
	\centering 		
\includegraphics[width=\columnwidth]{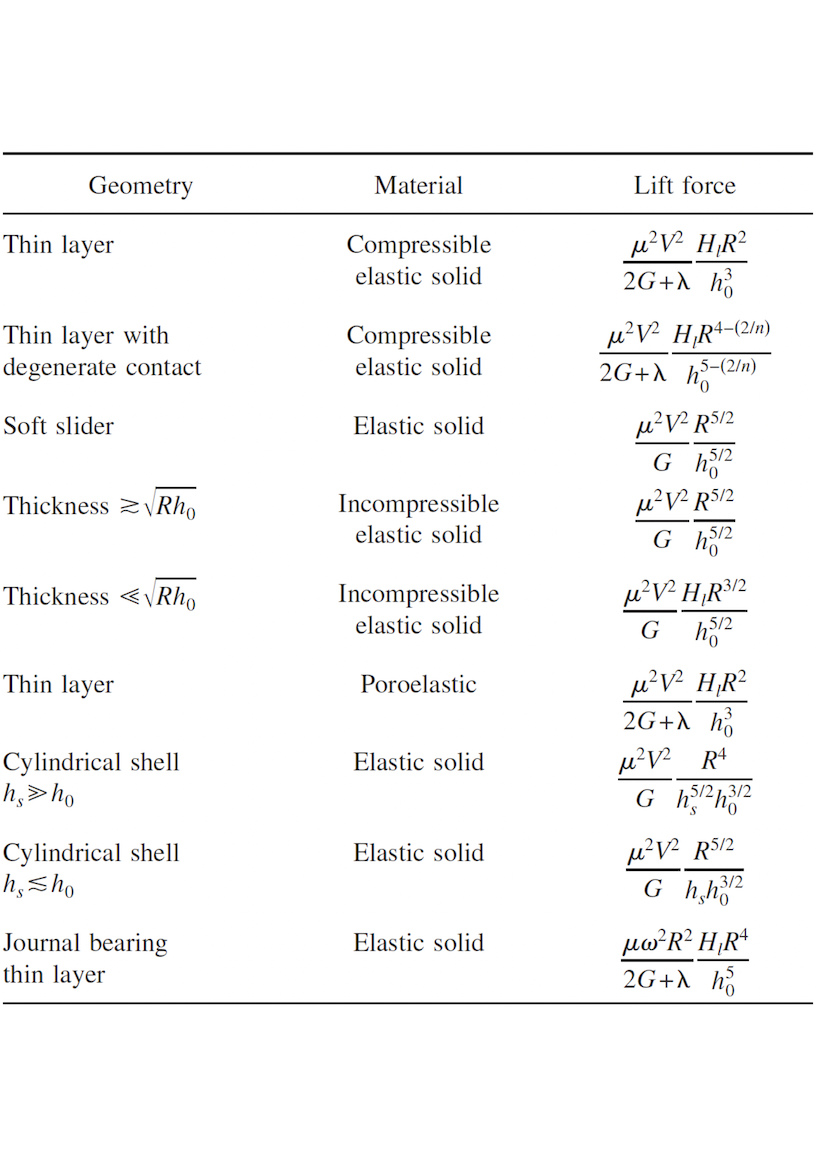}
	\caption{\textit{Different scalings of the soft-lubrication lift force for a sphere in 3D, for various geometries and rheologies of the elastic substrate~\cite{Skotheim2005}. Here, $\mu$ is the shear viscosity of the lubricant, $V$ the relative tangential speed, $\omega$ the angular speed, $n$ the contact-degeneracy parameter, $H_{\textrm{l}}$ the elastic substrate thickness, $R$ the sphere radius, $G$ the shear modulus of the elastic substrate, $\lambda$ its first Lam\'e coefficient, $h_0$ the fluid-gap thickness (noted $h$ in this manuscript) and $h_{\textrm{s}}$ the shell thickness. Note that notation $\lambda$ will be used for viscosity contrast in the rest of the manuscript.}} \label{tab:Maha}
\end{figure}
\begin{equation}
\delta(x) \sim \frac{1}{G}\int_{-\infty}^{\infty}\textrm{d}x'\,g(x-x')p(x')\ ,
\end{equation}where $g$ is the dimensionless elastic Green's function (in a 2D description here), that simply reduces to a Dirac distribution in the Winkler's case discussed above. Qualitatively, $\delta$ is still the linear response to the source $p(x)$ with a magnitude set by the compliance $1/G$ (and even a proportionality in Fourier space). Quantitatively, we expect differences depending on the exact Green's function characterizing the response. To go one step further along this line of thought, substrate viscoelasticity and poroelasticity are expected to add one or several new time scale(s) in the problem, rendering the response time-dependent, including memory effects. Similarly, large deformability and/or elastic nonlinearities may induce a saturation or even a non-monotonic behaviour of the force with gap distance, beyond the small-deformation scaling in Eq.~\eqref{lift}. Finally, adding non-Newtonian effects, or conservative surface forces, such as van der Waals forces and screened electrostatic interactions, is expected to lead to non-trivial effects and coupling with the EHD picture above. One thus realizes that there was room and need for further theoretical developments around Eq.~\eqref{lift}.

Perhaps the most emblematic example of such developments, is the series of work by \citet{skotheim2004,Skotheim2005}. Therein, a systematic exploration of various non-conforming and conforming contact geometries and elastic responses, in 2D and 3D, was carried out analytically and numerically. This is exemplified in Fig. \ref{tab:Maha}, with a collection of asymptotic lift-force scalings that are valid at large-enough distance (\textit{i.e.} weak deformation) for the particular case of a 3D sphere. This body of work applies the same soft-lubrication framework as the one introduced above, and employs as well numerical resolutions to go beyond the scaling expressions and the small-compliance limit. Interestingly, thanks to the numerical resolution, a maximum in the lift-force-vs-gap-distance behaviour was found in some cases. Note that a capillary version of soft lubrication, analogous to the elastic one at stake here, was not addressed therein, but was studied previously in the context of rising bubbles~\cite{smart91,sugiyama2010}.

\begin{figure}
	\centering 		
\includegraphics[width=\columnwidth]{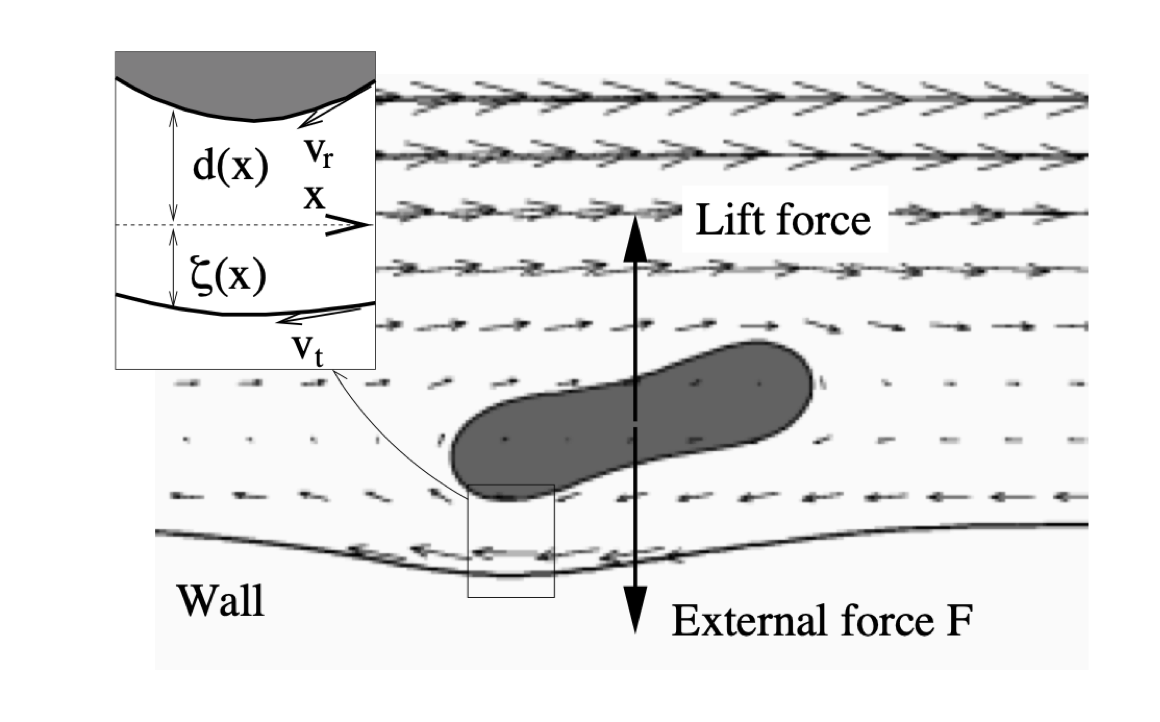}
	\caption{\textit{A deformable particle in a shear flow is elongated and tilted, resulting in a lift away from the wall. In the meantime, deformation of the neighboring elastic wall also induces lift, as in Fig. \ref{fig:Maha}. Figure adapted from~\cite{Beaucourt2004}.}\label{fig:beaucourt}}
\end{figure}

Nearly at the same time, and importantly, \citet{Beaucourt2004} understood the importance of such a lift force in a biophysical context. These authors addressed in particular the case of vesicles, as model biological elastic microparticles, during their motion in water near soft glycocalyx layers. Putting numbers on the lift expression, they found forces with magnitudes lying in the physiological range. This work thus highlights the potential importance of such soft-lubricated couplings for the dynamics of red blood cells, and thus biological processes that are essential to life. Note that this article is one of the few including both deformation of the substrate and deformation of the particle by the shear flow (Fig. \ref{fig:beaucourt}).

To go beyond  scaling symbols in the soft-lubrication lift expression for a sphere in 3D is a more intricate task. An elegant solution based on Lorentz's reciprocal theorem was sketched by Stone \textit{et al.} during an oral communication at the 2004 APS-DFD meeting~\cite{Stone2004}. It was later on systematically explored by~\citet{Urzay2007} for the problem of a sphere translating and rotating near a thin compressible elastic layer. Later on, \citet{Urzay2010} generalized the scope to the added role of DLVO intermolecular interactions. There, the competition of the hydrodynamic, intermolecular and deformation effects leads to forces which do not scale
linearly with the velocity, and produce a non-additivity of the intermolecular effects. Mainly, the intensity of the repulsive forces is reduced while the intensity of the
attractive forces is increased, collectively leading to an effective and reversible EHD adhesion scenario. Besides, a more exotic irreversible EHD adhesion regime was also found. Elastohydrodynamic corrections 
to the DLVO framework for the critical coagulation concentration of electrolytes were obtained too.

Beyond global quantities, such as the net normal lift force, local details on the contact shape matter as well, such as the fore-aft asymmetry in the space-dependent elastic deformation profile.  Furthermore, in 2D, the self-similar properties of the shape of the soft-lubricated contact zone in a high-loading case were investigated by~\citet{Snoeijer2013}. Asymptotic results for the liquid-gap thickness below a soft sphere in 3D pressed against a hard wall were shown to agree with both experimental and numerical data. Later on,~\citet{Essink2021} managed to obtain analytical scaling laws in the high-loading regime. In this work, the authors described various regimes of soft lubrication for two-dimensional cylinders in lubricated contact with compliant walls. They addressed and connected the limits of small and large entrainment velocities, near thin elastic coatings, both compressible
and incompressible. The analysis relies intimately on the introduction of an elastohydrodynamic boundary layer that appears at the edge of the contact region. Importantly, in order to identify the proper regime in a given experimental setting, this theoretical work reveals the importance of correctly estimating the ratio between the elastic deformation and the fluid-gap thickness, as well as the ratio between the thickness of the elastic material and the hydrodynamic radius.

So far, the problems studied mostly involved a time-independent fluid-gap thickness at zeroth-order in dimensionless compliance $\kappa$, and a transverse velocity. The case of a more general, time-dependent, but still prescribed, motion in 2D or 3D near a Winkler's foundation was addressed analytically and numerically by \citet{Weekley2006}. 
When the particle moves from rest towards the wall, fluid trapping beneath the particle leads to an overshoot
in the normal force on the particle, with trapping at
early times and fluid draining at late times. When
the particle is pulled from rest away from the wall, a transient adhesive normal force emerges. When a cylinder moves from
rest transversely along the wall, an overshoot in the transverse drag appears. However, the case of a free particle immersed in a viscous fluid and near a soft wall, with all degrees of freedom allowed, is relevant to experiments and needed to be addressed. The associated leading-order soft-lubrication interaction matrix was derived by \citet{Salez2015} in 2D, and later on by \citet{Bertin2021} in 3D. Interestingly, a counterintuitive set of fluid-inertial-like solutions emerges at low Reynolds number. These encompass: Magnus-like effects, enhanced sedimentation, adhesive-like EHD forces, roll reversal, oscillations, etc. In addition, the existence of a spontaneous soft-lubrication torque, at next (\textit{i.e.} second) order in $\kappa$, was revealed in 2D by~\citet{Rallabandi2017}, for compressible and incompressible settings.

We have focused on purely elastic materials in the description above. In such a framework, the softer the material, the larger the effect, until an optimum or saturation eventually occurs. This suggests to employ rather soft materials in practice. However soft gels and elastomers are inevitably prone to poroelastic and viscoelastic effects. While the former have been sketched in the lift context~\cite{Feng2000,skotheim2004,Skotheim2005}, the latter needed to be incorporated in details. \citet{Pandey2016} thus analyzed soft-lubricated contacts
with viscoelastic walls. In particular, the authors focused on three canonical viscoelastic descriptions, namely: Kelvin-Voigt,
standard linear, and power-law rheologies. They showed how viscoelasticity modifies the contact properties when the time scales of both the substrate and the driving become comparable. Mainly, they found modified asymptotic scaling laws for the lift force, indicating a decrease of the magnitude of the EHD effect due to inner viscous contributions within the viscoelastic material. Later on, \citet{Kargar2021} employed Lorentz's reciprocal theorem to derive a general integral relation between the soft-lubrication lift force and the linear response function of the soft substrate. They first analyzed the lift force as a function of Poisson's
ratio and thickness of the elastic material. Moreover, they found a superposition of steady and oscillating modes for a lubricated object moving near a viscoelastic material. The amplitudes and phases of these modes contain information about the
elastic and viscous components of the material response, thus opening the way to the fine characterization of the mechanical properties of materials via lift force measurements. Compared to normal mode excitations~\cite{Chan2009,Vakarelski2010,Leroy2011,Leroy2012,Villey2013,Kaveh2014,Wang2015,Wang2017b,Guan2017,Wang2018,Basoli2018,Wang2018b}, the interest of the transverse mode for such a purpose is rooted in Eq.~\eqref{lift}, where a $\sim V^2$ dependence appears. Therefore, with a sinusoidal excitation, a frequency doubling is expected, thus enabling the use of a region of the spectrum that is distinct from the one at the driving frequency. 

As introduced above, Winkler's foundation is the simplest linear and local elastic model (see Fig.~\ref{fig:winkler})~\cite{Dillard2018}. Since it avoids the complication of nonlocal responses associated with elastic materials, it is often used as a simplified model for thought. A natural question emerging from that is how valid such a model is to describe actual physical systems, with a particular focus on the lift problem. In particular, in the limit of strictly
incompressible and thin elastic layers, one expects an infinite
resistance to deformation, and hence the Winkler's approach breaks down. \citet{Chandler2020} provided an answer to such a question by formally deriving
a lift force that interpolates between the Winkler and
incompressible limits for thin elastic layers. They found that the
applicability of the Winkler model is
not determined by the value of the Poisson ratio alone,
but by some compressibility parameter that combines
the Poisson ratio with a measure of the layer
slenderness, which depends on the problem
under consideration. Essentially, for Poisson ratios strictly smaller than 0.5, the crossover to Winkler's model as the thickness is reduced is rooted in the elastic Green's function itself~\cite{Leroy2011,Kargar2021}. 
 
Finally, the effective compliance of a material, and the lift force as a consequence, can be increased tremendously by using slender geometries, such as membranes and plates. The EHD coupling in such systems was addressed by Daddi-Moussa-Ider and collaborators~\cite{Daddi2017,Daddi2018}. In the first article, the authors computed the leading-order frequency-dependent translational and rotational mobilities of an axisymmetric particle immersed in a viscous fluid and moving near an elastic cell boundary allowed to stretch and bend. The authors found that the translation-rotation coupling mobility is primarily determined by bending, whereas shearing (\textit{i.e.} wall-bounded shear flow) mostly affects the rotational mobility. In the second article, the authors derived the lift force exerted on a rigid spherical particle translating parallel to a finite-sized membrane. Specifically, the Lorentz reciprocal theorem was employed, as well as a perturbative expansion for small deformations of the membrane. The authors reported interesting attractive and repulsive regimes depending on the dominant elastic mode (\textit{i.e.} shearing vs bending) at play.

As a concluding remark, at leading order in dimensionless compliance, we expect no qualitative difference between the two dual situations of: i) a rigid particle near a soft wall; and ii) a soft particle near a rigid wall. This is reminiscent of the situation in dry elastic contacts~\cite{Johnson1985,Maugis2000}. 

 \subsection{Experimental pieces of evidence}
Despite the above abundant theoretical literature, experimental evidence for such a soft-lubrication lift force in soft matter is recent and scarce.

\begin{figure}
\begin{center}
\includegraphics[width=\columnwidth]{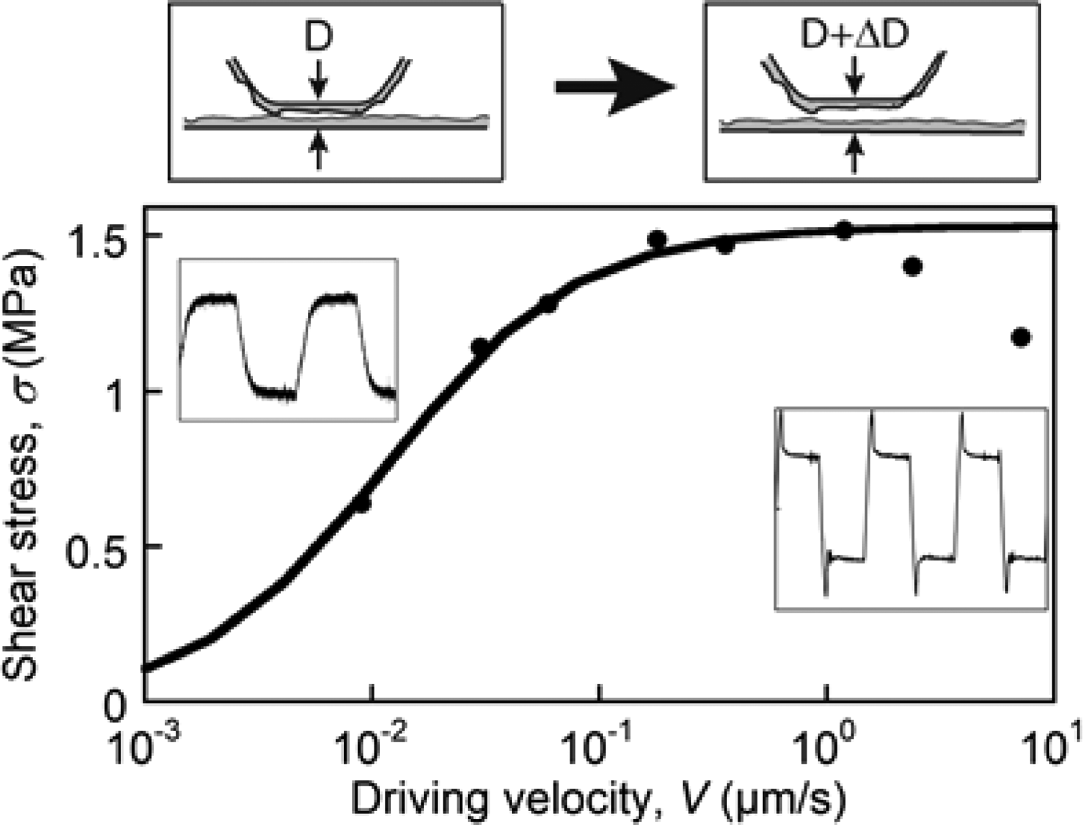}
\caption{\textit{Shear stress as a function of driving velocity measured with a SFA covered by strongly adhesive polyelectrolyte layers and in presence of a lubricant. Figure taken from~\cite{Bouchet2015}.}}
\label{bouchet}
\end{center}
\end{figure}

A possible preliminary qualitative observation may have been reported in the context of smart lubricants and adsorbed polyelectrolytes by \citet{Bouchet2015}. The authors investigated the lubricant properties of a strong polyelectrolyte, in aqueous solutions of
different salt concentrations. They first studied how the morphology of the adsorbed layer could be modified by increasing the salt
concentration. Then, a complex velocity dependence of the friction was observed, with a
maximum value at intermediate velocities and even some hysteresis. A progressive increase in separation between the rubbing
surfaces with velocity was also observed (see Fig.~\ref{bouchet}). 
These observations were qualitatively discussed in terms of an hypothetical indication of the presence of a soft-lubrication lift force. 

\begin{figure*}[t!]
\begin{center}
\includegraphics[width=15cm]{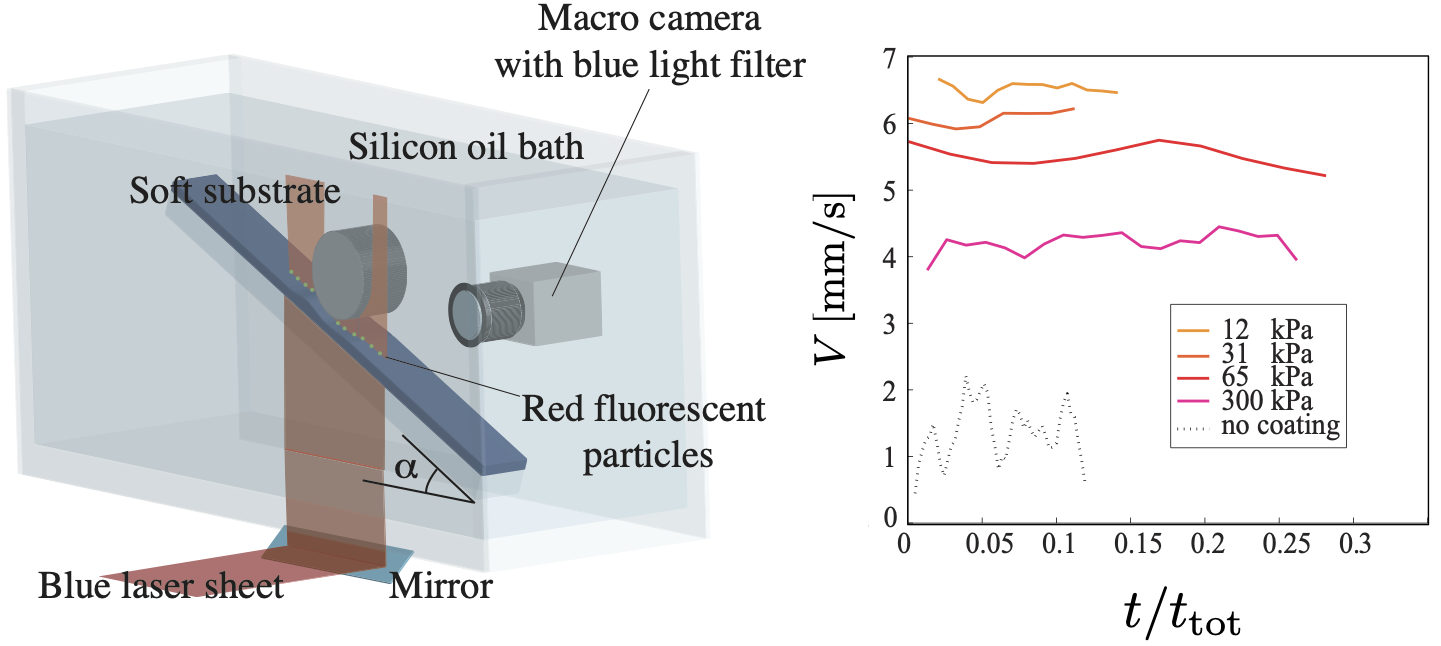}
\caption{\textit{(left) A rigid cylinder immersed in a viscous bath slides along an inclined plane covered with a thin elastic layer. Fluorescent particles embedded in the latter make it possible to observe its deformation using a laser and a camera. (right) Sliding speed $V$ as a function of normalized time $t/t_{\textrm{tot}}=tV_{\infty}/L$, for several shear moduli of the elastic coating. Here, $V_{\infty}$ is the time-averaged steady-state sliding speed, $L$ is the total length of the substrate, and $t$ is the time. The dotted line corresponds to the case of a bare glass substrate. Figure adapted from~\cite{Saintyves2016}.}}
\label{saintyves}
\end{center}
\end{figure*}

A first quantitative study, by \citet{Saintyves2016}, showed an effective reduction of friction induced by the soft-lubrication lift force. The authors employed a fluid-immersed negatively buoyant
macroscopic cylinder moving along a soft inclined wall. They observed a steady-state sliding regime
with an effective friction that was significantly
reduced relative to the rigid case (see Fig.~\ref{saintyves}). 
The observations were rationalized by invoking the soft-lubrication lift. This study was followed up by a work dedicated to the rotational motion of the cylinder~\cite{Saintyves2020}.
The authors experimentally quantified the steady spinning of the cylinder
and theoretically showed  that it is due to an aspect-ratio dependent combination of a soft-lubrication torque generated by the
flow and the viscous friction on the edges of the finite-length cylinder. We note that the contribution of edge effects within the lubricated motion of a cylinder moving near flat rigid walls was then revisited in more details~\cite{Teng2022}. The experimental results of~\cite{Saintyves2020} were consistent with a
transition from an edge-effect dominated regime for short cylinders to a gap-dominated
soft-lubrication regime when the cylinder is very long. A puzzling feature about these two studies is the fact that the Winkler's foundation describes best the observations, despite the rather incompressible character of the elastomers used. The answer to that puzzle might be given by \citet{Chandler2020} in the lift context. Indeed, for Poisson's ratios strictly smaller than $1/2$, an incompressible layer will eventually behave as a compressible one for small-enough film thicknesses.

\begin{figure*}[t!]
\begin{center}
\includegraphics[width=15cm]{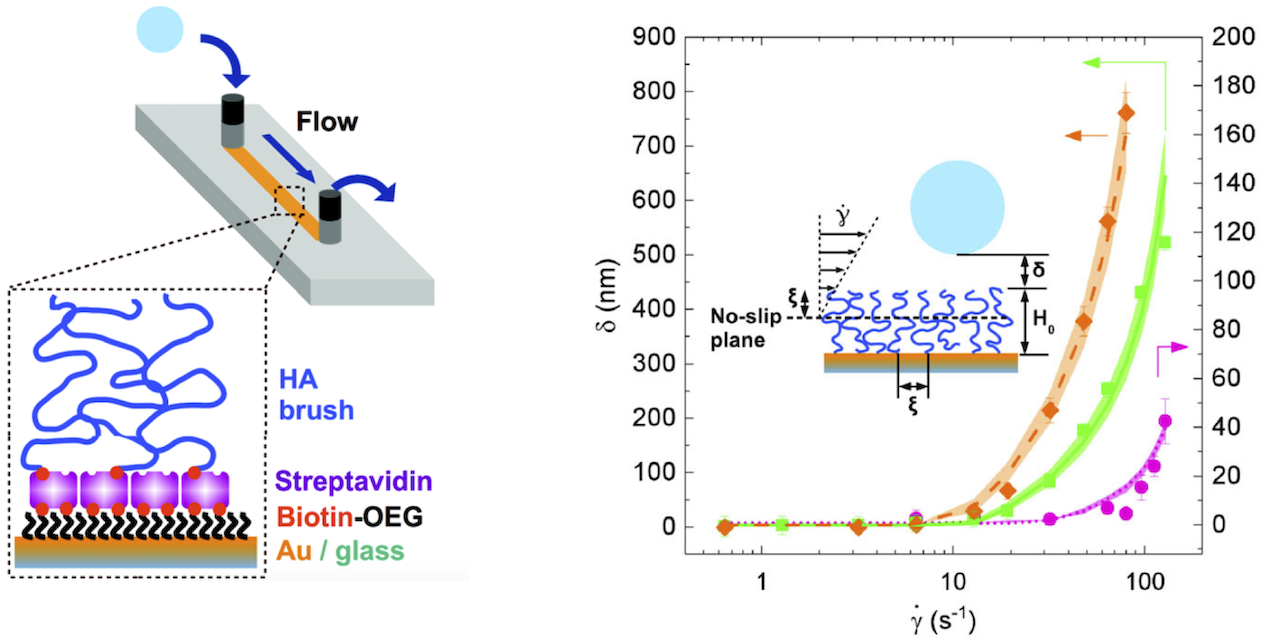}
\caption{\textit{(left) A glass microbead is advected in an aqueous environment within a microfluidic chamber whose walls are decorated by a biomimetic polymer brush. (right) The distance to the wall is measured versus the imposed shear rate, for three brush elastic moduli (increasing from orange to pink). Theoretical lines including the soft-lubrication lift contribution are fitting the data. Figure adapted from~\cite{Davies2018}.}}
\label{davies}
\end{center}
\end{figure*}

Subsequently, an experimental study by \citet{Davies2018} revealed the significance of the soft-lubrication lift force in biological and microscopic settings. The authors addressed the motion of glass microbeads in a linear shear flow close to a wall
bearing a thin soft biomimetic polymer brush. Combining microfluidics and optical tracking, they demonstrated
that the steady-state bead-to-surface distance increased with the imposed shear rate (see Fig.~\ref{davies}). 
The article is concluded by physiological estimates, indicating the potential relevance of the effect for the transport of red blood cells -- and thus for life processes.

\begin{figure*}[t!]
\begin{center}
\includegraphics[width=15cm]{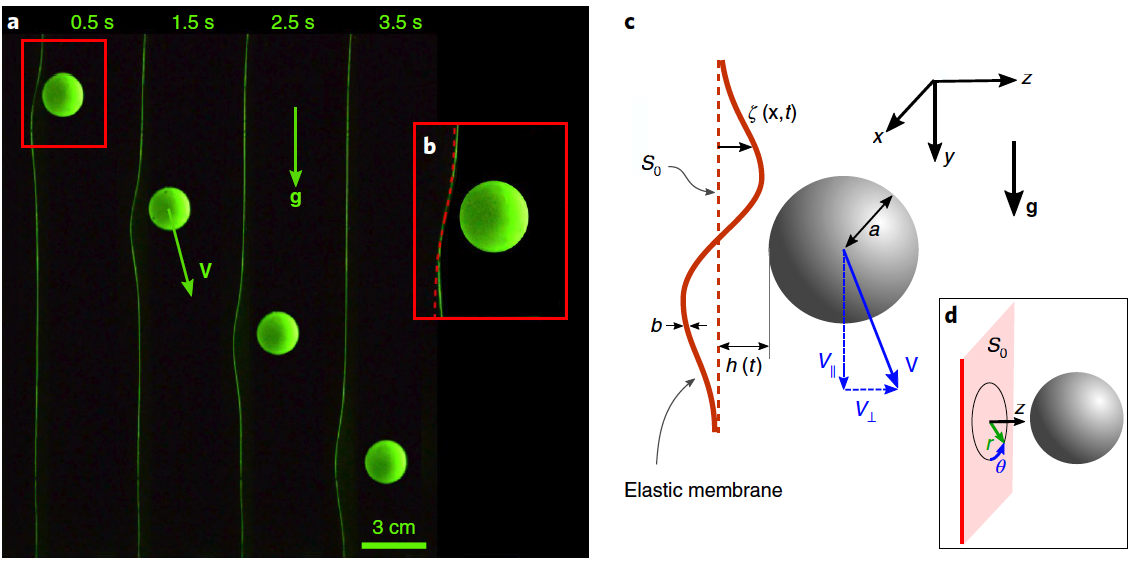}
\caption{\textit{Gravitational sedimentation of a macroscopic sphere immersed in a viscous fluid, along a vertical membrane under tension, exhibits an important normal drift induced by the soft-lubrication lift. Figure taken from~\cite{Rallabandi2018}.}}
\label{rallabandi}
\end{center}
\end{figure*}

The same year, a macroscopic study by \citet{Rallabandi2018} demonstrated the large amplification of the soft-lubrication lift for very compliant boundaries associated with slender geometries (see Fig.~\ref{rallabandi}). 
The authors combined theory and experiments in order to show that a small particle moving
along an elastic membrane through a viscous fluid is repelled from the membrane due to soft-lubrication forces. An analytic expression for the particle trajectory is derived, including a normal migration velocity of the particle that is quadratic in speed and depends on a combination
of the tension and bending resistances of the membrane. The quantitative agreement with the theoretical predictions
with no fitting parameter indicates once again the presence and relevance of the soft-lubrication lift force. Furthermore, due to the slenderness of the membrane, the effective compliance is large and the effect is strong enough for separation and sorting
of particles on the basis of both their size and density. Once again, the relevance for biology -- where membranes are widespread -- is discussed.

\begin{figure*}[t!]
\begin{center}
\begin{minipage}[t]{7.7cm}
\centering
\includegraphics[width=\columnwidth]{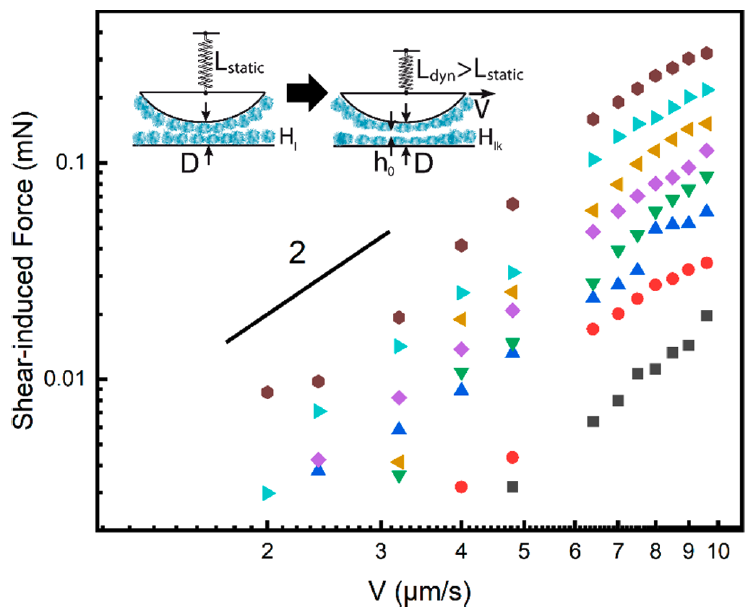}
\end{minipage}
\begin{minipage}[t]{8.cm}
\centering
\includegraphics[width=\columnwidth]{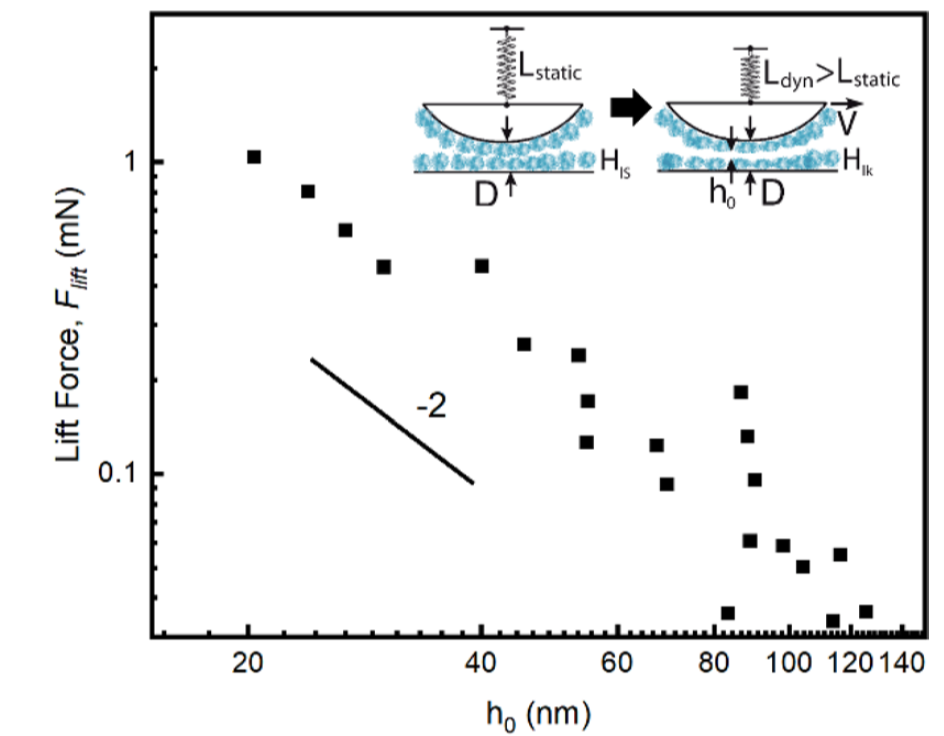}
\end{minipage}
\caption{\textit{(left) Shear-induced force as a function of driving velocity measured with a SFA covered by microgel layers and in presence of a lubricant, for various gap thicknesses increasing from brown (130 nm) to grey (300 nm) symbols. (right) Measured lift force as a function of lubricant thickness. Figure adapted from~\cite{Vialar2019}.}}
\label{vialar}
\end{center}
\end{figure*}

\begin{figure*}[t!]
\begin{center}
\includegraphics[width=15cm]{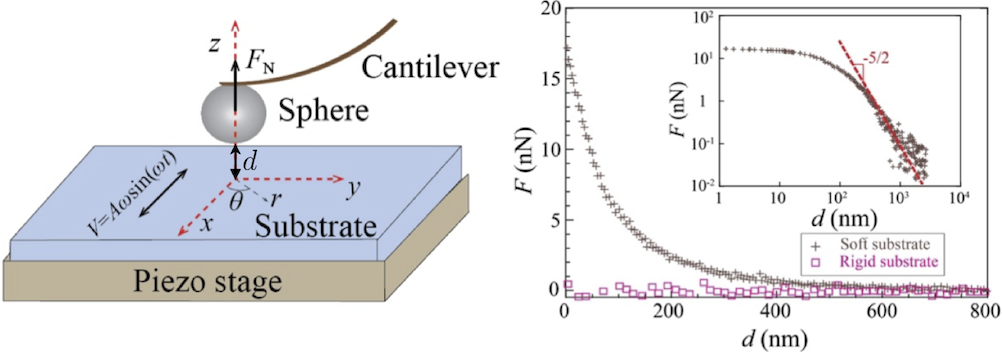}
\caption{\textit{(left) A soft substrate is fixed atop a rigid piezo stage that is transversally oscillated along time $t$, at angular frequency $\omega$ and with amplitude $A$. A rigid sphere is glued to an AFM cantilever and immersed in a viscous liquid lubricant near the substrate. The normal force $F_{\textrm{N}}$ exerted on the sphere at a given distance from the surface is directly measured from the deflection of the cantilever along $z$. (right) Temporal average $F$ of $F_{\textrm{N}}$ as a function of the gap distance $d$ to the substrate, for both rigid (silicon wafer) and soft polydimethylsiloxane (PDMS) substrates. The inset shows a log-log representation of the data for the soft substrate where the solid line indicates a $-5/2$ power law characteristic of an EHD lift force in the case of a semi-infinite incompressible elastic substrate~\cite{Skotheim2005}. Figure adapted from~\cite{Zhang2020}.}}
\label{zhang}
\end{center}
\end{figure*}

The above recent experimental literature provides confidence in the existence of the soft-lubrication lift force, as well as in its importance at small scales and for biology. However, in these works, the quantitative evidence for the soft-lubrication lift is indirect since only trajectories and effective friction coefficients are typically measured. Despite the fact that forces and trajectories are equivalent in steady Stokes flows, a direct measurement of the soft-lubrication lift force was thus missing. The first SFA and AFM direct force measurements of the soft-lubrication lift force at the nanoscale were performed by \citet{Vialar2019} and by \citet{Zhang2020}, respectively.
On the one hand, in the former SFA study, the authors investigated the behavior of mica surfaces coated with microgels under
shear and compression. The
emergence of velocity-dependent, shear-induced normal forces was observed and quantified (see Fig.~\ref{vialar}). 
Moreover, the scaling of data is in agreement with the soft-lubrication lift force but revealed a counterintuitive  value of the microgel elastic modulus. On the other hand, Zhang \textit{et al.} employed an AFM colloidal probe near an horizontally-oscillated elastomeric layer and measured the average lift force as a function of the gap size (see Fig.~\ref{zhang}), for various driving velocities, viscosities,
and stiffnesses. 
The results
are in agreement with a quantitative model developed from the soft-lubrication theory for small compliances~\cite{Bertin2021}. For larger compliances, or equivalently
for smaller confinement length scales, an empirical scaling law for the observed saturation of the lift force was 
proposed and discussed. This high-loading conjecture should be compared in future to recent theoretical developments~\cite{Essink2021}. 

 \section{Elastohydrodynamic lift in external flow}
 \label{sec:extlift}

Even when not directly deformed by the presence of a wall, deformable particles will exhibit a specific dynamics when interacting with flow velocity gradients. This will in turn create an additional flow which is often compatible with the asymmetric boundary conditions in the  direction normal to flow,  only if the particle has a normal motion relatively to the flow direction.

A reference situation is that of a particle in a simple shear flow above a wall (Sec. \ref{sec:liftwall}), but other configurations give also rise to  lift, among which the flow of a particle in a channel (Sec. \ref{sec:liftchannel}). The interaction between two flowing particles is also an example of lift, whose study is often of interest because it is the fundamental mechanism of non-Brownian diffusion in a suspension (Sec. \ref{sec:particleparticlelift}). Other, more marginal, situations for the lift of a single particle are examined in Sec. \ref{sec:otherlift}.

Before going further, a quick overview on the dynamics of a soft particle under simple shear flow is needed here, with the aim to focus on the main information needed to understand the dependence of the amplitude of the lift force with the mechanical properties of the considered particle.

\subsection{Dynamics of deformable particles under simple shear flow}

\label{sec:unboundeddynamics}

 A non-spherical rigid particle cannot maintain a steady orientation under shear flow, due to the rotational component. By contrast, particles with liquid cores can maintain a fixed shape under shear flow while accommodating the  rotational stresses by a rotation of their inner fluid. The way they do so depends on the detail of their mechanical properties, and a combined motion with whole rotation, coupled with shape oscillations, can be met.

In this review, the term vesicle will be used to designate a drop of liquid encapsulated by an incompressible elastic layer. The sole elastic deformation energy of this layer is thus associated to bending. In particular, initially spherical vesicles are non-deformable, by virtue of incompressibility of the membrane and of the encapsulated fluid. Vesicles have long been considered as model systems to mimic the main behaviour of more complex cells such a red blood cells. By neglecting the elastic shear contribution of the underlying cytoskeleton, researchers  have gained simplicity, that was needed for theoretical modeling purposes and to reduce cost and complexity of numerical simulations. In parallel, models for extensible shells --- which are often named capsules --- have been developed. Finally, the last ten years have seen the development of more accurate models to describe cell membranes, that include the possibility for shear at constant surface in the membrane plane.

The literature on the dynamics under shear flow of liquid drops \cite{rallison84,stone94}, vesicles  (see the  review \cite{vlahovska-cargese} and some subsequent articles \cite{farutin10,biben11,zabusky11,farutin12_1,farutin12_2}), capsules (see  the  review \cite{Barthes-biesel11} and also \cite{skotheim07,walter11,foessel11,dupont13,dupont16,barthes-biesel16,Zhang2020_3}), or (models of) red blood cells  (see a thorough introduction in \cite{minetti19} and some subsequent articles \cite{Guglietta2020,Mignon2021,gallen2021})  is extended and is still being enriched by studies of increasing refinement in the modeling and in the experimental approach. Adding more complexity to the rheological properties of the membrane leads in general to more complexity in the  diagrams of dynamical states. The parameters of the system are usually combined
in a set of dimensionless numbers such as reduced volume $\nu$ (characterizing initial deflation of the object), viscosity contrast $\lambda$ defined as the ratio between the viscosities of the inner and the outer fluid,
and capillary number(s) $Ca$, that compare hydrodynamic stress with either surface tension or
bending rigidity or shear elasticity of the membrane \footnote{Capillary number $Ca$ is thus defined either as  $Ca=\mu |\dot{\gamma}| R/\sigma$, where $\sigma$ is the surface tension, or, as used for vesicles, $Ca=\mu |\dot{\gamma}| R^3/\kappa$, where $\kappa$ is the bending modulus of the membrane, or $Ca=\mu |\dot{\gamma}| R/G$, where $G$ is the surface shear modulus of the membrane, as used  for capsules.}.

An oversimplified picture is that for low $\lambda$ and high $Ca$, particles adopt a drop-like behaviour, called tank-treading (TT), where the particle keeps a constant angle $\theta$ (see Fig. \ref{fig:abkarian}  for notation)  relatively to flow direction. For high $\lambda$ or low $Ca$, particles with rigid enough membrane behave more like solids and tumble under flow, with a periodic evolution of $\theta$. In between, a rich variety of motions has been described, from small oscillations around a given angle to off-plane orbital motion. The oscillations of the main axis of the particle is often accompanied by shape oscillations of more or less important amplitude. For ellipsoidal rigid particles, orbits have been exactly described by \citet{jeffery22}; this description is also a good proxy for tumbling motion of not too deformable particles like red blood cells \cite{minetti19}  or capsules \cite{dupont13} under moderate shear flow.

\subsection{Particle in a simple shear above a wall}

\label{sec:liftwall}

Once the dynamics of a particle in unbounded flow is solved, the flow produced by this dynamics  can be calculated in presence of the wall, and the resulting lift velocity determined.  Since the influence of the wall on the dynamics is not included in this approach, a question that arises naturally is what is the domain of validity of this approximation in the particle-to-wall distance domain ? To understand how the correction terms emerge, it is convenient to  introduce  the boundary integral formalism to solve the Stokes flow due to the force distribution on the particle surface.  From this formalism will also naturally emerge the notion of stresslet, the key ingredient to quantify far-field lift. In order to focus on the underlying physics we leave the technical details to the appendix.

To facilitate the notations and summations, we introduce the position vector $\mathbf{x}=(x_1,x_2,x_3)$, where $x_1$ corresponds to the flow direction, and $x_3$ to the direction perpendicular to the wall (located at $x_3=0$), \textit{i.e.} to $z$ in inset of Fig. \ref{fig:miseenbouche}.

We consider particles made of a 2D interface delimiting their interior, filled with a fluid of viscosity $\mu'\equiv \lambda \mu$. In that case, the flow field at any point $\mathbf{x}_0$ outside the particle reads \cite{Pozrikidis92}

\begin{eqnarray} u_j(\mathbf{x}_0)&=& u_j^\infty(\mathbf{x}_0)-\frac{1}{8 \pi\mu}\int_S \Delta f_i(\mathbf{x}) G_{ij}(\mathbf{x},\mathbf{x}_0) dS  \nonumber \\
&+& \quad \frac{1-\lambda}{8 \pi}\int_S u_{i}(\mathbf{x}) T_{ijk}(\mathbf{x},\mathbf{x}_0) n_k(\mathbf{x}) dS. \nonumber \\ &&
\label{eq:flowfield-lambda}\end{eqnarray}

The imposed flow is $\mathbf{u}^\infty$  and  $\Delta\mathbf{f}= \mathbf{f}^{ext}-\mathbf{f}^{int}=(\mathbf{\sigma}^{ext}-\mathbf{\sigma}^{int})\cdot\mathbf{n}$ is the discontinuity in the interfacial surface force.  The tensors $\mathbf{\sigma}$ are the  fluid stress tensors; in order to account for the presence of body forces, they can be replaced by the modified stress tensors such that $\sigma^{MOD}_{ij}=\sigma_{ij}+\rho \mathbf{g}\cdot\mathbf{x}\delta_{ij}$ \cite{Pozrikidis92}. Here, $\mathbf{g}$ is the acceleration field, like gravity, and $\rho$ is the associated quantity, like fluid density. Keeping this in mind, we will drop the $MOD$ superscript from now on. Finally $\Delta\mathbf{f}$ can be written as  $\Delta\mathbf{f}= (\rho^{ext}-\rho^{in}) \mathbf{g}\cdot\mathbf{x} \,\mathbf{n} + \Delta \xi$, where $\Delta \xi$ is the discontinuity in the surface force that depends only on the interface mechanical properties. For a given model of particle (\textit{e.g.} a drop, a vesicle, a capsule), and in the absence of significant inertia of the membrane, it can be calculated according to the chosen constitutive law for the surface, as it must equal the opposite of the membrane load.

$\mathbf{G}$ is the Green's function that is adapted to the boundary condition of the problem and $\mathbf{T}$ is the associated stress tensor. 

 Regarding numerical simulations, Eq. \eqref{eq:flowfield-lambda} can be implemented to compute the particle dynamics according to a two-step process where first the displacement of the particle membrane is calculated according to Eq. \eqref{eq:flowfield-lambda}, after what the force $\mathbf{\xi}$ can be calculated in this new configuration, and so on. This so-called boundary integral method has given rise to several developments regarding numerical schemes to be used, following the seminal work of \citet{pozrikidis2001}. In particular, it has successfully  been used to describe the motion of drops \cite{Uijttewaal93_2}, vesicles \cite{sukumaran01,coupier08,messlinger09,zhao11,farutin13} or capsules \cite{walter11,nix14,nix16} in the vicinity of walls. To do this, Green's function that are adapted to the considered boundary conditions must be used, which we describe below. Note that Eq. \eqref{eq:flowfield-lambda} does not provide a direct expression for the velocity as it appears on both sides of the equation, when $\lambda\ne 1$. This requires to implement adapted numerical schemes to ensure convergence.

For an unbounded domain, the Green's function $\mathbf{G}^\infty(\mathbf{x},\mathbf{x}_0)$ is called the Stokeslet and describes the flow field created at position $\mathbf{x}_0$  by a point force located at position $\mathbf{x}$.  This flow field is anisotropic and decreases as the inverse of the distance to the point force.

The Green's function we need here is that satisfying the no slip condition on the wall. \citet{blake71} proposed a calculation of this semi-infinite Green's function, using Fourier transform. It can be thought as the Green's function associated with other point singularities located at the reflection point  $\mathbf{x}^{IM}=(x_1,x_2,-x_3)$ of the initial force. Interestingly, in  \citet{blake71}, this interpretation in terms of singularities is obtained a posteriori, after the direct calculation has been performed. An alternative, more intuitive construction of  the fundamental solutions of Stokes flow in a half-space have been proposed by \citet{gimbutas2015}.

The semi-infinite Green's function reads $\mathbf{G}=\mathbf{G}^\infty+\mathbf{G}^w$, where the wall Green's function $\mathbf{G}^w(\mathbf{x},\mathbf{x}_0)$ is a function of $\mathbf{x}^{IM}-\mathbf{x}_0$. It introduces a velocity field in $\mathbf{x}_0$ that has several contributions that depend on the inverse of the distance $|\mathbf{x}^{IM}-\mathbf{x}_0|$ to the power 1,2 and 3 (see Appendix). Similar decomposition exists for the stress tensor $T_{ijk}=T^\infty_{ijk}+T^w_{ijk}$.

 The lift velocity $U_L$ can be then thought as the averaged velocity in the $x_3=z$ direction over the particle volume, which can be transformed into a surface integral for incompressible particles, noting that $\mathbf{\nabla} \cdot {(x_3 \mathbf{u})}=u_3$: \begin{equation} U_L=\frac{1}{V}\int_V u_3 dV=\frac{1}{V} \int_S x_3 \mathbf{u}\cdot\mathbf{n}\, dS,\label{eq:liftfromu}\end{equation} where $\mathbf{n}$ is the unit vector normal to the surface.
 
\paragraph{Far-field contribution and corrections}

In the absence of walls, the dynamics of the particle is governed by Eq. \eqref{eq:flowfield-lambda} where $\mathbf{G}=\mathbf{G}^\infty$ and $\mathbf{T}=\mathbf{T}^\infty$. This leads to the dynamics described in section \ref{sec:unboundeddynamics}.

The semi-infinite  Green's function and the associated stress tensor introduce a dependence of the velocity field with the distance to the wall. This velocity field may induce a displacement of the particle far or towards the wall, but also modify its shape dynamics. Therefore, the velocity field created by the presence of the particle will be modified, hence possibly the lift velocity.

We discuss below how these potential contributions scale with the distance $z$ of the particle  to the wall, in view of identifying the domains of validity of the far-field formalism and of the soft-lubrication regime. 

Following \citet{nix14} who proposed it in the case $\lambda=1$, we decompose the lift velocity $U_L$ into two contributions: (a) a self term $U^s$ that arises (through Eq. \eqref{eq:liftfromu}) from the flow  $\mathbf{u^s}$ created  by the particle, obtained from Eq. \eqref{eq:flowfield-lambda} by considering only $\mathbf{G}^\infty$ and $\mathbf{T}^\infty$, and a wall-term term $U^w$, obtained when considering $\mathbf{G}^w$ and $\mathbf{T}^w$.

In an infinite simple shear flow, the whole configuration has a point symmetry with respect to the centre of the particle.  Considering two opposite points on the membrane, one can see that the $ \mathbf{u^s}\cdot\mathbf{n}$ terms in Eq. \ref{eq:liftfromu} are equal while the $x_3$ term has opposite sign, whence  $U^s=0$, as expected. Note that this argument holds only in the absence of some symmetry-breaking mechanism, like buckling, which we are not aware of in unbounded simple shear flow, unless the particle itself presents some heterogeneities \cite{Liu2017}. The self term $U^s$ can be non zero if point symmetry is lost, in particular if the particle is deformed by the wall.

We now consider the leading order of the wall term, which is often the only one considered in models. Far from the wall, this velocity $U^w$ may be approximated by that of its center, that we set to be located at position $\mathbf{x}_0=(0,0,z)$. This far-field velocity $U^{w,ff}$ thus reads \begin{eqnarray} U^{w,ff} &=&-\frac{1}{8 \pi \mu}\int_S \Delta f_i(\mathbf{x}) G^w_{i3}(\mathbf{x},\mathbf{x}_0)\, dS \nonumber \\ &&+ \frac{1-\lambda}{8 \pi}\int_S u^0_{i}(\mathbf{x}) T^w_{i3k}(\mathbf{x},\mathbf{x}_0) n_k(\mathbf{x}) dS, \nonumber\\ &&\end{eqnarray}where $\mathbf{u}^0$ is the leading order term in the velocity on the particle surface. For $|\mathbf{x}-\mathbf{x_0}|\ll z$, one can expand $G^w(\mathbf{x},\mathbf{x}_0)$ and  $T^w(\mathbf{x},\mathbf{x}_0)$ around $\mathbf{x}_0$. In the absence of external forces and torque and considering symmetry properties of the tensors as well as fluid incompressibility, one eventually finds (see Appendix for details):

\begin{equation} U^{w,ff} = -\frac{9}{64 \pi \mu} \frac{\Sigma_{33}}{z^2}=\frac{A R^3 |\dot{\gamma}|}{z^2}, \label{eq:Uwff}
 \end{equation}
where
\begin{strip}
\begin{eqnarray}\Sigma_{ik}= &\int_S \Big[\frac{1}{2}(\Delta f_i (x-x_0)_k +\Delta f_k (x-x_0)_i )-\frac{1}{3}\Delta f_j (x-x_0)_j  \delta _{ik}\Big] \, dS \nonumber\\
&+(\lambda-1) \mu \int_S (u^0_{i}(\mathbf{x}) n_k(\mathbf{x})+u^0_{k}(\mathbf{x}) n_i(\mathbf{x}) )\,dS \label{eq:defstressletbis}\end{eqnarray}
\end{strip}is called the stresslet, while the second expression in Eq.~(\ref{eq:Uwff}) provides a convenient dimensionless form to discuss results obtained for different particles.  It should be noted that the second term vanishes not only for particles with no viscosity contrast but also for rigid particles \cite{batchelor1970}. Equation~(\ref{eq:Uwff}) was also proposed, following a more straightforward approach, by \citet{smart91}. By coherence with the leading order approximation we made here, one must keep in mind that the stresslet $\Sigma_{33}$ is that created by the interaction with the external flow, in the absence of wall. Checking the validity of this approximation (and the associated domain in the $z$-axis) can be made through full numerical simulations or experiments. Doing so, one must keep in mind that while the theoretical approach of Eq.~(\ref{eq:Uwff}) through the determination of the stresslet will provide the lift velocity at a given position for a particle in its stationary dynamics, simulations or experiments provide full trajectories along which the shape at given position might not be the stationary one. Comparing both approach requires to ensure that the typical time needed for shape change is much smaller than the migration time. \textit{A priori}, this will be achieved for sufficiently high capillary numbers. In practice, this conditions holds in the situations we will describe below. Also, in simple shear flow, the shape depends only weakly on the position. The situation will be more complex in quadratic flows, which will be shown to trigger more complex couplings between shape and lift direction. An agreement between direct numerical simulations and Eq.~(\ref{eq:Uwff}) using an independently calculated stresslet was found for vesicles in \cite{zhao11,farutin13} and for capsules in \cite{nix14}.

An explicit discussion on the domain of validity of the far-field expression, as well on the scaling for the next order terms, was carried out by \citet{nix14},  through numerical simulations of a capsule in a quite narrow range of capillary numbers (of order 0.1-1), and $\lambda=1$. 

They first showed  that the self term $U_s$ is negative (that is, the particle is attracted towards the wall), and decays as $(z/R)^{-4}$, on a large range $z/R \gtrsim 1.2 $. This shows that the modification of the lift velocity due to shape  distortion under the influence of the flow created by the stresslet in interaction with the wall, which exhibits contributions scaling like  $(z/R)^{-2}$ as in Eq.~(\ref{eq:Uwff}), is not of the same order as the direct stresslet contribution.  We make the remark here that, to our knowledge, this result has never been derived formally and generalised to a larger class of particles.

Regarding the wall term, \citet{nix14}  highlighted that the far-field expression can be corrected by two terms; first, replacing the stresslet by its value at the considered distance from the wall enhances the lift velocity. Second, the overall contribution of the wall is shown to be smaller, due to the asymmetric deformation  (with respect to the point-wise symmetric shape) that has a negative contribution to the lift velocity, as for the self term. It has been shown that these two corrections of the wall term also varies like  $(z/R)^{-4}$ and that their overall contribution added to the self term is negative.

We re-analyzed some results of the literature, which points to a potential generality of this result. As shown already in Fig.~\ref{fig:miseenbouche}, the correction to the far-field expression is also negative for moderately deflated vesicles, with a $(z/R)^{-4}$ scaling. It becomes significant for $z/R \lesssim 3$. These features seem to hold also for drops, as we found by revisiting the data from \cite{Uijttewaal93_2}, Fig. 5.

These results help identifying the transition zone between the far field regime and the soft lubrication regime, where the lift results from modified flow due to the deformation of the interface because of the flow between the particle and the wall. We discuss in next sections how this transition depends on the mechanical parameters involved. Noteworthy, the soft lubrication correction seems to be always negative for sheared particles. 

\begin{figure}
	\centering 		
\includegraphics[width=\columnwidth]{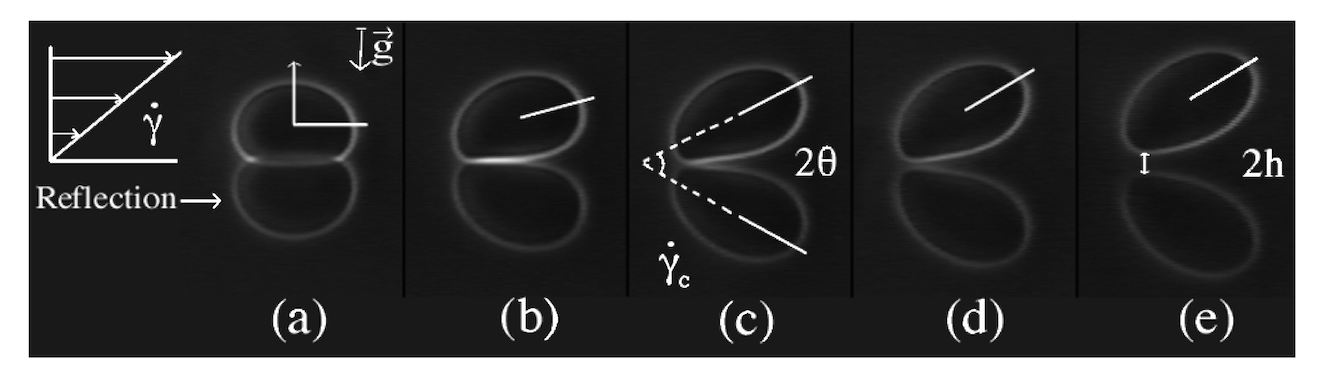}
	\caption{\textit{Sequence of shapes taken by a vesicle as it unbinds and lifts away from the wall under simple shear flow. Figure taken from~\cite{abkarian02}.\label{fig:abkarian}}}
\end{figure}

 We propose a geometrical interpretation of this negative contribution, based on the observed shape sequences upon unbinding, which are illustrated in Figs. \ref{fig:miseenbouche} or \ref{fig:abkarian} but also seen in \cite{sukumaran01,nix14}. Far from the wall, the particle adopts an orientation that roughly follows that of the elongational component of the flow, \textit{i.e.} 45$^\circ$ relatively to the flow direction. So does its bottom membrane, on average. Closer to the wall, the shape of the bottom end of the particle, which is close to the wall, is controlled by the local interaction with the wall, which creates, as in the soft-lubrication framework depicted in Fig. \ref{fig:Maha}, a quasi-horizontal gap with a small opening angle. Compared to the high opening angle induced by the bulk flow, this reduces the fore-aft asymmetry of the particle, hence a negative correction to the lift velocity. In future works, it would be interesting to examine this hypothesis by comparing the amplitude and scaling of this negative contribution to the lift force of a similar particle moving along the wall without external shear flow (i.e., in the typical soft-lubrication configuration).

In the soft lubrication framework, results are often presented in terms of lift force; on the other hand, for particles deformed under external flow, lift velocities is the usual output. This probably relates to the main motivation of these studies, which is to calculate flux of particles. These two quantities can be related through the drag acting on the particle. It is insightful to examine how this can be achieved in practice. To that aim, we first review the case of vesicles, which is interesting for comparing results from different studies as there are only few parameters involved. Furthermore, results focusing on either lift force or lift velocity are available.

\paragraph{Lift velocity and lift force, the vesicle case}

For large enough capillary numbers such that the hydrodynamic stress overcomes bending forces, the dynamics of vesicles depends only on their initial deflation and on the viscosity contrast \cite{farutin10}. In the range of parameters where tank-treading motion occurs, the shape and angle of inclination of the vesicle are therefore independent of the shear rate, so is the prefactor $A$  in Eq. \eqref{eq:Uwff}.  We are not aware of studies on lift of vesicles at small capillary numbers, when wrinkles appear due to hydrodynamic forces or Brownian fluctuations and modify the nature and the transition between dynamical regimes in unbounded shear flow \cite{noguchi05_3,deschamps09,abreu13}.

Two different sets of experiments are available in the literature that describe the lift of vesicles of radius of order 10 microns: \citet{callens08} studied the lift velocity of vesicles in the absence of gravity, while  \citet{abkarian02} and \citet{abkarian05} examined the close wall lift force by balancing it by the vesicle weight. 

Experiments for lift velocity were performed in parabolic flights, allowing for successions of normal gravity phases and low gravity phases. In the first phase, sedimentation of vesicles on the bottom plate of a shear chamber allowed for the creation of a well-defined initial condition, while the lift velocity of the vesicle could be measured in the low gravity phase, without being screened by sedimentation. Thanks to this, distances to the wall of up to 7 times the vesicle radius could be explored. Vesicles with inner fluid having the same viscosity inside and outside were studied in \cite{callens08} while more viscous inner fluids were considered in \cite{bureau17}. In the range $3\lesssim z/R \lesssim 7$, the distance to the wall is found to scale with time to the power $1/3$, indicating agreement with the far-field scaling. The authors also found a prefactor $A$ that is independent from the shear rate, that was varied by a factor 10. Eventually, the prefactor $A(\nu,\lambda)$ has been determined for reduced volumes $\nu \gtrsim0.95$ and $\lambda=1$ \cite{callens08}, 4 and 6.5 \cite{bureau17}. The reduced volume $\nu\le 1$ characterizes the deflation of the vesicle, therefore its ability to get deformed, and reads $\nu=3\mathcal{V}/4 \pi (\mathcal{A}/(4 \pi))^{3/2}$, where $\mathcal{V}$ and $\mathcal{A}$ are the vesicle volume and surface area, respectively. They are both constant due to the inner fluid and membrane incompressibility.

In the experiments by  \citet{abkarian02} and \citet{abkarian05}, vesicles with no viscosity contrast but with density contrast, and $\nu \gtrsim 0.92$, are sheared close to the wall, with shear rates varying by a factor 5 such that different equilibrium positions can be scanned. It is found that at equilibrium the gap $h$  between the vesicle and the wall scales linearly with the shear rate, suggesting the following expression the lift force $F_L$: \begin{equation} \label{eq:Abka} F_L=B(\nu) \mu R^3| \dot{\gamma}|/h. \end{equation}

Simulations of vesicles with no viscosity contrast  were proposed in several studies by using boundary integral method  \cite{sukumaran01, zhao11,farutin13}. 
All simulations, as well as  far-field theoretical calculations led  in \cite{vlahovska07} or \cite{farutin13}, consider mostly vesicles with reduced volumes higher than 0.95. This corresponds, \textit{e.g.}, to a prolate ellipsoid  of long axis 1 and short axis 0.63.

Using both the results  of \citet{callens08} and \citet{abkarian02} to make comparisons with these existing simulations and theories requires first to establish a link between lift velocity and lift force, \textit{i. e.} an adequate formulation of the drag.

\begin{figure}
	\centering 		
\includegraphics[width=\columnwidth]{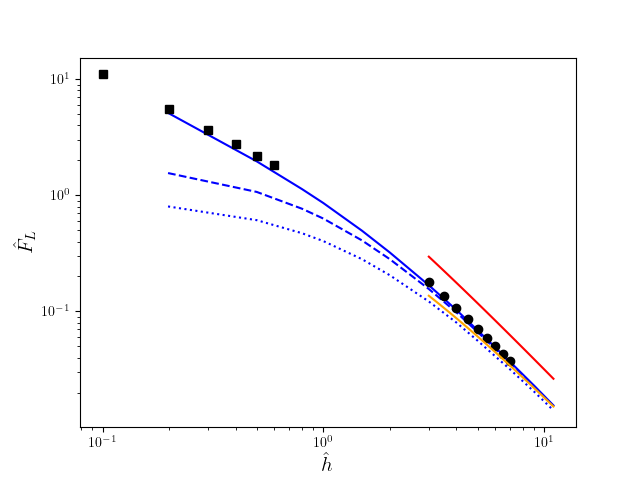}
	\caption{\textit{Dimensionless lift force $\hat{F}_L=F_L/(\mu |\dot{\gamma}| R^2)$ on a vesicle characterized by  $\nu=0.97$ and $\lambda=1$ sheared above a wall, as a function of reduced gap size $\hat{h}=h/R$ between the vesicle and the wall. $R$ is the radius of a sphere having the same volume as the vesicle. Lines and dots extend on a range where the law has been established in the different studies. $\blacksquare$: $\hat{F}_L=1.1/\hat{h}$ according to the close-wall experiments  of \citet{abkarian02}, Fig. 8. Other data are obtained through $\hat{F}_L=D\hat{U}_L$ where $\hat{U}_L$ is the  dimensionless lift velocity $U_L/(|\dot{\gamma} |R)$ and $D$ the dimensionless drag coefficient, which is taken either as  $D_0=6 \pi$ (unbounded domain), $D_1=6 \pi (1+9/[8 (\hat{h}+1)])$ (first order approximation of Brenner's drag, see Eq. \ref{eq:firstbrenner}), or $D_{\infty}=6 \pi  \Lambda(\cosh^{-1}[\hat{h}+1])$ (Brenner's drag, see Eq. \ref{eq:brenner}). $\bullet$:  $D=D_\infty$ and $\hat{U}_L= A(0.97)/(\hat{h}+1)^2$ with $A(0.97)=0.11$ according to the  far-field experiments of \citet{callens08}, Fig. 7; Red line: the same with $A(0.97)$  given by Olla's model \cite{Olla97} (ellipsoid of aspect ratio 0.705); Orange line: the same with $A(0.97)$ given by the theoretical model of Farutin {\it et al.} (Eq. 5 in \cite{farutin13}, for $\Gamma=0.021$, $\Gamma$ being defined in the referred article). Note that in \cite{callens08}, the volume and reduced volumes of the vesicle are known only indirectly through the measurement of its short axis in the vorticity direction and of the projection of its long axis parallel to the $z$ direction. These data are converted into volume and reduced volume through the direct calculation of \citet{farutin13}, which have shown excellent agreement with these experiments \cite{bureau17}. Blue lines:  $\hat{F}_L= D  \hat{U}_L$, where the lift velocity is obtained from direct simulation in \cite{zhao11}, Fig. 2 and $D=D_0$ (dotted line), $D_1$ (dashed line), or $D_\infty$ (full line).}\label{fig:comparliftvesicle}}
\end{figure}

In presence of a body force (which is often, in practical cases, gravity), the lift motion is modified and can even vanish if this body force acts opposite to the lift force. Regarding boundary integral framework, the presence of the body force induces an additional term $U^g$ in the migration velocity  $U^g=\frac{1}{V} \int_S x_3 \mathbf{u}^g\cdot\mathbf{n}\, dS$, where \begin{equation}
u^g_j(\mathbf{x_0})=-\frac{1}{8 \pi \mu}\int_S (\rho^{ext}-\rho^{in}) \mathbf{g}\cdot\mathbf{x} \,n_i G_{ij}(\mathbf{x},\mathbf{x}_0)\, dS  \label{eq:Ug}
    \end{equation}
is the flow field created by the particle due to the presence of a density difference across its membrane. For a sphere of radius $R$ in an unbounded flow, with a gravity field in the direction $-x_3$, solving Eq. \eqref{eq:Ug} would lead to the well-known Stokes law $6 \pi \mu R U^g=-P$, where $P=\frac{4}{3}\pi R^3  (\rho^{in}-\rho^{ext})$ is the weight of the particle minus the Archimedes force. In presence of a wall at $x_3=0$, the expression of $G_{ij}^w$ shows that the first order correction to the velocity would scale like $1/z$. For a sphere settling towards a wall, the full correction for the modified drag force as a function of distance to the wall has been solved by  \citet{brenner61}. \begin{eqnarray}
    && 6 \pi \mu R U^g \Lambda(\cosh^{-1}[z/R])=-P, \qquad \mbox{ with} \nonumber\\  
    \Lambda(\xi)&=&\frac{4}{3}\sinh(\xi)\sum_{n=1}^{\infty}\frac{n(n+1)}{(2n-1)(2n+3)}\label{eq:brenner}\\
    && \Big[\frac{2 \sinh((2 n+1) \xi)+ (2 n+1) \sinh(2 \xi)}{4 \sinh^2((n+1/2) \xi)-(2 n+1)^2 \sinh^2(\xi)}-1\Big].\nonumber
\end{eqnarray}

The first order of this correction is given by
\begin{equation}
    \Lambda(\cosh^{-1}[z/R])\simeq 1+\frac{9}{8}\frac{R}{z}, \label{eq:firstbrenner}
\end{equation}
 indicating, in agreement with the intuition, that the presence of the wall increases the drag on the particle. We hypothesize that for deformable particles, this expression of the drag force could be considered as a first approximation.

In the presence of a body force that acts against lift, an equilibrium position in the $z$ direction is found by the particle, corresponding to $U^g+U_L=0$. When the lift force $F_L$  balances the weight $P$, one would eventually find, that \begin{equation} 
F_L \simeq  6 \pi \mu R U_L  \Lambda(\cosh^{-1}[z/R]), \label{eq:liftforce} \end{equation}
where the lift velocity is given by the appropriate expression, and the correction on drag is assumed to be that for a sphere. This expression will serve us as a basis for discussion.

We can now address the question of the comparison of simulations, theory and experiments, while also considering the question of the link between lift force and lift velocity. To that aim, we focus on vesicles with no viscosity contrast and a reduced volume of 0.97, which is documented in experiments \cite{callens08,abkarian02}, simulations by \citet{zhao11}, an historical far-field model by \citet{Olla97} that assumes an ellipsoidal shape and the far-field model by \citet{farutin13} that makes no assumption on the shape and calculate directly the stresslet\footnote{Under the same hypothesis as \citet{farutin13}, \citet{vlahovska07} proposed an expression for the stresslet whose numerical prefactor differs by some 20\%. Since the model by \citet{farutin13} is more recent and is in perfect agreement with their own numerical simulation, we hypothesize that their result is more likely to be correct.}. For all above mentioned studies, but that of  \citet{abkarian02}, the lift velocity as a function of distance $z$ between the vesicle centroid and the wall is given, in the absence of gravity. For the sake of comparison with the results of \citet{abkarian02}, we make the assumption that $z=h+R$ (see inset of Fig. \ref{fig:miseenbouche}) and calculate the lift force through Eq. \eqref{eq:liftforce}. In order to support discussion, the lift force arising from the simulations of Zhao {\it et al.} is also calculated using the 1st order approximation of the drag coefficient  (Eq. \eqref{eq:firstbrenner}) as well as 0th order, \textit{i.e.} the drag of a sphere in an unbounded flow. The results are presented in Fig. \ref{fig:comparliftvesicle}.

Regarding far-field behaviour, simulations by \citet{zhao11}, modeling by \citet{farutin13} and experiments by \citet{callens08} show very good agreement. By contrast, the lift velocity predicted by  \citet{Olla97} is about 30\% larger. Comparison between Fig. 7 in \cite{callens08} and Fig. 1 in \cite{farutin13} shows that a vesicle whose shape is not prescribed \textit{a priori}   has a  (long axis) / (short axis)  ratio that is smaller than that given by the prolate ellipsoid assumption, made by Olla. This may qualitatively explain why, for a given reduced volume, the lift predicted by Olla is larger than that measured in other studies.

Remarkably, the close-wall measurements of the lift force by \citet{abkarian02}  match perfectly the simulations of Zhao {\it et al.}, who scanned the same range of particle-to-wall distance, providing the drag contribution is that of Brenner for a sphere of equal volume. This agreement is far from being reached if one considers only the 0th (as used in \cite{sukumaran01}) or 1st order of the drag force.

The conclusions of this novel aggregation of data are first that the two experimental sets of data by \citet{callens08} and \citet{abkarian02} both fall on the same master curve and second that the $1/\hat{h}$ scaling for the drag force, proposed by Abkarian {\it et al.}, is recovered, in the close-wall range, by multiplying the multipolar drag coefficient of Brenner with a lift velocity that is composed of a repulsive stresslet contribution that scales as $(\hat{h}+1)^{-2}$ and of an attractive contribution that is the consequence of wall-induced deformation. This contribution scales like $(\hat{h}+1)^{-4}$. This proves that using Brenner's drag to switch between lift forces and velocities is relevant in this context.

\citet{zhao11}  also ran simulations including gravity and found good agreement with the experimental data of \citet{abkarian02}. In the same article, they compared their simulations of the lift force with  that calculated from the lift velocity times a first order expression for the drag  (as in  Eq.~(\ref{eq:firstbrenner})). Unfortunately, we are not able to comment on this approach by lack of definition of some parameters and of detail of calculation by the authors. Other direct simulations of vesicles under gravity have been performed in by \citet{messlinger09}, but they consider a 2D system. On a quite narrow range of distances, the authors found that the force scales like $1/z^2$ even for $z/R$ close to 1.

As a conclusion,  our review highlights the necessity to better understand interactions in the crossover zone $z\sim R$ between soft-lubrication effects and shear-induced lift.  While the main contribution to lift, the stresslet created by the unperturbed particle, has been widely commented, the correction terms did not benefit from such an attention. We have also highlighted the difficulty in comparing results focused on the velocity with that focused on the force. While using Brenner's drag for a rigid sphere seems satisfactory, studies focused on the drag exerted on a deformed particle whose dynamics is imposed by the external flow would certainly be helpful.

\paragraph{Dependence with mechanical properties}

The relative contribution of each of the $1/z^2$ and $1/z^4$ terms depends on the detail of shape and dynamics of the particle.

Increasing the viscosity contrast seems to always make the lift velocity decrease, for vesicles \cite{messlinger09,zhao11, farutin13,bureau17} or for capsules, for which  \citet{singh2014} showed that this is the direct result of the increase of the absolute value of the second term in the stresslet expression  (Eq. \eqref{eq:defstressletbis}), although the absolute value of the first term decreases. It should be noted that for capsules, we have not found any experimental study regarding their lift under simple shear flow.

At least for vesicles, the range of validity of the far-field expression increases with the viscosity contrast, a consequence of the increasing cost for interface deformation under the influence of the wall \cite{zhao11}.

The lift velocity also depends on the ability of the particle to deform, which is essentially related to reduced volume for vesicles, and to capillary number for drops and capsules. It is natural to start our discussion with the simplest of these objects, \textit{i.e.} drops.

\citet{smart91} measured experimentally the lift on drops of viscosity contrast 0.083. Since then, we have found no record of another attempt to measure directly this lift in a simple shear flow; other works include drop-drop interactions, that will be discussed further. They found that the far-wall velocity follows Eq. \eqref{eq:Uwff}, with the prefactor $A$ being proportional to the capillary number $Ca$. This indicates that upon an increase of the flow stress (compared to the elastic restoring force), the drop elongates more, thus inducing a larger lift force. This result was theoretically derived by a direct calculation of the stresslet,  both by  \citet{chan79} and \citet{smart91}, under the hypothesis of  small deformation (hence the linear dependence with $Ca$). They both proposed an expression for the prefactor, which depends on the remaining characteristic parameters $\lambda$, the viscosity contrast between the inner and the outer fluids. Both expressions are different but vary by at most 2 \% on the whole range of viscosity contrast. This prefactor was found to be an increasing function of $\lambda$, yet with small variations, from 0.58 when $\lambda\to 0$ to 0.69 when $\lambda \to \infty$. These theoretical results do not match the experimental results, where the lift amplitude is found to be almost 2 times larger. 

Numerical simulations generally find a lift velocity that is slightly smaller than in theoretical works, therefore they agree even less with experiments \cite{Uijttewaal93_2,uijttewaal1995,Kennedy1994,singh2014}.

We mention here that with drops, another configuration of interest has been studied by \citet{smart91}, that of a free surface instead of a rigid wall. In this case, the overall expression for the far-surface lift velocity (Eq.~(\ref{eq:Uwff})) is unchanged, but a multiplicative factor $2/3$ is introduced. As the stresslet is that given by the bulk flow, we expect the expression given  by the authors to be valid not only for the drops they study. Experimentally, as for the lift due to a rigid wall, their experimental results are above the theoretical prediction.

\begin{figure}[t]
\begin{center}
\includegraphics[width=0.5\textwidth]{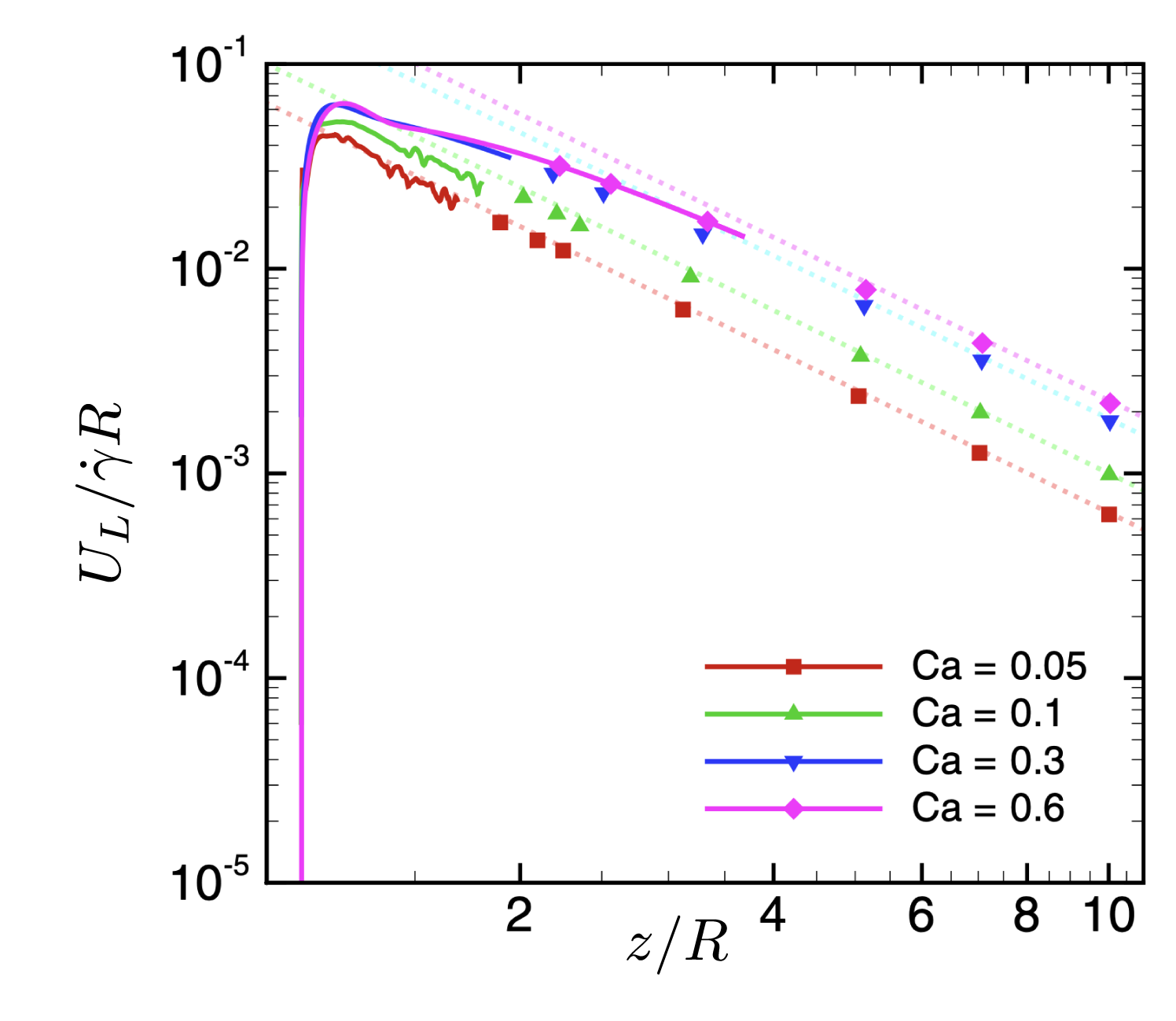}
\caption{\textit{Lift velocity in simple shear flow for a capsule with a membrane following Neo-Hookean law. Dots and full lines correspond to simulations while the dotted line is the far-field approximation (Eq. \eqref{eq:Uwff}). Figure adapted from \cite{nix14}.}\label{fig:nix}}
\end{center}
\end{figure}

An increase of the absolute value of the stresslet with the capillary number, therefore of the lift velocity, is also mentioned for capsules in  \cite{nix14}  (see Fig. \ref{fig:nix}). Regarding the relative contribution of the far-field term and of the first soft lubrication correction term, \citet{nix14} showed that, while the far-field approximation is valid even for particles in contact with the wall at low capillary number, this approximation is valid only when $h/R\gtrsim 3$ for a capillary number one order of magnitude larger (Fig. \ref{fig:nix}). This impact of the capillary number on the locus of the transition zone is similar to that previously discussed for vesicles of different viscosity contrast: less ability to deform lead to a smaller contribution of the soft-lubrication term.

Considering capsules with various viscosity contrasts, \citet{singh2014} evidenced the same year a similar range of distances where corrections to the far-field regime must be considered.  In this range, a phenomenological single scaling is proposed  by the authors.
 
 For a vesicle with no viscosity contrast, the more deflated the vesicle, the faster it migrates \cite{abkarian02,callens08,zhao11,farutin13}.  However, for high enough viscosity contrast, increasing asphericity can lead to transition towards tumbling motion, which is preceded by a decrease of the inclination angle, together with the lift velocity \cite{Olla97,bureau17}. The lift velocity eventually vanishes when vesicles are aligned with the flow. To our knowledge, this decrease has not been observed experimentally, but similar phenomenon has been exhibited in Poiseuille flow \cite{coupier08}, as will be discussed later on.

In their 2D simulations, \citet{messlinger09}  have considered the case of tumbling vesicles and shown a non-zero lift force on average, by contrast with the case of a purely rigid object. The reason stems from the elongational component of the flow that makes the vesicle be more elongated when its long axis is along this component than when it is orthogonal to it. Over one rotation period the mean shape is therefore not fore-aft symmetric (which, at least for vesicles, seems to lead to vanishing lift velocity).  Still the mean lift velocity remains much smaller than in the tank-treading regime. A more formal discussion on this aspect can be found in \cite{olla00}. Similar results were found for capsules by \citet{hariprasad14}, who  maintained a 2D capsule that would tumble in an unbounded flow  close to a wall by a gravity force of varying intensity. The proximity with the wall induces a change in the dynamics that switches from tumbling to tank treading as the force is increased. As a result, the lift force increases, such that, in the rather narrow range of parameters explored by the authors, a quasi constant equilibrium height is observed while the force is multiplied by a factor 4.

\paragraph{Red blood cells}

While more complex than the above mentioned particles, red blood cells exhibit similar lift properties.

In physiological conditions, red blood cells flow in vessels where the maximal shear rate ranges from 20 to 1500 s$^{-1}$. In these conditions, cells will essentially exhibit a tumbling-like motion that couples with strong deformations \cite{minetti19,mauer18}.

As seen in \cite{messlinger09} for tumbling vesicles, isolated red blood cells may still migrate. This has been confirmed experimentally in simple shear - like flow (in a large pipe in reality) in the 70' by Goldsmith, who highlighted a 4 $\mu$m transverse drift for cells having travelled 1 cm in the flow direction \cite{goldsmith71}.  More than 40 years later, a study in microgravity conditions, similar to that carried out for vesicles in \cite{callens08}, has quantified the lift of  red blood cells in a narrow range of shear rates $10\le\dot{\gamma}\le 50$ s$^{-1}$ \cite{grandchamp13}. The far-field scaling has been confirmed, with a prefactor $A R^3=0.36$ $\mu$m$^3$ (following the notations of Eq. \eqref{eq:Uwff}) in physiological conditions (\textit{i.e.} an external fluid of viscosity 1.4 mPa.s close to that of plasma). This factor is increased by a factor 15 if the viscosity is multiplied by 9, indicating the impact of flow stress on the deformation of the cells, that are more elongated and can even make a transition towards a tank-treading like motion \cite{fischer13,minetti19}. Considering that red blood cells have a mean volume of 90 $\mu$m$^{3}$ \cite{Baskurt1997}, whence $R=2.8\,\mu$m, this leads to a prefactor $A$ between 0.016 (tumbling-like regime) and 0.15: this latter value is comparable to that found for a vesicle in tank-treading regime.

The value for the lift velocity found in physiological conditions has been calculated to be compatible with the pioneering result of \citet{goldsmith71}. It should be noted that, at a distance \textit{e.g.} 4 $\mu$m from the wall the lift velocity is about 2 $\mu$m/s at a shear rate of 100 s$^{-1}$, which is comparable to sedimentation velocity \cite{matsunaga2016}. This illustrates the difficulty in measuring experimentally far-field lift velocities. 

While numerical simulations of red blood cells under flow are now numerous, we have found no records for this geometrically simple configuration of simple shear flow near a wall, as far as realistic 3D simulations of cells are concerned. We shall come back to this point later on while discussing collective effects.

\subsection{Lift in a channel flow}

\label{sec:liftchannel}

Lift in channel flow is the other configuration explored in the literature, as it is relevant for particle handling in microfluidic devices, and to understand biological flows such as blood flow especially.

These flows are characterized by the increased presence of walls but also by linear variations of the shear rate. For large channels (compared to particle size) and in order to gain fundamental understanding of the lift mechanism, it is insightful to consider first the case of an unbounded Poiseuille flow that is, a parabolic velocity profile with no walls imposing a condition of zero velocity.

\paragraph{Lift in unbounded Poiseuille flow}

In the soft lubrication framework, the necessary breaking of the symmetry relatively to the flow direction in terms of pressure is obtained by the presence  of a wall. The pressure gradient that is created between the particle and the wall is different from that on the other side of the particle, leading to different deformations patterns on each side of the particle, eventually leading to the fore-aft asymmetry that induces an overpressure below the particle and makes the appearance of the lift force possible.

This sequence is indeed also possible in the configuration of an unbounded Poiseuille flow where a particle is surrounded by different shear rates on both sides, except when it is located on the central line of the flow. This situation can create an asymmetry in the deformation patterns along the flow direction, making thus possible the appearance of a net lift force. Notably, while in the soft-lubrication framework the creation of a gap resembling that of the Reynolds slider makes it intuitive the sign of the lift force, the deformation arising on both sides of the particle leads to a less clear situation.

Indeed, several scenarii have been predicted. For a drop in a 2D parabolic flow, \citet{chan79} and \citet{leal80} have predicted that a drop would migrate outward for a viscosity ratio $0.71 \lesssim \lambda \lesssim 11.35$ but towards the central line for other values of $\lambda$. For an axisymmetric Poiseuille flow, the interval for outward migration becomes $0.56 \lesssim \lambda \lesssim 10.2$.

\begin{figure}[t]
\begin{center}
\includegraphics[width=0.5\textwidth]{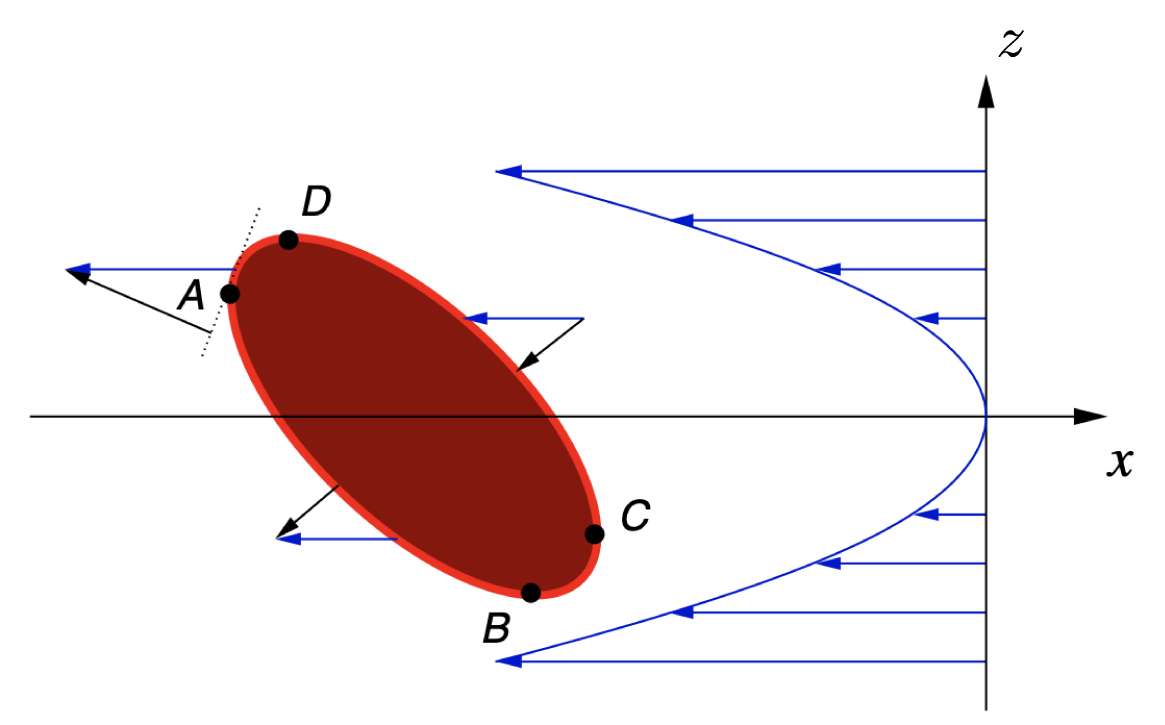}
\caption{\textit{Creation of an inward lift force in Poiseuille flow for a tank-treading vesicle located at a position $y>0$ from the central line, according to \citet{danker09}. The local velocity field can be decomposed into a local shear that dictates the shape of the vesicle and local quadratic correction (blue arrows). The velocity
component normal to the vesicle surface points downward on segments AB and CD and upward on the two other segments, which are smaller. Hence a negative lift force acting on the particle. This argument is debatable as it does not take into account either the part of the flow that modifies the particle shape but not its position,  or the relative intensity of the normal contribution on each segment. In addition, it should also apply \textit{per se} for liquid drops, and would contradict the finding of \citet{chan79}. This illustrates the difficulty in getting an intuitive picture for migration in quadratic flows. Figure taken from \cite{danker09}.}\label{fig:danker}}
\end{center}
\end{figure}

Theoretical and numerical studies on vesicles have shown that vesicles apparently behave differently: \citet{kaoui08} ran 2D numerical simulations of vesicles with $\lambda=1$ and evidenced inward migration at almost constant velocity along the trajectory but at the very end, when the particle meets the central line. \citet{danker09} confirmed these results  through a theoretical approach valid in the small deformation approximation. For all $\lambda$ such that a tank-treading motion takes place, an inward migration is predicted. A tentative physical argumentation in favor of this migration has been given in this article, which is reported in Fig. \ref{fig:danker}. Similarly, \citet{helmy82} theoretically showed that capsules migrate towards the centerline, in the limit of small deformation (small capillary number).

\citet{farutin13,farutin14} performed more systematic theoretical studies for a 3D, axisymmetric flow. They revealed a much more complex situation, for a vesicle of given reduced volume $\nu=0.9$. The behaviour strongly depends on the capillary number $Ca$ and on the viscosity ratio $\lambda$. For $\lambda=1$, vesicles migrate towards the center at high capillary number, at a constant velocity but in the vicinity of the center. This situation corresponds to the case most easily studied by theory as a stationary shape can be considered (due to large $Ca$), as in \cite{danker09,farutin13}. At lower $Ca$, \textit{i.e.} when the particle has no time to adjust its shape to the surrounding local flow before being advected further, an equilibrium position away from the center is found. For larger viscosity contrasts, outward motion is observed at large $Ca$ whereas metastability was observed at smaller $Ca$, the vesicle migrating outward or toward a position close to (but not on) the center, depending on its initial position.

\citet{kaoui09_2} reached similar conclusions for  2D vesicles with no viscosity contrast: upon a decrease of capillary number a deflated enough vesicle does not converge towards the center but stays at a finite distance from it, adopting an asymmetric shape described as a slipper shape. This situation is also favored by a deflation of the vesicle. A tentative explanation for this phenomenon is proposed by the authors, based on their numerical observation: the transition towards an off-center, asymmetric shape is accompanied by a reduction of the lag, which is anticipated to be a favorable configuration by the authors, despite the increase in internal dissipation inside the particle not being symmetric anymore.

However, arguments based on a minimization  of energy or of dissipation are not supported by any fundamental principle in this viscoelastic problem with moving boundaries. Indeed,  \citet{farutin14} discard both possibilities by showing they are not compatible with their numerical observations --- though in the meantime arguments of that kind are still being used in \cite{tahiri13}. Improper use of arguments based on dissipation consideration is also discussed in  \cite{dasanna21}.

A question naturally arises: does this complex behaviour survive in more realistic situations where walls are present ? Walls induce additional lift forces but also additional space dependence of the shape. A secondary question is, how do lift forces due to flow curvature and lift forces due to the presence of a wall compare to each other in intensity ? An attempt to answer partly this question can be found in the 2D numerical simulations of \citet{kaoui09_1} where a vesicle with no viscosity contrast is placed in a semi-bounded parabolic flow, \textit{i.e.} where only one wall is present (say, at position $z=-z_0$ while the centerline is at $z=0$). In this case, a vesicle placed at a distance $z>0$ from the center larger than its typical radius migrates outwards, while it would migrate inwards in the absence of the opposite wall. This indicates, at least in this specific situation, that the lift due to the presence of the wall overcomes that due to the flow curvature, at a distance from the center large than a particle radius.

\citet{nix16} carried out a more detailed study regarding this question, with capsules. They quantified the ratio of the  contributions arising from the shear gradient and from the presence of a wall, that grows as the particle approaches the center of a channel. An interesting output of their study is that the drift velocity due to shear gradient hardly depends on the viscosity ratio (in the explored range $1<\lambda<5$) while the effect of the wall diminishes with increasing $\lambda$ -- as already discussed here. As a result, the effect of shear gradient is predominant on a larger area within the channel for more viscous particles.

\paragraph{Migration in a channel}

In a channel, using boundary integral method to solve the flow field requires to incorporate more complex Green's function. Even in a 2D case where only two opposite walls are needed, this requires to incorporate multiple image systems \cite{thiebaud2013, naitouhra18}.  An alternative method consists in considering walls as soft boundaries of known rheological property, such that additional integrals must be considered, with the advantage to handle only the Green's functions for unbounded flow \cite{kaoui11}. In all cases, the strong impact of continuous shape evolution due to non-homogeneous shear rates makes it difficult to exhibit a simple scaling for the lift velocity: one cannot simply plunged a particle of given shape into the desired geometry. Yet, several experiments and simulations tend to prove that a scaling $U_L \propto |\dot{\gamma}(z)|/z^{\alpha}$, with the exponent $\alpha$ close to 1, holds for several types of particles. In the following, we will denote by $r$ the radius of the channel, and $\hat{r}=r/R$ its dimensionless form, that accounts for the degree of confinement of the particle.

\begin{figure}[t]
\begin{center}
\includegraphics[width=0.5\textwidth]{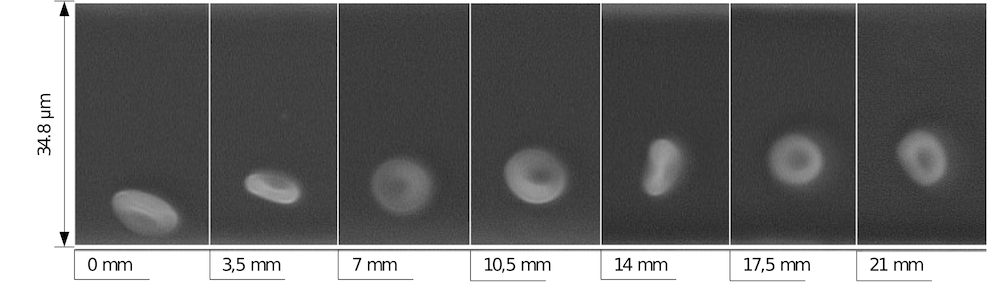}
\caption{\textit{Lift of a red blood cell in a microchannel: snapshots at different positions along the  channel. Figure taken from  Ref. \cite{losserand19}.\label{fig:losserand}}}
\end{center}
\end{figure}

\citet{coupier08}  first proposed such a scaling for lipid vesicles. By varying experimentally the confinement ($2\lesssim \hat{r} \lesssim 9$), they showed that, in the range of viscosity contrast $1\le\lambda\le 10$ and of reduced volume  $\nu\ge 0.92$, trajectories from the wall to the center are well described by the law \begin{equation}
  \dot{z}= \xi \frac{R^{\delta+1}|\dot{\gamma}(z)|}{(z-z_w)^{\delta}},
  \label{eq:liftpoiseuille}
\end{equation}where $\delta $ is close to 1 and $\xi$ a dimensionless parameter that depends on the vesicle properties, similar to $A$ for the drift under simple shear rate (Eq. \eqref{eq:Uwff}). $z_w$ is the position of the center of mass when the particle is as close as possible to the wall. For the quasi-spherical vesicles considered in \citet{coupier08}, $z_w\sim R$ but in general, it may depend on particle deformability. 2D numerical simulations have provided a similar scaling \cite{coupier08}. Having $\delta$ close to 1 can somehow be viewed as an intermediate case between the unbounded parabolic flow (where $\delta=0$) and the shear flow in presence of a wall (where $\delta=2$).
 
 The alternative law 
\begin{equation} 
  \dot{z}= \xi \frac{R_0^{\delta+1}|\dot{\gamma}(z)|}{z^{\delta}},
  \label{eq:liftpoiseuillegeneral}
\end{equation}is formally simpler and has been proposed in subsequent articles (\cite{qi17} then \cite{losserand19})  to allow for comparison between different situations with no need to take into account the detail of the near-wall interactions.
 
Simulations of red blood cells yet having a non-physiological viscosity contrast of 1 (and therefore in tank-treading regime) have highlighted an exponent that is essentially in the range 1.2-1.3 for $\hat{r}=6$ and 8.8 \cite{qi17}. 

Similarly to what was found for simple shear flows, the more complex dynamics followed by red blood cells in physiological conditions does not prevent them from following similar law. \citet{losserand19} found experimentally  through in vitro experiments that on a large range of confinements  ($1.5\lesssim \hat{r} \lesssim 10$),  Eq. \eqref{eq:liftpoiseuillegeneral} was followed with an exponent $\delta\simeq 1.3$ (Fig. \ref{fig:losserand}). They also mentioned that a fit of experimental data by a trajectory obtained through Eq. \eqref{eq:liftpoiseuillegeneral} poses practical issues as parameters $\xi$ and $\delta$ are strongly correlated: several pairs of values for these parameters indeed lead to reasonable fits. Then discussions on the exact value of exponent $\delta$ should probably be considered with care.

Regarding the dependence with the particle mechanical properties, the overall picture is that an increase in deformability (through an increase of capillary number or a decrease of the viscosity ratio) leads to an increase in migration velocity towards the center, be it for capsules \cite{Doddi08,qi17}, vesicles \cite{coupier08} or red blood cells \cite{losserand19}. For vesicles, varying the reduced volume of viscous enough particle have an interesting effect: starting from the sphere, deflating the vesicle allows for the deformation that leads to migration, but below a given reduced volume, the more elongated particle aligns in the flow (approaching then the tumbling transition) and the lift velocity then drops to 0 \cite{coupier08}.

The above mentioned studies focus on the migration from the wall towards the center. As pointed out in \cite{nix16,kaoui09_1}, this migration is dominated by the wall effect in its vicinity. When the particle approaches the center, shear gradient contribution will become dominant. As discussed previously, the direction of the transverse migration might be reversed. In addition, since the shear rate decreases as the particle approaches the center, the capillary number decreases therefore the particle shape is not any more in a quasi-steady configuration, leading to a more complex coupling between shape and migration. This aspect is discussed in particular in the last pages of \cite{li10}.

\paragraph{Shape-lift coupling and instability in channels}

\begin{figure}[t]
\begin{center}
\includegraphics[width=0.5\textwidth]{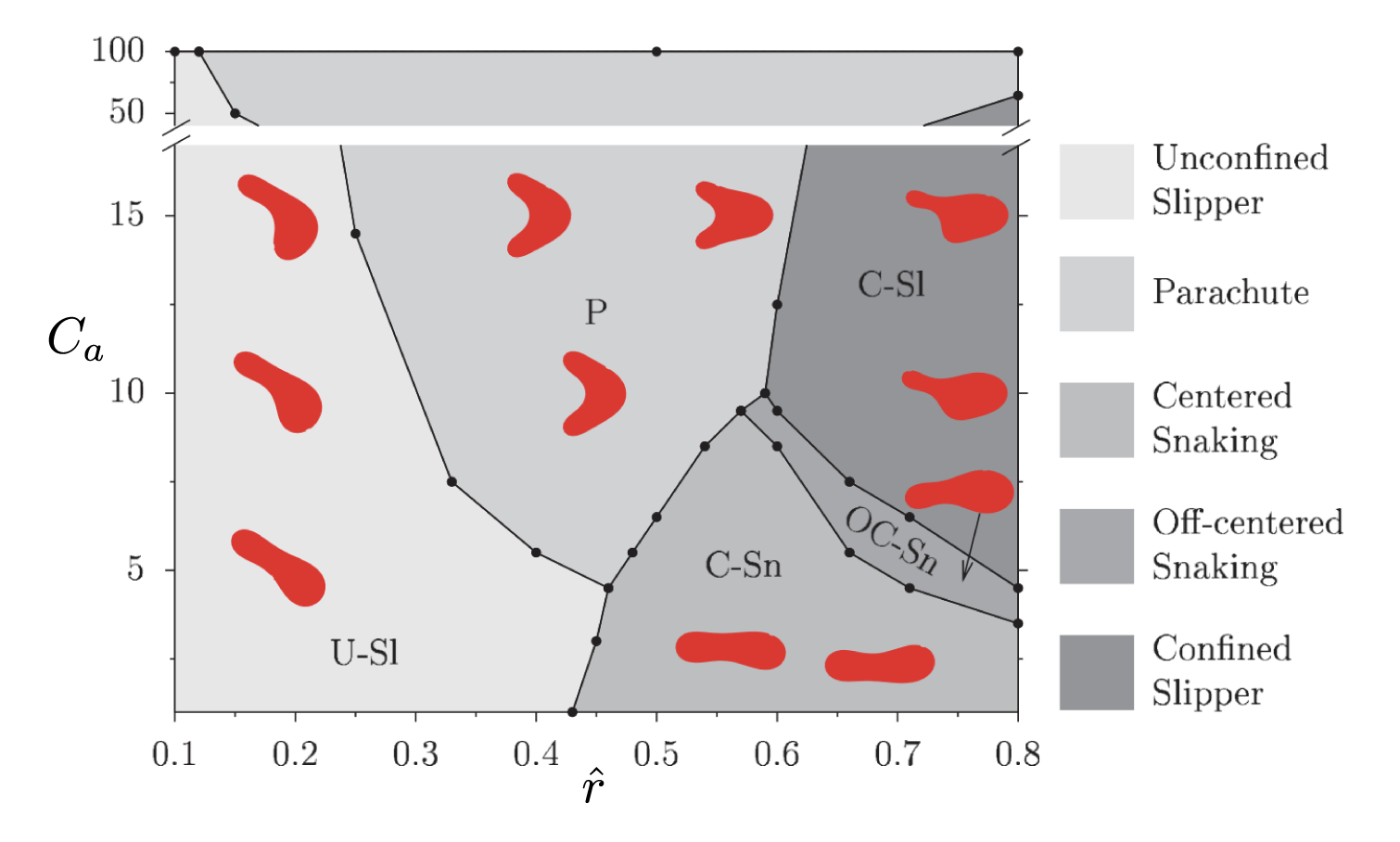}
\caption{\textit{Diagram for the behaviour of 2D vesicles in a channel, as a function of confinement ratio $\hat{r}$ and capillary number $Ca$. Figure taken from \cite{kaoui11}.}\label{fig:kaoui11}}
\end{center}
\end{figure}

While particles approach the centerline, the presence of walls seem not to destroy the complexity seen in unbounded Poiseuille flow. It rather complexifies the picture, at least for deformable enough particles. \citet{kaoui11} considered 2D vesicles with no viscosity contrast, and scrutinized their behaviour while the confinement and the capillary number are varied. As shown in Fig. \ref{fig:kaoui11}, increasing the confinement leads to the appearance of another kind of behaviour, which is an oscillation in the lateral position, which can be centered or not, which is called snaking. The possibility to de-stabilize this state towards a stationary shape through a time-varying flow has been explored by \citet{boujja18}.

Remarkably, while a transition from symmetric, centered shape to off-center slipper shape is observed upon a decrease of the capillary number, a transition from symmetric shape towards slipper is also observed upon an increase of the capillary number, as long as a more viscous inner fluid  is considered: in Ref. \cite{tahiri13}, a 2D vesicle with a viscosity contrast of 5 exhibits such a behaviour, which encourages the authors to establish a similitude with experimental observations on red blood cells. The latter indeed exhibit the appearance of slipper shapes upon an increase of flow velocity, in very confined situations \cite{tomaiuolo09,Guckenberger18}.

This behaviour for high viscosity contrast particles was later on confirmed by 3D numerical simulations of red blood cells \cite{Guckenberger18,takeishi21,dasanna21}, but also of vesicles \cite{agarwal2020} --- thus disregarding shear elasticity as an important parameter in this problem. \citet{Guckenberger18} explored the full range of parameters relevant for microcirculation, and furthermore showed  that most configurations in the parameter space lead to bistability between the centered, parachute-like shape and the slipper shape, whose existence depend on the initial condition. Noteworthy, the snaking behaviour observed in 2D simulations, but also in another 3D study \cite{fedosov14} does not seem to take place in their study. Similarly, a study of a vesicle placed in a confined simple-shear flow has exhibited similar centered/off-centered transition without snaking dynamics. \citet{Recktenwald2022} however highlighted slight oscillation in lateral position of slipper-shaped red blood cells through experiments and numerical simulations. They could be interpreted as a signature of off-centered snaking. These oscillations are obtained after the flow velocity has been increased progressively, and they tend to disappear after a while, on a time scale that depends on the viscosity contrast. These results suggest that the discussions on the existence of stable states should include the question of the relaxation time needed to exit or enter a given state, which calls for longer simulation times and also makes the comparison with experiments more delicate.

More recently, several experimental developments have taken place for better characterization of shape categories, including careful design of chips to control initial conditions \cite{reichel19}, 3D tomography of flowing cells \cite{Simionato21}, and machine-learning based methods for high throughput classification \cite{Kihm18,Martin-Wortham21,Simionato21}.

While the behaviour of these confined cells may be used as a tool to characterize their individual  mechanical properties, it should also be noted that it has direct consequences on the collective behaviour, since the cell shape will directly influence the flow pattern around it, therefore the aggregation-disaggregation dynamics of a train of cells \cite{McWhirter09, ghigliotti12,Tomaiuolo12,claveria16, takeishi17, aouane17, yaya21}.

\subsection{Other configurations}
\label{sec:otherlift}

\paragraph{Curved streamlines}

Curved channels are frequent in microsystems. While more marginally studied, this configuration has attracted some attention, in particular because this geometry leads to interesting features when inertia comes into play (Dean vortices).

Before considering this complex geometry, \citet{ghigliotti11}  first considered a model configuration with an unbounded flow consisting in circular streamlines. When placing a 2D vesicle in this flow, they observed that a tank-treading vesicle migrates towards the center while a tumbling one hardly migrates. They demonstrated that the inward migration velocity is proportional to $N R^2 \dot{\gamma}/(r-R)$, where $r$ is the radial position of the vesicle, and $N$ is the normal stress difference, that is related to the cell mechanical properties. \citet{chan79} also predicted such a result for a drop. In real systems, be it a curved channel or a Couette device, a wall would be present at some point, therefore inducing outward migration. 

Ebrahimi, Balogh and Bagchi have recently demonstrated that indeed a capsule would converge towards an intermediate position between the wall and the centre line, for a channel of circular \cite{ebrahimi2021_1} or rectangular \cite{ebrahimi2021_2} cross section. Being the result of two competing effects based on the same mechanism, the resulting position is independent from the capillary number. Higher curvature leads to a final position closer to the inner wall.

These studies are, to our knowledge, the sole ones accounting for migration in curved channels at zero Reynolds number. It remains to be determined whether this would greatly affect the flow of particles in channels, where the curved part has necessarily a finite length. In \cite{ebrahimi2021_2}, Fig. 3b, a capsule starting on the central line of the channel is shown to have moved by 5\% of the distance to the inner wall after the channel has turned by 180$^\circ$, for a very sharp turn of curvature radius of 5 times the cell radius. While this will probably lead to negligible effect in most channels of interest, one may still use this effect to induce particle separation by considering channels in spirals. Such a geometry may also be used to validate the above mentioned numerical studies.

\paragraph{Oscillating flows}

The case of near centerline migration has shown us that the interplay between migration  and shape leads to complex behaviour when a stationary shape cannot be reached. Another way to produce a time lag between shape relaxation and particle migration is to force changes in the applied flow. These changes can be triggered by time varying boundary conditions or by the geometry, a typical situation being structured microchannels.

Following a series of studies on particle dynamics under oscillating unbounded flows \cite{nakajima90,kessler09,dupire10,noguchi10_1,zhao11_3,matsunaga2015}, this more recent field of research has now been explored through  several kind of particles and geometries and will probably meet growing interest in the next years, for the rich behaviours that emerge and the potential applications that could be developed.

We first consider a particle placed in a time-periodic harmonic shear flow bounded by a wall. In this problem, a new dimensionless parameter must be added, which is the ratio $\hat{\omega}$  between the oscillation pulsation and the maximal shear rate. At high enough capillary number such that the particle shape relaxation time is set by the shear rate, the intuitive picture is that, if $\hat{\omega}$ is increased from 0 (corresponding to stationary flow), the particle will face situations where it does not have enough time to re-orient itself after flow reversal, such that it will migrate towards the wall. On average though, the net migration should be positive. In the large $\hat{\omega}$ limit, the picture is that of a fixed shape in a time varying flow, leading to no lift velocity on average.

\citet{Zhu2015} considered  3D simulations of a capsule  with no viscosity contrast. In addition to confirming the decrease of the mean migration velocity with $\hat{\omega}$, they also highlighted a nonmonotonic evolution of this mean velocity, when rescaled by the typical flow velocity, with the capillary number, at given $\hat{\omega}$. This is due to the plateauing of the mean deformation  upon an increase in capillary number (in practice, upon an increase in maximal shear rate), because the capsule fails to reach its potential maximal deformation, due to flow reversal. It is found that the optimal capillary number scales linearly with $\hat{\omega}^{-1}$, in line with the idea that at high capillary number the deformation is not limited by its own deformability but rather by the time $\propto \omega^{-1}$ during which the shear is applied in a given direction.

It is interesting to observe that the notion of time lag between shape deformation and surrounding flow is sufficient to create an effective asymmetry  leading to migration even in an unbounded shear flow, providing the particle presents an initial asymmetry, as discussed in \cite{Laumann2017} where a wide class of particle is considered.

We are not aware of studies with flow reversal in a Poiseuille flow. Periodic spatial modulation of the channel have instead attracted some attention in the last 10 years, but most focus was on shape changes of centered vesicles or red blood cells \cite{noguchi10_2,braunmuller11}. Yet, such a sawtooth channel is argued to be an efficient way to center cells in a microfluidic devices in \cite{amirouche20}. However, the picture might turn out to be more complex, according to recent numerical simulations of vesicles in a wavy channel, that have exhibited off-centered equilibrium positions in a configuration where the same vesicles would be centered if the channel was straight \cite{Laumann2019_2}.

\section{Elastohydrodynamic interactions between particles}

\label{sec:particleparticlelift}

\begin{figure}[htbp]
	\centering 		
\includegraphics[width=\columnwidth]{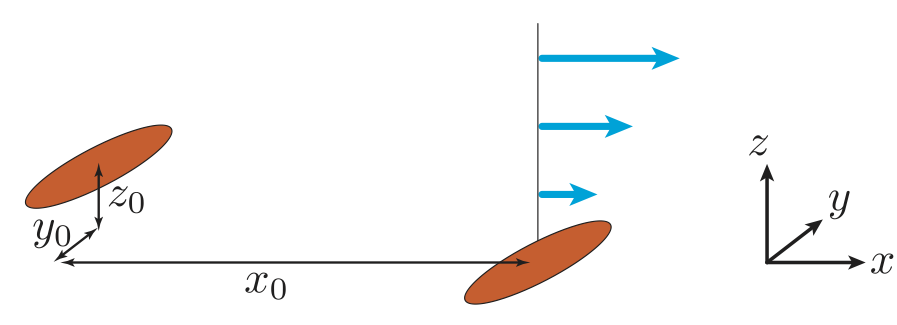}
	\caption{\textit{Fundamental configuration for pair interaction studies}}\label{fig:pairconfig}
\end{figure}

As two particles cross each other in a flow, they may experience a lift force of similar nature as that induced by the presence of a neighbouring wall. The induced normal displacement has been documented by several experimental and numerical studies. Such fluid-mediated scattering events in a suspension induce an effective diffusion in all directions. This effective diffusion has two consequences: mixing in the suspension, and flux along concentration gradients. Contrary to Brownian diffusion, these two phenomena are characterized by coefficients that are \textit{a priori} independent \cite{dacunha96}, the down-gradient diffusion coefficient being expected to be several times larger than the self-diffusion coefficient \cite{dacunha96,hudson03,grandchamp13}. They can, in principle, be deduced from the knowledge of the displacement map of the two particles as a function of their initial relative position \cite{dacunha96,loewenberg97}. However, this approach poses convergence issues due to the slow decrease of the interaction force with lateral distance between particles \cite{loewenberg97,Wang98}.

\begin{figure*}[t!]
\begin{center}
\includegraphics[width=2\columnwidth]{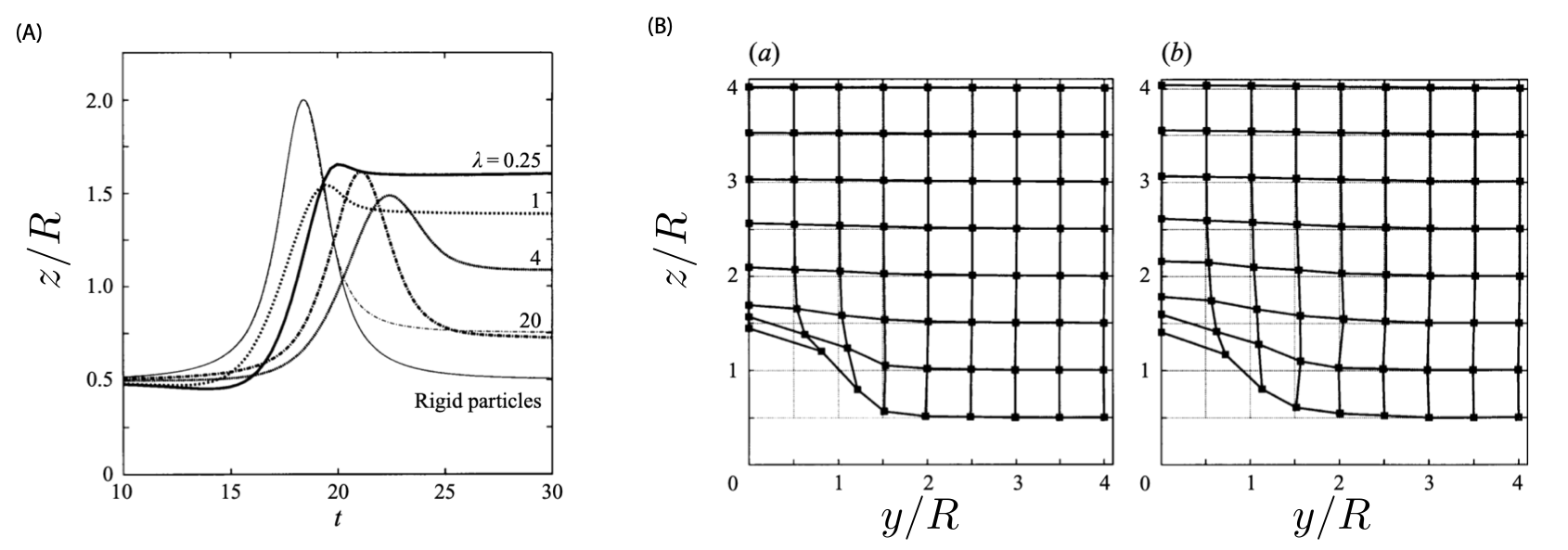}
\caption{\textit{Pair interaction between two droplets in unbounded shear flow. (A) Evolution of the shift $z/R$ in the shear direction versus time, for a fixed capillary number and various viscosity contrasts $\lambda$ indicated on the figure. Particles are located in the same shear plane ($y_0=0$). The fine solid curve corresponds to rigid sphere, and illustrates the absence of net lift, for symmetry reasons. The fine dash-dotted curve refers to a configuration not relevant here. The representation in (B) is commonly used in these problems to depict the final position in the $z$ (shear gradient) and $y$ (vorticity) directions as a function of initial position. The vertices of the light gray grid indicate the initial positions that were tested, while the vertices of the thick, deformed,  one indicate the final positions far from the reference particle. Here, $\lambda=1$ and in (b) the capillary number is larger than in (a). In (b), for an initial position $y/R=2,z/R=1$, the grid is slightly deformed towards smaller $y$, indicating an attraction in the vorticity direction. Adapted from \cite{loewenberg97}.} \label{fig:loewenberg}}

\end{center}
\end{figure*}
\subsection{Pair interaction}

By comparison with the lift of a particle close to a wall, the finite extent in all directions of the interacting particles makes this interaction problem even richer.  As they approach each other also in the flow direction, soft lubrication effects will also take place when they collide. We will first consider two identical particles  in an unbounded shear flow of flow direction $x$, shear direction $z$ and vorticity direction $y$,  one  of them placed at an initial position $(x_0,y_0,z_0)$ from the other, whose center of mass is taken as the origin, with $x_0<0$ of large absolute value (see Fig \ref{fig:pairconfig}). If $z_0>0$, the two particles will eventually cross each other, which may result in a net displacement $(\Delta y, \Delta z)$ in the two directions perpendicular to flow. This displacement depends \textit{a priori}   on both initial coordinates $(y_0,z_0)$ of the first particle. Along the flow direction, an additional displacement $\Delta x$ will also be found. Compared to the differential displacement between the particles due to advection, it is however quite small and is seldom commented.

Before turning to deformable particles, it is worth mentioning that the finite duration of the interaction between particles makes it possible to obtain a net separation between solid particles. \citet{dacunha96} proposed a model for the interaction between rough spherical particles, assuming that the approach phase builds up a repulsive force while the separation phase does not. However simple this assumption might seem, the existence of this separation effect was later on proved experimentally \cite{blanc11}.

\citet{loewenberg97} studied numerically pair interaction of identical drops in simple shear flow, for different values of viscosity contrast and capillary number. A representative set of  their data is shown in Fig. \ref{fig:loewenberg}. The relative trajectories shown in Fig. \ref{fig:loewenberg}A show that, for particles not separated in the vorticity direction, a significant shift of order one radius is observed, if the initial position in the shear direction is also of the order one radius. For drops, this shift decreases upon an increase of viscosity contrast, which marks a strong difference with the case of a drop above a flat wall, for which the dependence of the lift velocity with $\lambda$ is weak. 
The synthesis of the final positions reached depending on the initial positions shows several interesting features that illustrate the complexity of this problem. First, a larger capillary number does not necessarily induce a larger displacement. For small initial $z$, the contrary even occurs (see Fig. 5 in \cite{loewenberg97}). Again, this is in marked contrast with the results for a particle near a wall. This points to a more complex situation from the geometrical point of view, since the incoming particle does not  only flow above the other one but also hits it initially: \citet{loewenberg97} argued that the increased deformability reduces somehow the cross section for near contact interaction at collision.

Though not commented by the authors, a weak attraction in the vorticity direction can be seen for a drop initially located at position $(2R , R)$, in Fig. \ref{fig:loewenberg}B(b). This phenomenon also appeared later on in other studies. 

Fig. \ref{fig:loewenberg}B also shows that net displacements are much larger in the shear gradient than in the vorticity direction. This implies that effective diffusion due to collisions is strongly anisotropic, as will be discussed later. 

\begin{figure}[htbp]
	\centering 		
\includegraphics[width=\columnwidth]{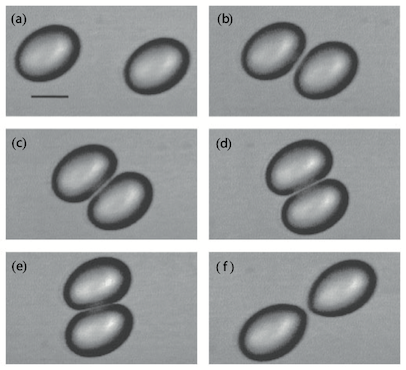}
	\caption{\textit{Succesive snapshots of interacting drops with $\lambda=1.4$ and $R=20\,\mu$m. The scale bar is $25\,\mu$m. Adapted from \cite{guido98}.} }\label{fig:guido}
\end{figure}

As for lift above a wall, experimental studies on droplet interaction are scarce. \citet{guido98} and \citet{wang16} showed some trajectories, and confirmed the typical trajectories shown in Fig. \ref{fig:loewenberg}A, and the weak net displacement as soon as the initial distance is larger than a few radii, as in Fig. \ref{fig:loewenberg}B. As can be seen in Fig. \ref{fig:guido}, the collision between the particles result in the creation of an extended and long-lasting contact between the particles, with the creation of lubrication film.  \citet{loewenberg97} exhibited different scalings for the duration of the approach sequence and have showed that the separation stage is much shorter.

Capsules are again widely ignored by experimentalists, while several numerical simulations shed light on the detail of their interactions under shear flow.   \citet{lac07} asked themselves whether the presence of a membrane modifies the drift observed for droplets. Their numerical simulations showed that for capsules with no viscosity contrast placed in the same $xz$ plane, the net displacement is smaller. They also found a weak effect of the capillary number and of the initial deflation. The same team published a complementary study where they studied the consequence of the capsules not being in the same shear plane (\textit{i.e.} $y_0\ne 0$) \cite{lac08}. By contrast with what is observed on drops, a clear attraction in the vorticity direction is observed for initial separation in the vorticity direction of order the capsule radius and small separation in the shear direction (i.e., small velocity difference). In this situation, the displacement depends strongly on the capillary number, and increases with it.  The authors provided no explanation for this phenomenon.

Lac {\it et al.} also highlighted the fact that even if the capsules are placed on the same streamline of the unperturbed flow, they may still interact and cross each other. This is due to the flow perturbation associated with the tank-treading motion of the membrane, which have the effect to shift the particles apart: the particle located at $x_0<0$ and $z_0=0$ will see its $z$ position increasing because of the clockwise tank-treading motion of the membrane of the particle located at the origin.

 \citet{singh15}  confirmed the difference between drops and capsules, the latter showing less cross-stream separation, for viscosity contrast equal to 1. The difference turns out to decrease upon an increase of this viscosity contrast.

\citet{le11} considered a refined model for capsules  that includes bending elasticity (while previous one only included shear elasticity). They also considered different initial shapes, one of them being the biconcave shape of red blood cells. Finally they considered a viscosity ratio of 4, closer to physiological values for red blood cells. The capsules were kept in the same shear plane. Instead of particle crossing, they observed for small initial $z_0$ a motion called spiraling, that consists in oscillations of the particle positions at finite distance from each other. In this article, the origin of this effect is particularly unclear, as the positions around which oscillations take place correspond to an equilibrium configuration induced by the periodic boundary condition. Moreover, the viscosity ratio apart, the configuration is quite similar to that studied in  \cite{lac07}, where no such motion was described.

For small capillary numbers, more deflated capsules exhibit more complex interaction patterns, like a swapping of positions as they collide, or a pairing followed by a rotation of the couple as a whole. This happens  in conditions where the capsule would tumble, if isolated in the flow. As these behaviours take place in the middle of the simulation box, they are probably more believable. This possibility of more complex interactions has been confirmed by other numerical simulations of capsules with no viscosity contrast \cite{hu20}: for an initial position defined by small enough $|x_0|$, $z_0$ close to 0  and $y_0\ne 0$, the initial shift in the $z$ direction due to the flow induced by the reference capsule is not large, essentially because the studied capsule can flow straight. However, the attraction after interaction, as observed in Fig. \ref{fig:loewenberg}A, is maintained, such that the sign of $z$ is changed, implying a backward motion of the capsule and a new crossing. In the meantime, it is, on average, attracted in the $y$ direction (as already seen in \cite{lac08}). Depending  on deformability, the interaction might end up there (thus resembling the swapping motion described by \citet{le11}), or go on for one or more additional interaction, leading to what the authors called minuet motion. In agreement with \citet{le11}, swapping or multiple swapping (\textit{i.e.} minuet) is favored by low capillary number.

Pair interaction of vesicles has been studied through experiments, numerical simulations and theory. Using the far field perturbation due to one vesicle (which is proportional to its stresslet and decays as the inverse of the distance squared), \citet{farutin13} proved that, for weakly deformed vesicles with no viscosity contrast, placed in the same shear plane, the net displacement scales as the inverse of the initial position $z_0$ squared. They also provided an expression for the prefactor, as a function of reduced volume. Their result agrees quantitatively with their own simulations. They did not study the case of initial offset in the $y$ direction. Following the same theoretical framework, \citet{Gires12} showed that vesicles with high viscosity contrast (but not tumbling)  exhibit attraction in both directions as soon as $|y_0|>|z_0|$. The range of validity of this theory makes it however weakly amenable to experimental check. 

 \citet{gires14} studied the case of vesicles with no viscosity contrast not initially placed in the same shear plane through numerical simulations. They found that for vesicles with initial position $y_0$ larger than a threshold which is of order the vesicle radius, attraction in the vorticity direction takes place, while almost no net displacement is observed in the shear direction. The author notes that, while in \cite{Gires12} the attraction can be interpreted in terms of contribution of the far field perturbation due to the other vesicle, here the small distance between the vesicles makes it necessary to consider additional forces due to the fluid flow in the thin film created between the vesicles,  that are deformed by their interaction. This soft-lubrication approach may follow the guidelines of \citet{loewenberg97}. Numerically, the additional pressure that builds up was particularly discussed in the study of capsules dancing menuet~\cite{hu20}.

\begin{figure}[htbp]
	\centering 		
\includegraphics[width=\columnwidth]{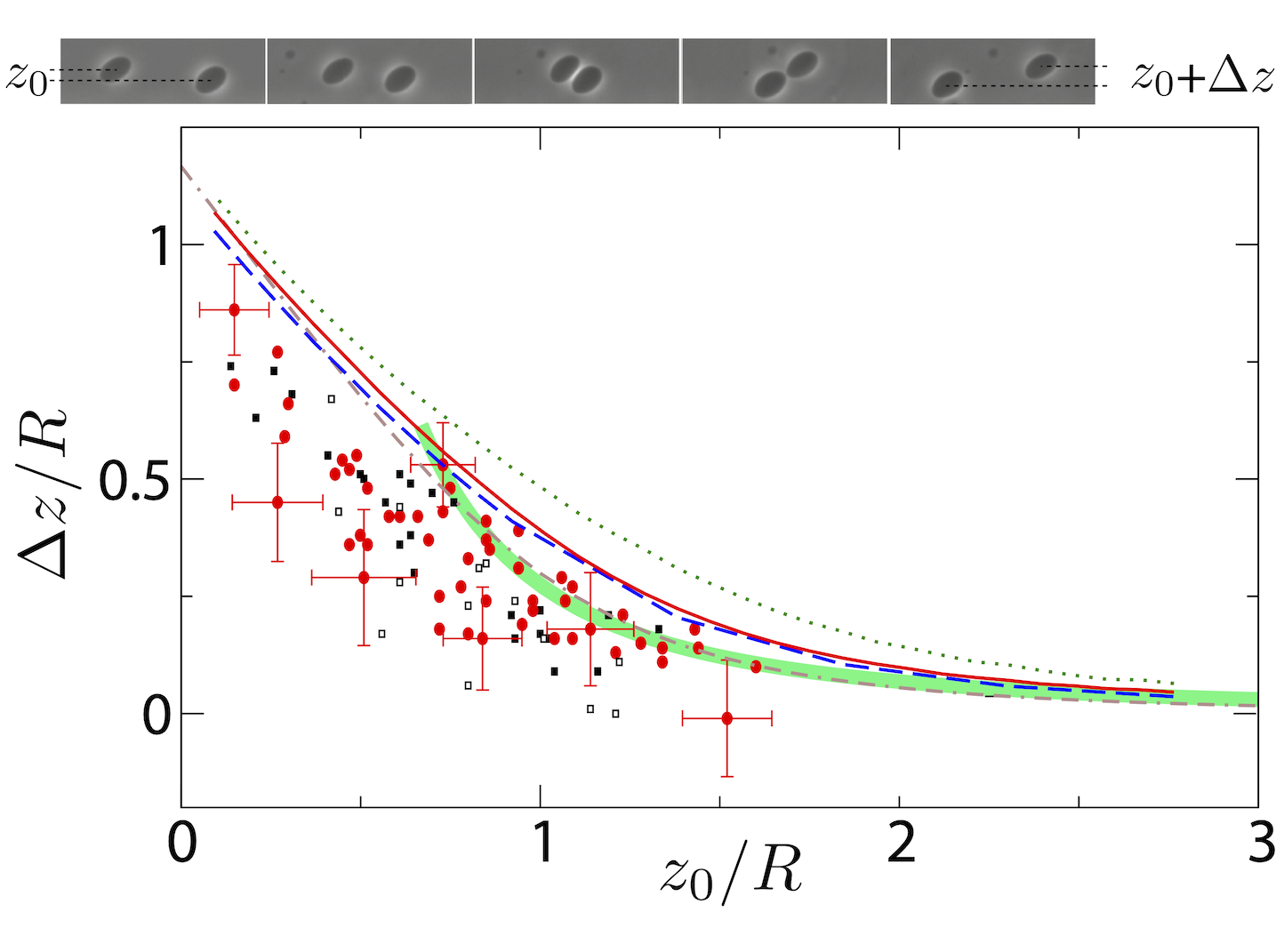}
	\caption{\textit{Net displacement $\Delta z$ as a function of initial offset $z_0$ for vesicles placed in the same shear plane. Dots correspond to experimental data for vesicles with $0.98>\nu\gtrsim 0.75$ and $\lambda=0.28 $ (black squares), $\lambda=1$  (red disks), and $\lambda=3.8$ (open squares). $Ca$ lies between 10 and 100. Dashed line ($Ca=10$), full line ($Ca=50$) and dotted line ($Ca=100$): simulations for $\lambda=1$ and $\nu=0.95$. Dash-dotted line: simplified model (Eq. \eqref{eq:interaction_roughmodel}). All data adapted from \cite{gires14}. Full thick green line: far-field theoretical prediction from \cite{farutin13}. Snapshots of two interacting vesicles of similar sizes and reduced volume taken from \cite{gires14}.}}\label{fig:gires}
\end{figure}

\citet{gires14} also presented experiments for vesicles with $0.28<\lambda<3.8$, and reduced volume $\nu\gtrsim 0.75 $ placed in the same shear plane. Their main results are shown in Fig. \ref{fig:gires}. As for most previous studies, a maximal shift of the order one radius is found. Remarkably, and in the limit of the experimental uncertainty the net displacement does not seem to depend much either on the reduced volume or on the viscosity contrast, though both are varied in a large range. The experimental results match well with simulations of vesicles with no viscosity contrast. 

In an attempt to rationalize this weak dependence on the mechanical properties of the vesicles, the authors proposed to model the interaction between the vesicles as the lift of one vesicle above a wall of finite length $2R$. Assuming the vesicle starting at $z_0$ moves with velocity $\dot{\gamma}z$ relatively to the vesicle of reference, Eq. \eqref{eq:Uwff} leads to $dz/dx=A R^3/z^3$. Integration along the trajectory leads to the net displacement \begin{equation}  \Delta z=(z_0^4/R^4+8 A)^{1/4}-z_0/R.\label{eq:interaction_roughmodel} \end{equation}

A fit of experimental data with this rough model, shown in Fig. \ref{fig:gires}, showed good agreement with single fitting parameter $A$ close to the typical values found by Olla. The overall amplitude of the interaction curve is set by the maximal displacement $(8A)^{1/4}$. This $1/4$ exponent explains why the variations of $A$ with cell mechanical property are smoothed out when net displacement is considered. Note that the long distance limit $z_0/R \gg 1 $ of Eq. \eqref{eq:interaction_roughmodel} is $\Delta z \sim 2A R^3/z^3$, which is not in agreement with more accurate theoretical developments that predict a $1/z^2$ decay \cite{farutin13} (thick line in Fig. \ref{fig:gires}). Yet, this law can serve as a good proxy for estimating the drift due to interaction. \\

In channels, the interaction between particles becomes more complex, due to the varying shear rate. Also, the presence of walls strongly modifies the perturbing field creating by one particle, and attraction/repulsion in the flow direction becomes a dominant feature, as it exists also for rigid spheres \cite{zurita-gotor2007}. This leads to complex  structuring mechanisms in the flow direction \cite{McWhirter09, ghigliotti12,Tomaiuolo12,claveria16, takeishi17, aouane17, yaya21}, which we will not study here.\\

Finally, the collision between particles of different properties is of great interest to understand segregation mechanisms within a suspension.

\citet{kumar11,kumar14, singh15} studied pairs of capsules with different rigidities; the key finding is that the stiffer particle is more displaced, although the relative displacement remains quasi constant. In line with the result for similar capsules, the net displacement depends only weakly on the capillary number. \citet{zavodsky2019} simulated interactions between red blood cells and platelets (modeled as smaller and 10 times stiffer particles than red blood cells). The displacement of the red blood cell is found to be negligible, while the platelets can be displaced by around 2 times their radii.

 \begin{figure}[ht]
	\centering 		
\includegraphics[width=\columnwidth]{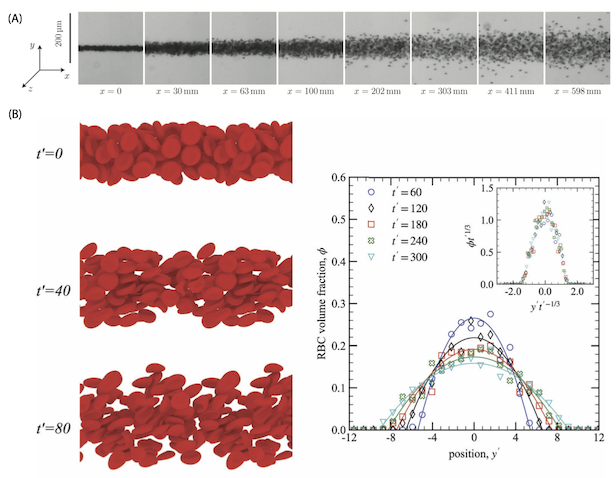}
	\caption{\textit{Shear induced diffusion of interacting particles. (A) A jet of red blood cells injected in a flat channel diffuses in the $y$ direction because of shear in the $z$ direction, thus allowing to determine the $f_3$ coefficient experimentally. Picture taken from \cite{grandchamp13}. (B) Simulations of red-blood-cell-like capsules diffusing in a simple shear flow, allowing to determine the coefficient $f_2$. The cross-stream concentration profile is the truncated parabola expected from the model. Proper rescaling shows a collapse indicating that the width increases as a function of time with exponent $1/3$.  Picture taken from \cite{Malipeddi2021}.}\label{fig:diffusion}}
\end{figure}

 \subsection{Effective diffusion in suspensions}
 
In a  suspension, the effect of multiple pair interactions with random initial relative positions, is to give birth to a diffusive-like flux.  This pair interaction is the dominant mechanism in semi-dilute suspensions (such that interactions involving more than two particles can be neglected).

This flux is one of the ingredients that can be incorporated in continuous models to describe the distribution of concentrations across a flow chamber. It will tend to oppose to the advection due to wall repulsion. When a pair interaction is sufficient for lateral displacement to occur, the diffusion constant depends linearly on the concentration of diffusing species\footnote{For spherical particles, pair interaction does not lead to net displacement, so triplet interaction is the dominant mechanism, leading to a quadratic dependency of the diffusion coefficient with the particle concentration.}. Furthermore, this effective diffusion is anisotropic, due to the inequal displacements in shear and vorticity direction.  This results in anisotropic non-linear advection-diffusion equations which can only be analytically solved in some few cases. The type of solutions strongly depends on the boundary conditions of the problem.

For instance, in \cite{rusconi08}, an initial step function of concentration of asymmetric particles is flowing in a channel where its interface diffuses. The authors found that the widening of the interface is characterized by an exponent $1/2$, as for Brownian diffusion. By contrast, it was shown in several studies  that a narrow stream of red blood cells \cite{grandchamp13,Malipeddi2021}, or droplets \cite{malipeddi2019} injected in a channel diffuses with an exponent $1/3$ (Fig. \ref{fig:diffusion}). The difference between the two experiments lies in the  boundary conditions: a step function that becomes smoother and smoother can be considered, as long as diffusion has not reached its edges, as a function with fixed maximal amplitude. By contrast, an initially narrow distribution sets a condition of constant integral, and not of constant maximal value. In this case, the subdiffusive behaviour is accompanied by the existence of a different family of self-similar distributions, which are truncated parabolas (Fig. \ref{fig:diffusion}). This subdiffusion can be understood by the fact that the more the cells diffuse, the less they interact, so the less lateral displacement is produced.

For particles placed in a simple shear flow, the down-gradient diffusive flux for a suspension of local concentration $\phi$ can be written as \begin{equation} \mathbf{J}= - f R^2 |\dot{\gamma}| \phi \mathbf{\nabla} \phi. \label{eq:diffusiveflux} \end{equation}

The term $|\dot{\gamma}| \phi$ accounts for the frequency of collisions. The dimensionless prefactor $f$ is related to the detail of the interaction, as described previously for different kind of particles. If the gradient of concentration is in the shear direction, this coefficient is often denoted $f_2$, and $f_3$ if the gradient is in the vorticity direction. Although it is possible to describe locally the flow in a channel as a simple shear flow (whose axis depend on the position in the channel) \cite{grandchamp13}, this description in terms of two different coefficients is certainly not sufficient to account for effective diffusive processes in channels, where shear gradients must also be taken into account. For that reason, most characterization of fluxes were run in simple shear flows, to the exception of the experiments in channels of \cite{grandchamp13}, where simplifying hypothesis had to be made.

While looking at the time-evolution of concentration profiles allows to determine these unknown coefficients $f_i$, this method becomes more complex if other effects have to be considered, in particular that of the presence of walls. If the time-evolution equation becomes difficult to solve analytically, it is still possible to solve for the resulting stationary distribution, assuming the effective diffusive flux and convective flux due to the wall-induced lift can be simply added. If the lift velocity due to the walls is known, this distribution is a function of $f_2$ only \cite{hudson03,podgorski11,bureau17,malipeddi2019}.

For droplets, a coefficient $f_2$ of order 0.2 was experimentally found by \citet{hudson03}, for drops of viscosity ratio close to 0.2. The dependence with the capillary number was not studied. In the numerical simulations of  \citet{malipeddi2019}, $f_2$ is a nonmonotonic function of the capillary number taking values between 0.2 and 0.45, a maximal value being reached for intermediate capillary number. This behaviour agrees with the calculation of \citet{loewenberg97} for the self-diffusivity coefficient, and can be understood as follows: for small capillary numbers, the drops stay spherical and do not diffuse. For large capillary numbers, their deformation is so strong that they elongate in the flow, which results also in a quasi-symmetric situation. This discussion is similar to that held for vesicles near a wall, where such a nonmonotonic behaviour was observed as a function of reduced volume, that controls their ability to deform.

Coefficient $f_2$ was also experimentally determined  for slightly deflated lipid vesicles with no viscosity contrast, and a coefficient $0.06\pm 0.02$ was found by two different methods \cite{bureau17}.

For red blood cells, experiments for cells under moderate shear rate --- such that they are in a tumbling-like regime --- have led  to $f_3\simeq0.2$ and $f_2\simeq2.7$ \cite{grandchamp13}\footnote{The values given here were re-calculated from the original article \cite{grandchamp13} where the authors use for $R$ the maximal radius 3.6 $\mu$m; in this review, $R=2.8$ $\mu$m is based on the cell volume.}. The latter value is strongly different from that found for drops or vesicles; however, a renormalization of the concentrations by considering the effective volume occupied along time by these tumbling cells lead to find closer results, though the effective diffusion of red blood cells still appears as stronger. In their numerical simulations, Malipeddi and Sarkar found that $f_2$ increases from 0.3 to 0.6 as the capillary number increases and allows for transition between a tumbling-like to a tank-treading regime \footnote{The values given here were re-calculated from the original article \cite{Malipeddi2021} where the authors use for $R$ the maximal radius 4 $\mu$m; in this review, $R=2.8$ $\mu$m is based on the cell volume.}. A small decrease is however observed as the cell transits between the two regimes. These values are much smaller than that found experimentally. A potential explanation could be that the experiments were run in a Poiseuille flow and simplifications in the modeling had to be made to lead to equations that could be solved. Also, the authors mainly studied the non-physiological case $\lambda=1$, but they showed on selected cases that the diffusion coefficient does not vary much with $\lambda$. Considering that strong modifications in a red blood cell dynamics are expected upon a transition to physiological to unity viscosity contrast \cite{fischer13,minetti19}, this point would deserve to be further elucidated.

Regarding self-diffusivity, the difficulty in tracking particles among others makes numerical methods the tool of choice for the determination of diffusion coefficients. Conclusions on the effect of mechanical properties are in line with the previous discussions, see e.g \cite{Malipeddi2021}.

Finally, we remark that $f_3$ coefficient has, in general, seldom been measured. In particular, consequences of attraction in the vorticity direction has never been observed, nor introduced in models.  As it would reinforce concentration gradients rather than smooth them  out, it may lead to interesting problems where initially homogeneous suspensions could become unstable. \\

Interactions between particles of different sizes or mechanical properties may  lead to segregation effect within the suspension. A key example is that of blood, where platelets and white cells are often met in the edges of the channels. This multi-parameter problem is complex and lies beyond the scope of this review. It has been studied, mainly through numerical simulations, by different groups, and would probably deserve a review in the next years \cite{crowl_fogelson_2011,kumar11,kumar12,fedosov12,kumar14,kruger16,Muller16,rivera16,Chang18,ye2019,zavodsky2019,Zhang2020_2}. As for effective diffusion of a single type of particle, it can also be addressed through continuous models involving cross terms between the different types of particles, as in \cite{rivera16}.

\subsection{Creation of cell free layers in blood flow}

The flow of red blood cells in microcirculation is marked by the existence of a cell-free layer (CFL)  near the walls \cite{fedosov10,narsimhan13,katanov15,sherwood12}, which has first been observed by  \citet{poiseuille1835} almost two centuries ago. This CFL has been acknowledged to be at the origin of the decrease of the apparent viscosity referred to as F\r{a}hr\ae{}us-Lindquist effect \cite{Lindqvist31} as well as the decrease of the hematocrit in small vessels compared to large ones \cite{Fahraeus1929,popel05}. In-vitro, the existence of this CFL can be used to separate red blood cells from other components, including plasma \cite{abkarian08,Li2020_Ye}.

The shear-induced lift of red blood cells is reckoned
as the main origin of the creation of this depleted layer. In a first approach, one can quantify this depletion layer by zeroing the sum of the advection flux $\phi U_L$ and of  the effective diffusive flux (Eq. \eqref{eq:diffusiveflux}), for a given mean volume fraction.

Using such a model, one can calculate an analytical relationship between mean concentration and thickness of the CFL,  in a simple shear flow where the lift velocity is assumed to be the sum of the lift velocities due to each wall. Doing so,  \citet{rivera16} proposed a fit of several data coming from previous simulations or experiments. This result was obtained with a fit parameter $A/f_2$ of order 0.5 (where $A$ is the constant of Eq. \eqref{eq:Uwff} and $f_2$ that of Eq. \eqref{eq:diffusiveflux}). This value deserves a comment: in \cite{rivera16}, agreement is found in particular with numerical simulations run by the same group \cite{kumar14}, where capsules are considered,  whose characteristics are such that they are in a tank-treading regime. For vesicles in tank-treading regimes, the $A/f_2$ ratio is of order $0.1/0.06\sim 1.7$ which is indeed of the same order of magnitude as the ratio obtained from the fit. At that point, the picture is clear. On the other hand, red blood cells in microcirculation are clearly not in such a regime, when isolated. Indeed, for red blood cells, the ratio becomes $0.016/2.7\sim0.006$ \cite{grandchamp13}, which is much lower and would lead, when used in the theoretical model, to the absence of CFL. The effective $A/f_2$ parameter of order 0.5 that is needed to account for the presence of the CFL highlights the complexity of modeling structuring effects by continuous models. This questions the relevance of such a modeling whose goal is indeed to establish a micro-meso link between cell mechanical properties and structure of the suspension, unless additional ingredients are considered.

In particular, the simple model above  neglects several features: the modification of cell-cell interactions in the vicinity of walls, the screening of lift forces by neighboring cells, and the modification of cell dynamics due to the presence of neighboring cells; indeed neighbors tend to prevent tumbling-like motion, which would favor an increase of the lift parameter $A$.  In this spirit, an attempt to determine the lift force on a cell under an external force directed towards the wall that mimics the effect of neighboring cells can be found in \cite{hariprasad14}.

Another ingredient may also be considered:  in a Poiseuille flow, collisions between red blood cells
induce a transverse flow because of the concentration gradient, but also because of the shear rate gradient, which also makes the collision probability asymmetric. One can show that the associated flux reads \cite{rivera16} \begin{equation} -(f_2-2 f_{2s}) R^2 \Phi^2  \frac{\partial | \dot{\gamma}|}{\partial z}, \end{equation} where $f_{2s}$ is the $f$-coefficient associated with self-diffusion. As $f_2$ is always greater than $2 f_{2s}$ \cite{dacunha96}, this flux is directed towards the center of the channel. In a channel of radius $r$, the ratio $\zeta$ between the convective flux and this new effective diffusive flux reads 

\begin{equation} \zeta= \frac{\xi}{R^{1-\delta} (f_2-2 f_{2s})}\times\frac{ r-z}{z^\delta \Phi} \simeq 0.007 \frac{ r-z}{z^\delta \Phi}, \end{equation}where the last equation was obtained using $R=2.8\,\mu$m, $\xi=1.1\times 10^{-2}$ and $\delta = 1.3$ \cite{losserand19}, $f_2=2.7$ and $f_2/(f_2-2 f_{2s}) \simeq 9/7$ \cite{grandchamp13}. For a channel radius of order some tens of microns, a cell even quite close to the wall ($z \simeq R$) and a volume fraction of some 10\%, this ratio is $\lesssim 1$, meaning yt the effect of asymmetric collision due to shear gradient cannot be omitted and may deserve to be considered as a contributor to the creation of cell free layers.

Finally, modeling the core of the suspension, where the highest concentrations are expected, as a suspension where only pair interactions take place, is probably not relevant. In addition, the modification of the local rheology due to this concentration leads in practice to a plug flow with high shear region near the walls (see, \textit{e.g.} \cite{roman16}). \\

The agreement between the numerical simulations of \citet{kumar14} --- which are not based on red-blood-cell-like objects --- and experimental observations on red blood cells, as far as CFL thickness is concerned, leads to question the ability of these to predict other phenomena impacted by the cell mechanical properties. More generally, benchmarking of numerical methods on the behaviour of the particles under flow is often partial. Regarding capsules, this can be explained by the lack of experiments quantifying lift, but experimental results on red blood cells under flow do exist \cite{yao01,dupire12,grandchamp13,lanotte16,minetti19,losserand19,amirouche20}. Nevertheless, numerical methods are often validated only through quasi-static standard configurations like micropipette aspiration or optical tweezers stretching --- as in  \cite{Malipeddi2021} or in  \cite{Balogh17} which is used by \citet{balogh19} to set exhaustive discussion on the dynamics of creation of the CFL in complex networks --- or by considering simpler objects like quasi-spherical capsules --- as in  \cite{Doddi09} --- or through the observation of a collective behaviour --- as in \cite{Balogh17,Fedosov11,ye2019} --- which may hide several offsetting issues. 
More precisely, \citet{siguenza17} showed  that agreement on quasi-static load is not sufficient. Agreement with experiments under flow would therefore be a plus, keeping in mind that, quoting  \citet{nicoud19}, "this is in fact not always sufficient as the robustness of the numerical results to physical/numerical parameters may be so large that a good agreement may be reached by chance". Efforts in running comparison with single cell dynamics results has been noticed in the recent literature \cite{zavodsky2019,Zhang2020_2}.\\

As a concluding remark, dynamic interactions between soft particles exhibit a rich variety of behaviours, including attraction, whose impact at the level of a suspension have not yet been discussed. As noted above, the creation of a cell-free layer in a blood stream (and, by extension, in any other confined flow of deformable particles) has not yet been modeled in a framework that relies on what is known about cell-cell and cell wall interactions. Layering effects have been reported in such suspensions, which are still unexplained \cite{thiebaud14,Shen16,shen17,zhou20,feng21,audemar2022}. Another issue is that of the modeling of interactions in the vicinity of the flow centerline, where the shear rate is zero, hence a vanishing diffusive flux. This leads to unphysical accumulation of particles in the centerline, if one uses continuous effective diffusion models \cite{Philipps92}. While some ad-hoc corrections can be introduced to account for this finite-size effect,  there is certainly a long way to go before establishing a comprehensive link between local mechanisms and  effective diffusive flux in such a configuration, where the  longitudinal attraction/repulsion mechanisms should also be considered.

\section{Flow-induced electrokinetic lift}
\label{sec:eklift}
\subsection{Context}
In relation to the flow properties of fluid-suspended objects mentioned before, it is of interest to note that, for suspensions of charged particles in an electrolyte, a phenomenon known as the ``primary electroviscous effect'' has been identified since the 50s (see \cite{Hinch1983} and references therein), which points to the importance of the coupling between flow and ionic transport near the surface of the particles, resulting in a modified lubrication drag~\cite{Rodriguez2022}, and an enhanced viscosity of charged suspensions compared to uncharged ones. Along this line, we describe in the following section the electrokinetic effects that give rise to lift forces at play at low Reynolds numbers with rigid objects bearing surface charges. Such forces have, in recent years, been mainly described and exploited in the context of particle manipulation (separation, focusing) in microfluidic applications, in which an external electric field is applied parallel to the channel walls. These applications, and their theoretical foundations, have been recently reviewed by \citet{Xuan2019}. Therefore, it is beyond our scope here to cover the rich corpus of observations and predictions made on cross-stream particle motion in the presence of an electric field (we briefly come back on these in the concluding part of this section). Rather, we limit our discussion to the case where no external electric field is applied, with the aim of (i) emphasizing that such electrokinetic effects do arise in flow situations with no electrical driving, (ii) providing the reader with a concise review of their modeling, and (iii) evaluating whether such effects are important to account for in aqueous-based suspensions.

\begin{figure}[htbp]
	\centering 		
\includegraphics[width=\columnwidth]{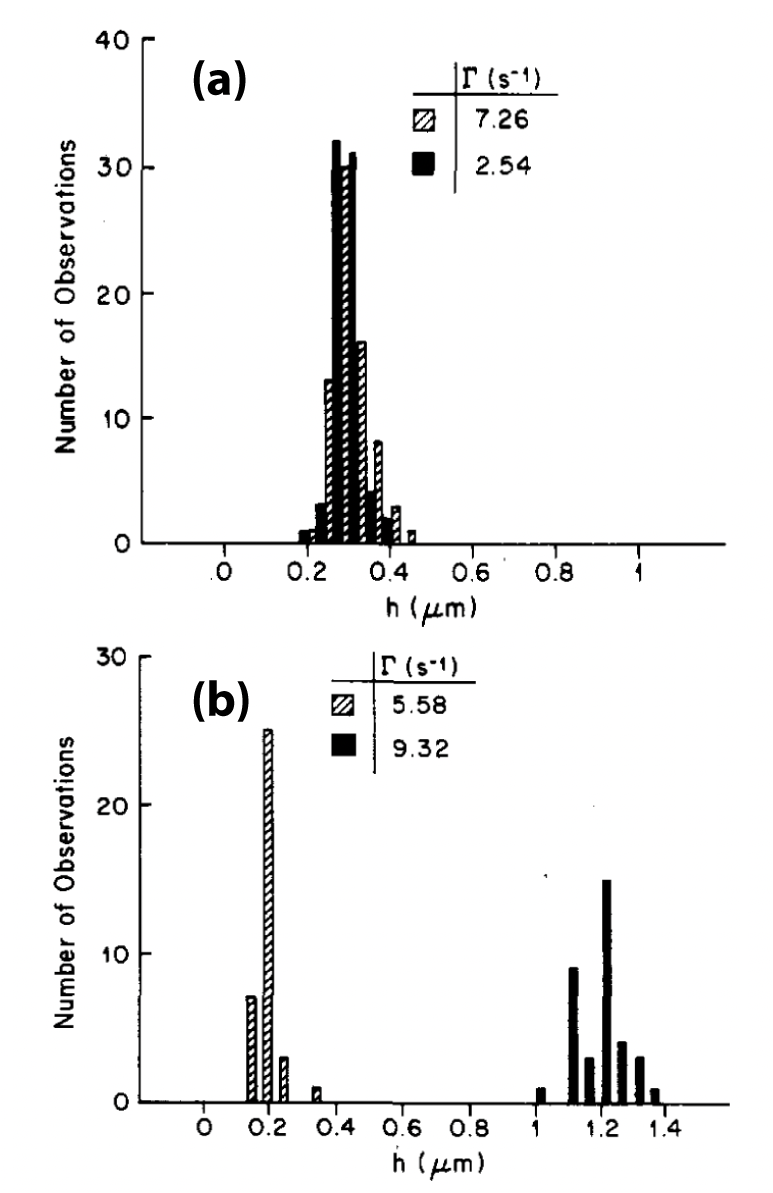}
	\caption{\textit{Histograms of bead/wall distances obtained at two different shear rates  $\Gamma$ in water/glycerol mixtures of (a) $\mu=2.9\times10^{-3}$ Pa.s$^{-1}$ and $K=2$ $\mu S$.cm$^{-1}$, and (b) $\mu=630\times10^{-3}$ Pa.s$^{-1}$ and $K=0.046$ $\mu S$.cm$^{-1}$. Adapted from  \cite{Alexander1987}.}}\label{fig:electro1}
\end{figure}

\subsection{Experimental observations}

\citet{Alexander1987} described an experimental method designed to determine the interaction potential between a colloidal particle and a surface. Their approach consisted in measuring the temporal fluctuations of the translation velocity of a bead driven by a shear flow near a flat wall, and to rely on theoretical results established previously by \citet{Goldman1967} in order to infer, from their velocity measurements, the distance between the bead and the wall, using the following relationship between bead velocity $V$, shear rate $\dot{\gamma}$, bead radius $R$ and bead/wall distance $h$  \cite{Goldman1967}:

\begin{equation}\label{eq:GCB1}
V(h)\simeq |\dot{\gamma}|R\frac{0.7431\left(1+h/R\right)}{0.6376-0.2\ln \left(h/R \right)}.
\end{equation}

Doing so, they assumed that the shear flow did not perturb the equilibrium colloidal forces to be characterized (arising from electrostatic double layer interactions in their experiments). In order to validate experimentally such a hypothesis, they performed a series of measurements in which they varied the strength of the shear flow (the shear rate at the wall, $\dot{\gamma}$), and the viscosity of the suspending fluid (working with various water/glycerol mixtures). While they indeed measured no effect of $\dot{\gamma}$ in low viscosity fluids, they unexpectedly observed that, in liquids with high glycerol contents, the flowing beads travelled at a larger distance from the wall at higher shear rates (Fig. \ref{fig:electro1}). 

\begin{figure}[htbp]
	\centering 		
\includegraphics[width=\columnwidth]{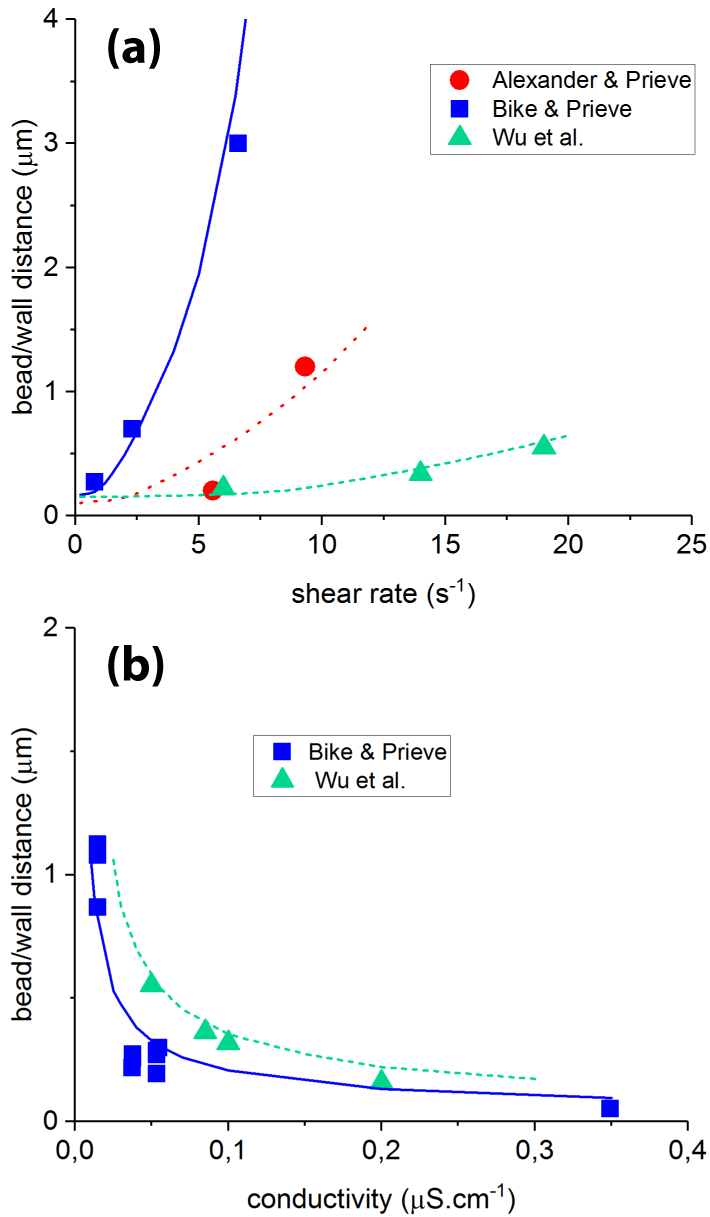}
	\caption{\textit{(a) Bead/wall gap distance as a function of imposed fluid shear rate, taken from references \cite{Alexander1987} (red filled circles), \cite{Bike1995a} (blue squares), and \cite{Wu1996} (green triangles). The three datasets have been obtained with beads of diameter $D=9.2 \,\mu$m and solution conductivity $K=0.046\, \mu$S.cm$^{-1}$ \cite{Alexander1987},  $D=10 \,\mu$m and solution conductivity $K=0.013\, \mu$S.cm$^{-1}$ \cite{Bike1995a}, and $D=5.1 \,\mu$m and solution conductivity $K=0.05\, \mu$S.cm$^{-1}$ \cite{Wu1996}. (b) Distance as a function of carrying fluid conductivity, taken from \cite{Bike1995a} (blue squares), and \cite{Wu1996} (green triangles). The two datasets have been obtained with $D=5.2 \,\mu$m and shear rate $\dot{\gamma}=6$ s$^{-1}$ \cite{Bike1995a}, and $D=5.2 \,\mu$m and shear rate $\dot{\gamma}=19$ s$^{-1}$ \cite{Wu1996}. In (a) and (b), the lines correspond to theoretical predictions using a lift force as computed from Eq. \eqref{eq:Taba}, as described in  the text. Theoretical curves in (a) were obtained with $\psi=-40$ mV and ionic strength $C_{\infty}= 2\times 10^{-4}$ M (red dotted line), $\psi=-45$ mV and  $C_{\infty}= 10^{-4}$ M (blue full line), $\psi=-40$ mV and  $C_{\infty}= 1.6\times 10^{-4}$ M (green dashed line). Curves in (b) were obtained with $\psi=-30$ mV (blue full line) and $\psi=-40$ mV (green dashed line), with $C_{\infty}$ varied in the range $10^{-4}-2.5\times10^{-3}$ M. }}\label{fig:electro2}
\end{figure}

This first observation was followed by more systematic studies by \citet{Bike1995a} and  \citet{Wu1996}, who investigated in more details the role of shear rate and suspending fluid composition on the observed lift of flowing particles. Their findings are summarized in Fig.  \ref{fig:electro2}: both groups of authors observed, as initially found by Alexander and Prieve, that the bead/wall distance increases as the shear rate is increased, this effect being much more pronounced in fluids of higher glycerol content.

\subsection{Origin}

The observed phenomenon being amplified in high glycerol content fluids, this rules out hydrodynamic inertial effects to be at the origin of the lift, as those would rather be weakened upon increasing the fluid viscosity, which is the case at increasing concentrations of glycerol. As noted already by \citet{Alexander1987}, high glycerol content fluids also exhibit lower conductivities, which rather hints to an electrokinetic origin, with a lift force associated to the streaming potential arising from the relative motion of two charged surfaces.

\begin{figure}[htbp]
	\centering 		
\includegraphics[width=\columnwidth]{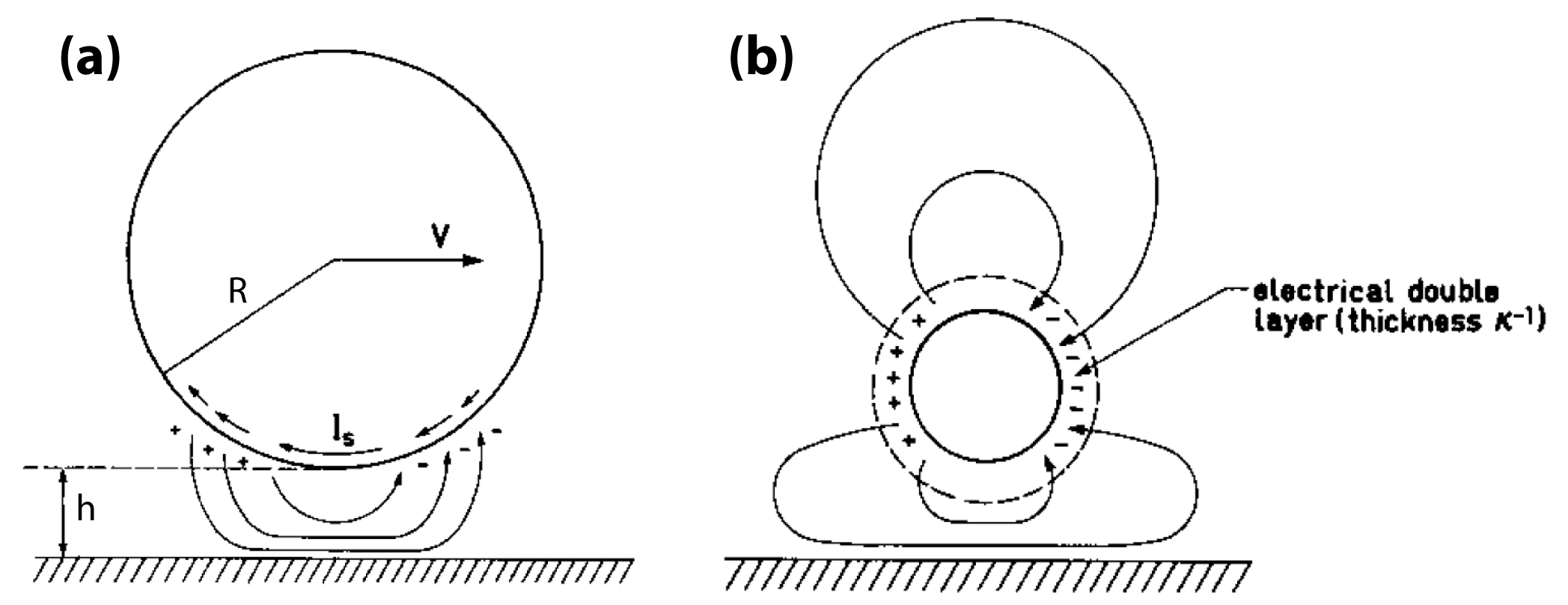}
	\caption{\textit{(a) A charged particle of radius $R$ translating in a fluid at velocity $V$ and at a distance $h$ of a flat surface. The bead/fluid relative motion sweeps charges within the Debye layer that screens the bead surface charges, resulting in near-surface currents ($I_{\textrm{s}}$). (b) The field associated to the dipole induced by the streaming currents displays a non-axial symmetry due to the proximity of the wall, which results in a force acting transversally to the bead motion.  Adapted from \cite{Vandeven1993}.}}\label{fig:electro3}
\end{figure}

Indeed, when a solid bearing surface charges is in motion relative to a polar liquid, the fluid flow associated with this motion induces currents of ions within the near-surface Debye layer that screens the surface charges from the electroneutral bulk liquid. Such a charge transport within the Debye layer is compensated for by the buildup of currents in the bulk of the surrounding fluid (Fig. \ref{fig:electro3}a). An electric field is induced by these streaming currents, which has two consequences: (i) it sets the Maxwell (electrical) stress acting on the body; and (ii) it creates an electro-osmotic flow that perturbs the initial driving flow. In addition, polarization of the ionic concentrations in the liquid surrounding the particle gives rise to a diffusio-osmotic flow perturbing further the driving flow. These osmotic phenomena thus contribute to the net hydrodynamic stress acting on the solid. For a charged sphere purely translating in an unbound polar fluid, all these effects result in an extra drag acting on the sphere, along the direction of motion, but no force acting transverse to the motion of the bead. Any factor breaking the axial symmetry of this situation will induce a force transverse to the motion, {\it i.e} a lift force: this can be for instance an angular velocity imposed to the bead \cite{Khair2019}, or the presence of another solid/liquid boundary (electrically charged or not) near the flowing particle (Fig. \ref{fig:electro3}b).

\subsection{Modelling}

Soon after the initial observations described above, several groups of authors have attempted to establish a theoretical description of the phenomenon for a bead of radius $R$ translating at velocity $V$ at a distance $h$ from a flat wall (see Fig. \ref{fig:electro3}). This formally amounts to solving  a set of equations consisting of (i) the Nernst-Plank equation describing the convection-diffusion of ionic species, (ii) the Poisson equation relating the electric potential to the density of charges in the fluid, (iii) the Stokes equation accounting for Coulomb forces, balancing pressure, viscous and electrostatic forces, and (iv) the continuity equation (fluid incompressibility). These are associated with boundary conditions imposing no slip, no normal current, and electric potential on each solid surfaces.

Such coupled electro- and hydrodynamic problems are mathematically quite involving. We skip here all the technical aspects related to solving, present qualitatively the assumptions made in  the various theoretical studies and provide the analytical expressions obtained for the lift force under these assumptions. 

As summarized by \citet{Cox1997}, the solutions of such a type of problems depend, in addition to the distance between the solids, their shape and relative motion, on the following parameters:
\begin{itemize}
\item the Peclet number $Pe=VR/D_1$, with $D_1$ the diffusion coefficient of (say) cations, comparing convection to diffusion effects,
\item $D_1/D_2$, the ratio of cation to anion diffusivities,
\item the Debye length $\kappa^{-1}=\sqrt{\epsilon k_B T/(2z^2e^2c_{\infty})}$, {\it i. e.} the extension of the ion cloud screening surface charges, with $\epsilon$ the fluid permittivity, $k_BT$ the thermal energy, $z$ the valency of the ions, $e$ the elementary charge, and $c_{\infty}$ the bulk (number) ion concentration,
\item the Hartmann number $\lambda_H=2c_{\infty}k_BTR/(\mu V)$, with $\mu$ the fluid dynamic viscosity, giving the relative importance of electrical body forces on fluid flow,
\item the particle, $\psi_p$, and wall, $\psi_w$, surface potentials (or their dimensionless forms $\tilde{\psi}_{p,w}=\psi_{p,w}ze/(k_BT)$
\end{itemize}

A number of attempts have been made in order to determine the normal force that could arise from electrodynamic and hydrodynamic couplings when a bead flows near a flat wall. 

\citet{Bike1990} employed the lubrication approximation combined with the assumption that the Debye layer $\kappa^{-1}$ is smaller than the gap $h$, {\it i.e.} $\kappa^{-1}\ll h \ll R$ (this is the so-called  ``thin Debye layer limit''), and computed an electrokinetic lift force $F_{BP1}$ reading:

\begin{multline}\label{eq:Fbp1}
F_{BP_1}=\left(\frac{\epsilon}{4\pi} \right)^3\frac{\pi RV^2}{K^2h^3}\left[ 0.384 \psi^2 \right. \\ 
\left. +0.181 \psi \Delta\psi +0.0242 (\Delta\psi)^2\right],
\end{multline}
where $\epsilon$ and $K$ are the fluid permittivity and conductivity, $\psi=(\psi_w+\psi_s)/2$ and $\Delta\psi=\psi_w-\psi_s$. 
The above expression was obtained by accounting only for the Maxwell stress arising from the streaming potential, and neglecting {\it a priori} other electro-osmotic perturbations of the driving flow.

The same authors also derived, in a subsequent article  in which they relaxed the lubrication approximation, an expression for the lift force that holds for $h\gtrsim R$  \cite{Bike1992}:

\begin{equation}\label{eq:Fbp2}
F_{BP_2}=\left(\frac{\epsilon}{4\pi} \right)^3\frac{27\pi R^2V^2}{16K^2(R+h)^4}\left( \psi_s+2\psi_w\right)\psi_s.
\end{equation}
This expression coincides with that obtained by \citet{Vandeven1993} when $\psi_w=0$.

Equations (\ref{eq:Fbp1}) and (\ref{eq:Fbp2}) both capture qualitatively the fact that the lift force, hence the bead/wall distance, is expected to be larger at larger shear rates (recalling that $V\sim \dot{\gamma}R$) and for lower solution conductivity $K$. However, when used with physically sound values for $\psi$, $\epsilon$ and $K$, none of the above expressions allows to quantitatively account for the steady-state bead/wall distances measured experimentally, with computed lift forces several orders of magnitude too low to explain observations \cite{Bike1990,Bike1995b,Bike1992,Vandeven1993}. 

The problem was tackled later by  \citet{Cox1997}, who pointed out that, in contrast to what was assumed in previous works, the dominant contribution is not due to the Maxwell stress alone but arises from the electro-osmotic flow generated by the streaming potential, which perturbs the driving flow. Cox derived a general solution scheme, using asymptotically matched expansions in $\delta=1/(\kappa R)$, which is valid in the thin-Debye-layer limit. This framework was employed by several authors in order to address the specific problem of a charged sphere translating at speed $V$ and rotating at angular velocity $\Omega$ in the vicinity of a charged wall \cite{Wu1996,Warszynski1998,Tabatabaei2006}. \citet{Wu1996} and \citet{Warszynski1998} made derivations for a cylinder/flat geometry, followed by the use of Derjaguin approximation to convert the obtained result to the sphere/flat situation, whereas the work reported in  \cite{Tabatabaei2006} was obtained directly for a sphere. We thus provide below the expression for the electrokinetic lift force derived by \citet{Tabatabaei2006} \footnote{The provided expression for $F_{Taba}$ is obtained by ``re-dimensionalizing'' the dimensionless forms reported in \cite{Tabatabaei2006} as equations (7.3) and (7.4). Doing so, we noted a series of misprints in the original article by Tabatabaei {\it et al.}: (i) dimensionless forces $\tilde{F}$ should read $\tilde{F}=F/(\mu V R)$ (and not  $\tilde{F}=F/(\mu V)$ as in eq. 2.2 in \cite{Tabatabaei2006}), and (ii) the dimensionless lift force of eq. 7.3 should read $\tilde{F_z}=4\pi\lambda_H Pe^2(\kappa^{-1}/R)^4(h/R)^{-2} f_z$ (and not $\tilde{F_z}=4\pi Pe^2(\kappa^{-1}/R)^4(h/R)^{-2} f_z$)}:

\begin{multline}\label{eq:Taba}
F_{Taba}=\frac{12\pi \epsilon^2 (k_B T)^3 R^2}{25(ze)^4c_\infty h^2} \\
\times \left\{ \left[ \left( \frac{G_p}{D_1}+\frac{H_p}{D_2}\right) + \left( \frac{G_w}{D_1}+\frac{H_w}{D_2}\right) \right]^2\left( V+R\Omega \right)^2  \right. \\
\left. -\alpha_3\left[ \left(\frac{G_p}{D_1}+\frac{H_p}{D_2}\right) - \left( \frac{G_w}{D_1}+\frac{H_w}{D_2} \right) \right]^2 \left( V^2-R^2\Omega^2\right)  \right\},
\end{multline}
with $\alpha_3\simeq -1.66678$, and the quantities $G_i$ and $H_i$ defined as:

\begin{equation}
G_i=\ln\frac{1+e^{-\tilde{\psi}_i/2}}{2}, \,\,\, H_i=\ln\frac{1+e^{\tilde{\psi}_i/2}}{2},
\label{taba2}
\end{equation}
where $i=(w, p)$ stands for wall and particle. 
The above expression was shown by the authors to hold valid for low and moderate (of order a few unities) Peclet numbers \cite{Tabatabaei2006}.

\begin{figure}[htbp]
	\centering 		
\includegraphics[width=\columnwidth]{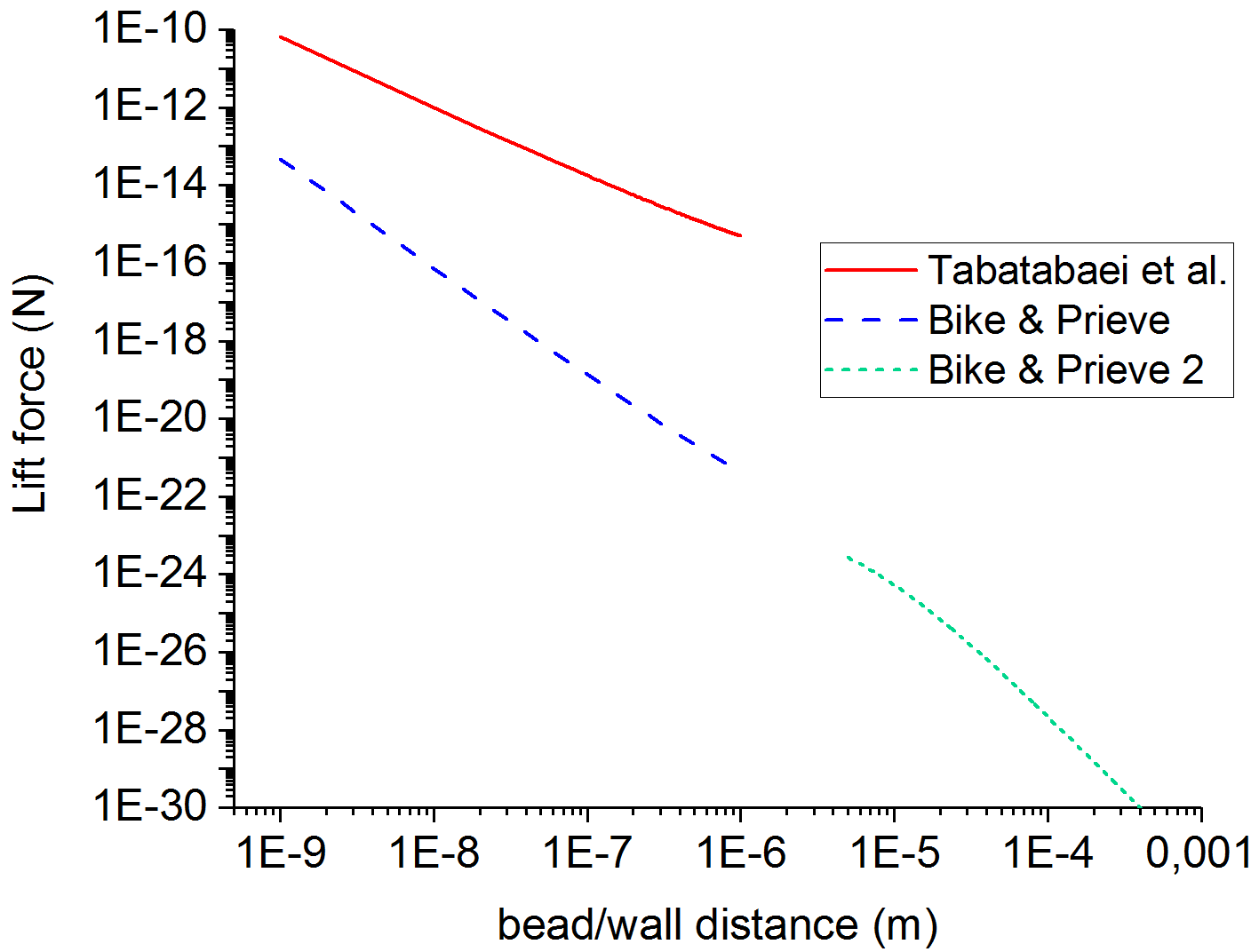}
	\caption{\textit{Comparison of lift force predictions computed using Eq. \eqref{eq:Taba} (Tabatabaei {\it et al.}, red line), Eq. \eqref{eq:Fbp1} (Bike \& Prieve, blue dashed line), and Eq. \eqref{eq:Fbp2} (Bike \& Prieve 2, green short-dashed line), in the case of a pure translation motion ($\Omega=0$ in Eq. \eqref{eq:Taba}). Computation were done using $R=5\,\mu$m, a salt concentration $C_{\infty}=10^{-5}$M (with the number concentration per cubic meter $c_{\infty}=C_{\infty}\times 10^3 \times N_A$), $T=300$ K, $\psi_w=\psi_s=-50$ mV, and $\epsilon=80\epsilon_0$ taken for aqueous suspending fluid. A shear rate $\dot{\gamma}=10$ s$^{-1}$ was used to compute $V(h)$ according to Eq. \eqref{eq:GCB1}. Diffusion coefficients were set to $D_1=1.33\times 10^{-9}$ m$^2$.s$^{-1}$ and $D_2=2\times 10^{-9}$ m$^2$.s$^{-1}$ (typical for Na$^+$ and Cl$^-$ in water), and solution conductivity $K$ estimated as $K=e^2c_{\infty}(D_1+D_2)/k_B T$.} }\label{fig:electro4}
\end{figure}

For the sake of comparison, we have plotted on Fig. \ref{fig:electro4} the electrokinetic lift forces predicted by Eqs. \ref{eq:Fbp1}, \ref{eq:Fbp2} and \ref{eq:Taba}, as a function of bead/wall gap distance $h$. It clearly appears that, in addition to the different $h$-dependence predicted by the theories, the lift force computed by Tabatabaei {\it et al.} using Cox's framework is several orders of magnitude larger than that computed by Bike and Prieve.

More recently, \citet{Schnitzer2011}  pointed out an inconsistency in Cox's solution scheme. These authors noted that the Hartmann, $\lambda_H$, and Peclet, $Pe$, numbers are not independent, but linked via $\lambda_H Pe\sim 1/\delta^2$, with $\delta=\kappa^{-1}/R$. Therefore, in the $\delta\rightarrow 0$ limit used in \cite{Cox1997}, $\lambda_H$ and $Pe$ cannot be both of order 1, contrary to what was assumed by Cox. Yariv {\it et al.} therefore revisited Cox's scheme in a series of articles treating the two cases \{$Pe\gg 1$, $\lambda_H= \mathcal{O}(1)$\} \cite{Schnitzer2011,Schnitzer2012b} and \{$\lambda_H \gg 1$, $Pe= \mathcal{O}(1)$\} \cite{Schnitzer2012a,Schnitzer2016} separately.

In the \{$Pe\gg 1$, $\lambda_H= \mathcal{O}(1)$\} limit, they find a lift force which, as assumed by Bike and Prieve, is governed by the contribution of the Maxwell stress, and derive an expression that reduces to Eq.~(\ref{eq:Fbp1}) above \cite{Schnitzer2012b}. In the opposite limit where $\lambda_H \gg 1$ and $Pe= \mathcal{O}(1)$,  \citet{Schnitzer2016} demonstrate that the leading contribution to electroviscous effects is due to the diffusio-osmotic flow resulting from salt concentration polarization, and derive an expression for the lift force that is identical to the one obtained by Tabatabaei {\it et al.}. It thus appears that, in spite of the improper assumption made by Cox, a fortuitous cancellation of errors in the solution scheme has led  Tabatabaei {\it et al.} to reach a valid expression for the lift force.

 \subsection{Comparison with experiments}
 
 Let us now estimate the order of magnitude of $\lambda_H$ and $Pe$ typically encountered in the experiments described in the first section: with beads of radius $R$ of micrometric size, flowing at a velocity $V$ being a fraction of $\dot{\gamma}R$, and an ionic diffusion coefficient in high viscosity solutions of $D\simeq 10^{-12}$ m$^{2}$.s$^{-1}$, one finds a Peclet number in the range $0.5-10$ for shear rates in the range $1-10$ s$^{-1}$. Conversely, with salt concentration of about 10$^{-4}$ M in solutions of viscosity $\mu\sim 1$ Pa.s, the Hartmann number falls in the range $100-1000$ for the same range of shear rate. Under such conditions, the \{$\lambda_H \gg 1$, $Pe= \mathcal{O}(1)$\} limit  identified by Yariv {\it et al.} seems appropriate for a direct comparison of theoretical predictions with experimental observations.
 
 As was done in previous studies \cite{Wu1996}, we compute the bead/wall distance at steady-state from the following force balance:
\begin{equation}
F_{lift}+F_{Debye}=F_{grav},
\label{eq:balance}
\end{equation} 
in which the electrokinetic force $F_{lift}$ and the double-layer force $F_{Debye}$ both repel the bead from the surface and balance the gravity $F_{grav}$ that brings the bead towards the wall. The latter merely reads:
\begin{equation}
F_{grav}=\frac{4\pi}{3}R^3g\Delta\rho,
\label{eq:grav}
\end{equation} 
with $g=9.81$ m.s$^{-2}$ and $\Delta\rho\simeq200$ kg.m$^{-3}$ for polystyrene beads in glycerol.

The repulsive double-layer force is given by \cite{Wu1996}:
\begin{multline}\label{eq:debye}
F_{Debye}=128\pi R k_B T c_{\infty}\kappa^{-1} \times \\
\tanh\left(\frac{ze\psi_w}{4k_B T}\right)\tanh\left(\frac{ze\psi_p}{4k_B T}\right)\exp(-\kappa h).
\end{multline} 

The lift force is computed from Eq.~(\ref{eq:Taba}), in which we substitute Eq.~(\ref{eq:GCB1}) for $V(h)$ and use the following result from  \citet{Goldman1967} in order to compute the angular velocity $\Omega (h)$:
\begin{equation}\label{eq:GCB2}
\Omega(h)\simeq \dot{\gamma}\frac{0.4218}{0.6376-0.2\ln \left(h/R \right)}.
\end{equation}
 
 We then solve Eq.~(\ref{eq:balance}) numerically for $h$, for a given set of parameters \{$R$, $\dot{\gamma}$, $T$, $c_{\infty}$, $\epsilon$, $z$, $D_1$, $D_2$, $\psi_p$, $\psi_w$\}. Quantitative comparison between predictions and observations is done by taking the values of $R$ and $\dot{\gamma}$ reported in the experimental studies, $T=300$ K, $\epsilon=43\epsilon_0$ for the permittivity of glycerol, and $z=1$ for monovalent salts. Diffusion coefficients of ionic species are estimated from their known values in water divided by the dynamic viscosity of the suspending fluid reported in the experimental studies, which leads to $D_1$ and $D_2\sim 10^{-12}$ m$^2$.s$^{-1}$ (see caption of Fig. \ref{fig:electro2} for detailed values). Once $D_1$ and $D_2$ are set, concentration $c_{\infty}$ is chosen in order to match the reported value of solution conductivity using $K=e^2c_{\infty}(D_1+D_2)/k_B T$. Finally, for the sake of simplicity we set $\psi_p=\psi_w=\psi$, and use $\psi$ as the only free parameter in the model. 
 
 Doing so, we find that the lift force derived by \citet{Tabatabaei2006} or  \citet{Schnitzer2016} allows us to quantitatively account for the various experimental observations, using sensible values for $\psi$ ranging from -30 mV to -45 mV. Such an agreement is illustrated on Fig. \ref{fig:electro2}.

 \subsection{Concluding remarks}
 
We have shown in the previous section that electrokinetics can indeed account quantitatively for the lift of a charged sphere flowing near a surface in a polar fluid. The recent theoretical work by Yariv {\it et al.}, revisiting the pioneer study of Cox, allows identifying the relevant mechanisms underlying the buildup of an electrokinetic lift force. It thus appears that the symmetry breaking of the linear Stokes flow in such problems is associated to the streaming potential that builds up when counterions in the Debye layer are swept by the flow. This potential gives rise to both a non linear Maxwell stress and to osmotic flows controlled by the non linear transport of charges in the vicinity of the flowing object, both contributing to the lift force, with weight depending on the Peclet number, {\it i.e.} on the relative importance of convection and diffusion of ions. 

 \begin{figure}[htbp]
	\centering 		
\includegraphics[width=\columnwidth]{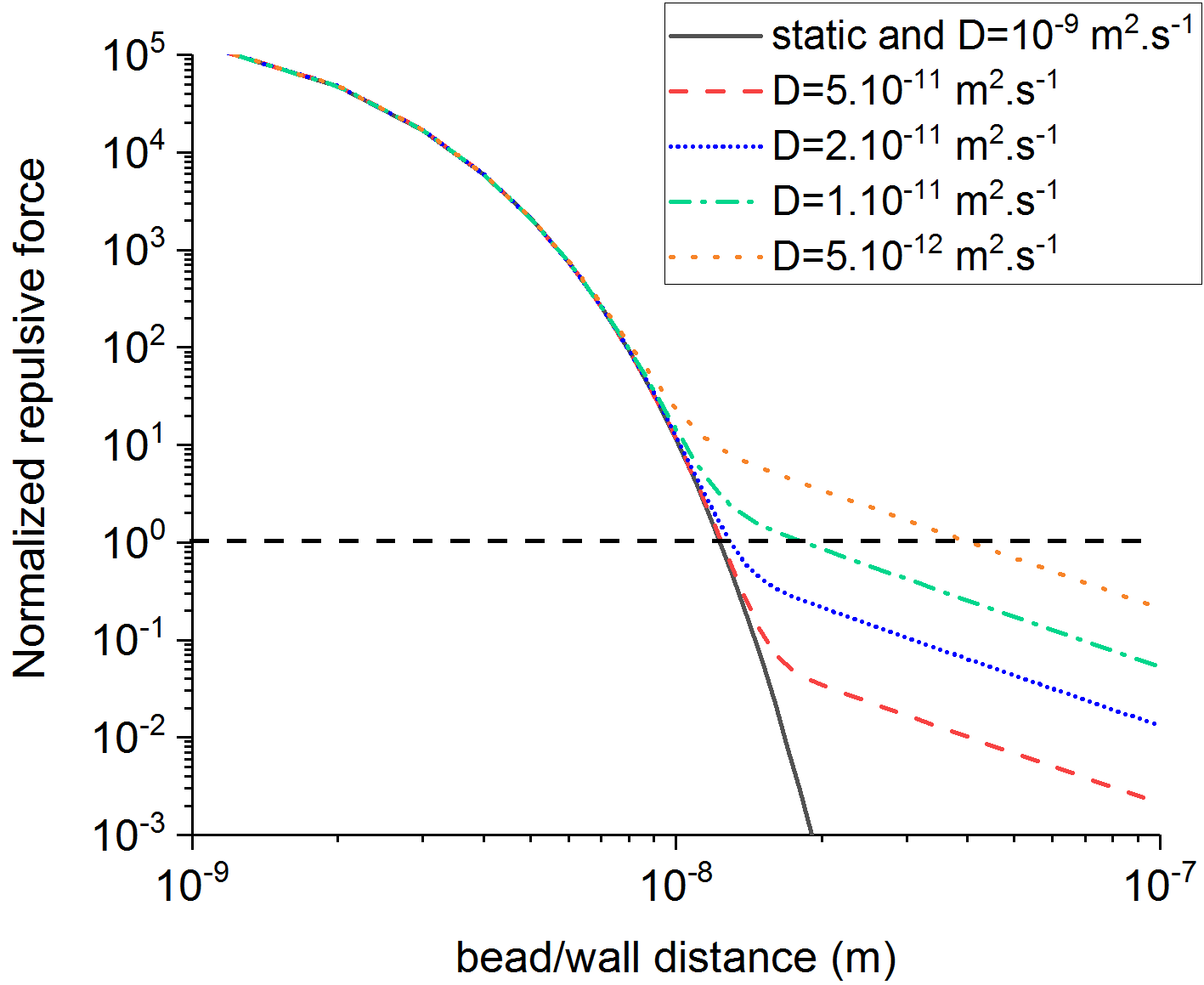}
	\caption{\textit{$(F_{Debye}+F_{lift})/F_{grav}$ as a function of $h$ for beads of $R=5\, \mu$m, $\dot{\gamma}=10$ s$^{-1}$, $C_{\infty}=0.1$ M (physiological range), for the range of diffusion coefficients indicated in the figure. Intersections of the curves with the horizontal line show the steady-state (or static equilibrium) value of $h$. No differences between static and $D=10^{-9}$ m$^2$.s$^{-1}$ (order of magnitude for sodium in water) are observed. Electrokinetic lift effects become sizeable only for  $D<10^{-11}$ m$^2$.s$^{-1}$.}} \label{fig:electro5}
\end{figure}

In the context of particle sorting, the study performed by Hollingsworth and Silebi directly points to the relevance of such flow-induced electrokinetic lift forces \cite{Hollingsworth1996}. The authors performed capillary hydrodynamic fractionation of submicron-sized particles suspended in low conductivity aqueous media, and showed  that a proper theoretical description of their measurements of separation factors required accounting for electrokinetic lift forces between the flowing beads and the walls of the capillary.
 
It is important to recall however that, in practice, such electrokinetic lift phenomena are of sizeable magnitude only in low conductivity fluids. This implies for example that in biological situations, at ionic strength $\simeq 150$ mM and $D\simeq 10^{-9}$ m$^2$.s$^{-1}$, electrokinetic lift of cell-sized objects is essentially not relevant. 
As an illustration of this, we have plotted on Fig. \ref{fig:electro5} a series of force/separation distance curves for a bead carried by a fluid containing 100 mM of monovalent salt, computed for various values of the ion diffusion constant. The steady-state distance between the bead and the wall can be read off the graphs as the point at which the normalized interaction force crosses the horizontal dashed line. It can thus be seen that deviations from the static equilibrium distance, due to electrokinetic lift, are observable only for diffusion coefficients below $10^{-11}$ m$^2$.s$^{-1}$. 

As a consequence, and as mentioned already in the introduction part of this section, fluidic applications exploiting electrokinetics for \textit{e.g.} particle separation/manipulation in aqueous medium do not rely on flow-induced electrokinetic effects but rather exploit non-inertial lift forces arising in the presence of an externally applied electric field. As initially pointed out theoretically by Young and Li \cite{Young2005} and Yariv \cite{Yariv2006}, the electrophoretic motion of a  spherical particle is affected by the presence of a nearby wall, due to local symmetry- and uniformity-breaking of the electric field. As a result, the bead experiences a net dielectrophoretic-like lift force ($F_{DEP}$), perpendicular to the applied electric field. This force has been shown theoretically to scale, in the limit where the bead/surface gap $h$ is large compared to the bead radius $R$, as $F_{DEP}\sim \epsilon R^6E^2/(R+h)^4$, where $E$ is the magnitude of the applied electric field \cite{Yariv2006,Yariv2016}. Such an electrically-driven lift force has been experimentally observed to be at play in flows induced by dc electric fields, for both micron- and sub-micron-sized beads \cite{Xuan2010,Yoda2011,Yoda2014,Lu2015,Liu2017b}. A semi-quantitative agreement have been obtained between such measurements and theoretical predictions accounting for $F_{DEP}$. Interestingly, it has been shown very recently, in an experimental study of the frequency-dependence of the electrokinetic lift under ac applied fields, that the dielectrophoretic force alone cannot fully account for the lift magnitude at low frequencies \cite{Fernandez2022}. The authors conclude that the dielectrophoretic lift force is responsible for high-frequency observations, but that it is dominated, at low frequencies,  by another phenomenon coined ``concentration polarization electro-osmosis'' (CPEO) \cite{Fernandez2021}. CPEO is associated with an electrically-induced quadrupolar stationary flow around the particle, where fluid is drawn to the particle in the field direction and expelled from the particle in the perpendicular direction, eventually yielding to hydrodynamic bead/wall repulsion \cite{Fernandez2022}. The structure of such flow, observed with dielectric particles, is analogous to that associated with ``induced charge electro-osmosis'' which has been described for conducting particles \cite{Bazant2010}. Overall, the above brief summary of electrically-driven lift shows a richness of phenomena that would deserve a review of their own, which is beyond the scope of the present article.

While we have, in this section, focused our attention on the generation of lift forces of electrokinetic origin when a bead moves parallel to a wall in a shear flow, it is worth mentioning that recent works have addressed, both experimentally and theoretically \cite{Mugele2018,Mugele2020,Rodriguez2022}, the issue of electrokinetic effects in squeeze-flow geometries, $i.e.$ when a bead moves perpendicularly close to a wall, and their role on the overall repulsion between the surfaces. Finally, it is of interest to note, in the framework of this review, that theoretical efforts have recently been made in order to provide a description of the combined effects of  electrokinetics and elastohydrodynamics in the emergence of lift forces \cite{Chakra2011,Chakra2017}.

\section{Conclusion and perspectives}

From the above review, we understand that there exist several mechanisms for lift forces at zero Reynolds number. They invariably involve viscous flows as well as soft or charged boundaries -- which are all widespread ingredients in the physics of transport at small scales. These mechanisms are thus highly-relevant to micro- and nanofluidics as well as for biological flows. In some cases, the magnitudes of these lift forces are comparable to surface and biological forces, and might thus have been overlooked in the interpretation of some results and phenomena. Besides, such effects might be controlled and employed in applications like \textit{e.g.} contactless rheology, optimized transport, drug delivery, cell filtering, confined chemistry, or even information processing within complex fluidic networks. In the remainder, we list a few elements of perspective. 

Soft-lubrication lift forces may be discovered to play some role in the fascinatingly-low and still-puzzling effective friction coefficients of mammalian cartilaginous joints, among other possible mechanisms~\cite{Jahn2018}. They might also allow for a smart tuning of the bulk and interfacial rheology of dense suspensions~\cite{Meeker2004}, including the shear-thickening effect. Indeed, if a lubricated-to-dry-contact transition~\cite{Wyart2014} is proved to be the microscopic mechanism of such a macroscopic manifestation, then the soft-lubrication lift between soft particles might repel/remove that transition. Besides, wall softness is expected to play a role too~\cite{rosti19}. More marginally, landslides are resulting from flows in poroelastic rocks and their mechanism remains a puzzle \cite{Campbell1989}. Elastohydrodynamic couplings may contribute there as well.

Also, as a symmetry-breaking mechanism is a central ingredient for the appearance of lift forces at zero Reynolds numbers, one could design in future new lift strategies independently of softness and charges. Slip inhomogeneities~\cite{Rinehart2020}, surfactant gradients \cite{Hanna2010,Pak2014} and thus Gibbs elasticity at capillary interfaces, as well as compressibility effects in gaseous layers, or moderate inertial contributions~\cite{Matas2004,Fouxon2020}, are possible examples among numerous others. Compound particles may also carry their own symmetry-breaking mechanisms and migrate transversally even in the absence of walls \cite{veerapaneni11,Liu2017}.

Beyond lift forces, other non-trivial EHD couplings have been revealed~\cite{Weekley2006,Urzay2010,Salez2015,Bertin2021,Noichl2022}, with important consequences including adhesive-like forces and enhanced sedimentation effects, among others. The experimental investigation of these scenarios is an important task for the future, and should enable the development of novel, efficient, contactless microrheological methods~\cite{Bar-Haim2017}. Similarly, looking for an active soft-lubrication lift~\cite{Trouilloud2008}, in addition to other active soft-lubrication couplings~\cite{Nambiar2022}, is an exciting perspective as \textit{e.g.} bacterial colony formation might be affected by it.

EHD couplings are effective ways to reduce or optimise frictional properties~\cite{Greenwood2020}. This has been investigated both experimentally and theoretically with rough or patterned substrates~\cite{Persson2009,Moyle2020,Hui2021} and might have important implications for soft robotics~\cite{Peng2021}. Moreover, prey capture by animals can be associated to lubrication through viscous adhesion~\cite{Brau2016}. A natural question emerges on if, and how, elasticity of the tongue/prey could play a role and modify the picture of the capture dynamics.

Finally, a route previously followed by solid-state physics, and then by hydrodynamics through nanofluidics and beyond, was to investigate the effects of system downscaling and hence the limits of the classical continuum description at small scales due to \textit{e.g.} surface forces, thermal fluctuations and eventually perhaps quantum effects. We expect a similar interest in investigating nanoconfined elastohydrodynamics, with fascinating perspectives for fundamental physics and biophysics. Eventually, lift at low Reynolds number might have key implications for life at low Reynolds number~\cite{Purcell1977}.

 \appendix
 \section{Boundary integral and stresslet formulation}

We present her with more  details the boundary integral formulation and the far-field approximation leading to the expression of the lift velocity as a function of the stresslet, as used in section  \ref{sec:extlift}.

We introduce the position vector $\mathbf{x}=(x_1,x_2,x_3)$, where $x_1$ corresponds to the flow direction, and $x_3$ to the direction perpendicular to the wall (located at $x_3=0$).

Following \cite{Pozrikidis92}, the flow field at any point $\mathbf{x}_0$ outside the particle reads \begin{eqnarray} u_j(\mathbf{x}_0)&=& u_j^\infty(\mathbf{x}_0)-\frac{1}{8 \pi\mu}\int_S \sigma_{ik}(\mathbf{x}) n_k(\mathbf{x}) G_{ij}(\mathbf{x},\mathbf{x}_0) dS  \nonumber \\
&+& \quad \frac{1}{8 \pi}\int_S u_{i}(\mathbf{x}) T_{ijk}(\mathbf{x},\mathbf{x}_0) n_k(\mathbf{x}) dS.
\label{eq:flowfield-appendix}\end{eqnarray}

Here, $\mathbf{u}^\infty$ is the imposed flow, $\mathbf{\sigma}$ is the fluid stress tensor such that  $\mathbf{f}^{ext}=\mathbf{\sigma}\cdot \mathbf{n}$  is the force distribution acting on the  surface. We recall  that $\mu$ is the viscosity of the fluid. $\mathbf{G}$ is the Green's function that is adapted to the boundary condition of the problem and $\mathbf{T}$ is the associated stress tensor. In order to account for the presence of body forces, $\mathbf{\sigma}$ can be replaced in the above expression by the modified stress tensor such that $\sigma^{MOD}_{ij}=\sigma_{ij}+\rho \mathbf{g}\cdot\mathbf{x}\delta_{ij}$ \cite{Pozrikidis92}. Here, $\mathbf{g}$ is the acceleration field, like gravity, and $\rho$ is the associated quantity, like fluid density. Keeping this in mind, we will drop the $MOD$ superscript from now on.

A more convenient expression can be obtained when one knows the specific mechanical properties of the particle boundary. A widely considered configuration is that of a 2D interface delimiting the interior of the particle, filled with a fluid of viscosity $\mu'\equiv \lambda \mu$ from the surrounding fluid. In that case, Eq. \eqref{eq:flowfield-appendix} becomes \cite{Pozrikidis92}

\begin{eqnarray} u_j(\mathbf{x}_0)&=& u_j^\infty(\mathbf{x}_0)-\frac{1}{8 \pi\mu}\int_S \Delta f_i(\mathbf{x}) G_{ij}(\mathbf{x},\mathbf{x}_0) dS  \nonumber \\
&+& \quad \frac{1-\lambda}{8 \pi}\int_S u_{i}(\mathbf{x}) T_{ijk}(\mathbf{x},\mathbf{x}_0) n_k(\mathbf{x}) dS. \nonumber \\ && 
\label{eq:flowfield-lambda-appendix}\end{eqnarray}

Here, $\Delta\mathbf{f}= \mathbf{f}^{ext}-\mathbf{f}^{int}=(\mathbf{\sigma}^{ext}-\mathbf{\sigma}^{int})\cdot\mathbf{n}$ is the discontinuity in the interfacial surface force. It can be written as    $\Delta\mathbf{f}= (\rho^{ext}-\rho^{in}) \mathbf{g}\cdot\mathbf{x} \,\mathbf{n} + \Delta \xi$, where $\Delta \xi$ is the discontinuity in the surface force that depends only on the interface mechanical properties. For a given model of particle (\textit{e.g.} a drop, a vesicle, a capsule), and in the absence of significant inertia of the membrane, it can be calculated according to the chosen constitutive law for the surface, as it must equal the opposite of the membrane load. 

For an unbounded domain, the Green's function is called the Stokeslet and describes the flow field created in $\mathbf{x}_0$ by a point force located in $\mathbf{x}$. We will denote it as $\mathbf{G}^\infty$ and it reads \begin{equation}
    G_{ij}^\infty(\mathbf{x},\mathbf{x}_0)=\frac{\delta_{ij}}{r}+\frac{r_i r_j}{r^3}, \quad \mbox{where $\mathbf{r}=\mathbf{x}_0-\mathbf{x}$}.\label{eq:freeGreen-appendix}
\end{equation}

The associated stress tensor is \begin{equation}
    T_{ijk}^\infty(\mathbf{x},\mathbf{x}_0)=-6\frac{r_i r_j r_k}{r^5}.\label{eq:freestress-appendix}
\end{equation}

The Green's functions we need here is that satisfying the no slip condition on the wall. \citet{blake71} proposed a calculation of this semi-infinite Green's function, using Fourier transform. It can be thought as the Green's function associated with other point singularities located at the reflection point  $\mathbf{x}^{IM}=(x_1,x_2,-x_3)$ of the initial force. 

The semi-infinite Green's function reads $\mathbf{G}=\mathbf{G}^\infty+\mathbf{G}^w$, where the wall Green's function $\mathbf{G}^w$ is given by \begin{eqnarray} G^w_{ij}(\mathbf{x},\mathbf{x}_0)&=&-G^\infty_{ij}(\mathbf{x}^{IM},\mathbf{x}_0)\\ &&- 2 x_3 G^{SD}_{ij}(\mathbf{x}^{IM},\mathbf{x}_0)+ 2 x_3^2  G_{ij}^{D}(\mathbf{x}^{IM},\mathbf{x}_0), \nonumber \label{eq:Gw-appendix}\end{eqnarray}where  \begin{equation}
 G^{SD}_{ij}(\mathbf{x},\mathbf{x}_0)=(1-2\delta_{j 3})\Big(\frac{\delta_{ij} r_3-\delta_{i3} r_j+\delta_{j 3} r_i}{r^3}-\frac{3 r_ir_jr_3}{r^5}\Big) \end{equation} 
 is a Green's function associated with a Stokeslet doublet  and 
 \begin{equation} G^D_{ij}(\mathbf{x},\mathbf{x}_0)=(1-2\delta_{j 3})\Big(\frac{\delta_{ij}} {r^3}-\frac{3 r_i r_j}{r^5}\Big) \end{equation} is a Green's function associated with a source doublet. By Green's function associated with a doublet, we mean the Green's function allowing for the calculation of the far-field velocity associated with a pair of singularities of opposite sign or direction located at a finite distance.
 
 Similar expressions exist for the stress tensor $T_{ijk}=T^\infty_{ijk}+T^w_{ijk}$, which can be found in \cite{Pozrikidis92}, p. 85.

Far from the wall, the velocity $U^w$ of the particle may be approximated by that of its center, that we set to be located at position $\mathbf{x}_0=(0,0,z)$. This far-field velocity $U^{w,ff}$ thus reads \begin{eqnarray} U^{w,ff} &=&-\frac{1}{8 \pi \mu}\int_S \Delta f_i(\mathbf{x}) G^w_{i3}(\mathbf{x},\mathbf{x}_0)\, dS \nonumber \\ &&+ \frac{1-\lambda}{8 \pi}\int_S u^0_{i}(\mathbf{x}) T^w_{i3k}(\mathbf{x},\mathbf{x}_0) n_k(\mathbf{x}) dS, \nonumber \\ & & \end{eqnarray}where $\mathbf{u}^0$ is the leading order term in the velocity on the particle surface.

$G^w_{i3}(\mathbf{x},\mathbf{x}_0)$ indeed represents the flow field created by the singularities from the image system, located  at $\mathbf{x}^{IM}=-\mathbf{x_0}$. For $|\mathbf{x}-\mathbf{x_0}|\ll R$, one can expand $G^w(\mathbf{x},\mathbf{x}_0)$ and  $T^w(\mathbf{x},\mathbf{x}_0)$ around $\mathbf{x}_0$, such that: 

\begin{eqnarray}  U^{w,ff} &=& -\frac{1}{8 \pi \mu} G^w_{i3}(\mathbf{x}_0,\mathbf{x}_0) \int_S \Delta f_i(\mathbf{x})\, dS   \nonumber\\
&& -\frac{1}{8 \pi \mu} \frac{\partial G^w_{i3}}{\partial x_k}(\mathbf{x}_0,\mathbf{x}_0) \int_S \Delta f_i(\mathbf{x})(x-x_0)_k \, dS\nonumber\\&&
+ \frac{1-\lambda}{8 \pi}  T^w_{i3k}(\mathbf{x}_0,\mathbf{x}_0) \int_S u^0_{i}(\mathbf{x}) n_k(\mathbf{x}) dS.\label{eq:flowfield1st-appendix}\end{eqnarray}

In the absence of external force (like gravity) the first term of the right hand side is zero.

The integral that appears in the second term is the dipolar tensor that characterizes the first moment of the force distribution on the particle surface. Depending on the authors, it is sometimes denoted as $D_{ik}$. We now turn to the usual decomposition of this tensor (see \textit{e.g.} \cite{yeomans2014}): \begin{equation} D_{ik}=\frac{1}{3} D_{jj} \delta_{ik} + S_{ik} + T_{ik}.\end{equation}.

The first term has no impact on the flow, as can be seen by inserting it in Eq. \eqref{eq:flowfield1st-appendix}: the resulting term is $\propto \partial G^w_{k3}/\partial x_k$, which is the divergence of the Green's function and is 0 (since this function represents a solution of the incompressible Stokes flow). The traceless symmetric tensor
\begin{strip}
\begin{equation} S_{ik}= \int_S \Big[\frac{1}{2}(\Delta f_i (x-x_0)_k +\Delta f_k (x-x_0)_i )-\frac{1}{3}\Delta f_j (x-x_0)_j  \delta _{ik}\Big] \, dS \label{eq:defstresslet-appendix}\end{equation} 
\end{strip}is often  called the stresslet and its asymmetric counterpart $T_{ik}$ is called the rotlet (or couplet, following \citet{batchelor1970}). The latter is proportional to the torque exerted on the particle and is therefore 0 in the absence of external torque. As the stresslet is symmetric, only the symmetrical part $\frac{1}{2}(\frac{\partial G^w_{i3}}{\partial x_k}+\frac{\partial G^w_{k3}}{\partial x_i})$ of the derivative of the Green's function eventually contributes to the lift velocity. Following \cite{nix14}, we call it $K^w_{i3k}$.

We now make the remark that $T^w_{i3k}(\mathbf{x}_0,\mathbf{x}_0)=-\delta_{ik}p_j(\mathbf{x}_0,\mathbf{x}_0)+ 2  K^w_{i3k}(\mathbf{x}_0,\mathbf{x}_0)$, where $\mathbf{p}$ is the  pressure vector associated with the Green's function \cite{Pozrikidis92}. Since the flux of $\mathbf{u}$ through $S$ is 0, its contribution to the lift is 0. As $K^w_{i3k}$ is symmetric, it will act only on the symmetrical part of the last integral of Eq. \eqref{eq:flowfield1st-appendix}.

Finally, in the absence of external force and torque, the lift velocity is given by the image system  of the  stresslet $\Sigma_{ik}$, acting on the center of the particle. It is given by

\begin{eqnarray}  U^{w,ff} &=& 
 -\frac{1}{8 \pi \mu} K^w_{i3k} (\mathbf{x}_0,\mathbf{x}_0)  \Sigma_{ik}, \label{eq:uw-1-stresslet}
\end{eqnarray}
where

\begin{equation}\Sigma_{ik} =S_{ik}+(\lambda-1) \mu \int_S (u^0_{i}(\mathbf{x}) n_k(\mathbf{x})+u^0_{k}(\mathbf{x}) n_i(\mathbf{x}) )\,dS. \label{eq:defstressletbis-appendix}\end{equation}

This last expression defines more generally the stresslet, for a larger class of particles than Eq. \eqref{eq:defstresslet-appendix}. It should be noted that the second term vanishes not only for particles with no viscosity contrast but also for rigid particles \cite{batchelor1970}.

An  expression for $K^w_{i3k}$ can be found in \cite{nix14}: \begin{equation}
   K^w_{i3k}=\frac{1}{8 z^2}(- 5 \delta_{ik}+ 9 \delta_{k3}\delta_{i 3} ).
\end{equation}
 
This leads to \begin{equation} U^{w,ff} = -\frac{9}{64 \pi \mu} \frac{\Sigma_{33}}{z^2}. \label{eq:Uwff-appendix}
 \end{equation}
 
 By coherence with the leading order approximation we made here, one must keep in mind that the stresslet $\Sigma_{33}$ is that created by the interaction with the external flow, in the absence of wall. 
 
\paragraph{Acknowledgments}
The authors thank Y. Amarouchene, N. Bain, O. B\"aumchen, V. Bertin,   M. Cloitre, D. D\'ebarre, C. Drummond,  E. Dufresne, M. Essink, A. Farutin, G. Ghigliotti, N. Fillot,  B. Kaoui, F. Lequeux,   S. Losserand,  A. Maali, L. Mahadevan, J. D. McGraw, S. Mendez, C. Misbah, A. Pandey, T. Podgorski, B. Rallabandi, B. Saintyves,  A.-V. Salsac, K. Sekimoto, J. H. Snoeijer, H. Stone, C. H. Venner, and    V. Vitkova  for fruitful discussions. They acknowledge financial support from the European Union through the European Research Council under ERC Consolidator grant n°101039103 \textit{EMetBrown}. Views and opinions expressed are however those of the authors only and do not necessarily reflect those of the European Union or the European Research Council. Neither the European Union nor the granting authority can be held responsible for them. They also acknowledge financial support from the Agence Nationale de la Recherche (ANR-21-ERCC-0010-01 \textit{EMetBrown}, ANR-21-CE06-0029 \textit{SOFTER}, ANR-21-CE06-0039 \textit{FRICOLAS}). Finally, they thank the Soft Matter Collaborative Research Unit, Frontier Research Center for Advanced Material and Life Science, Faculty of Advanced Life Science at Hokkaido University, Sapporo, Japan.

\paragraph{Author contribution statement}

 All authors contributed equally to the paper.
 
 \paragraph{Data availability statement}

This manuscript contains no data other than that extracted from the cited literature.
 

\end{document}